\documentclass[twocolumn, twocolappendix]{aastex631}

\usepackage[shortlabels]{enumitem}
\usepackage{bm}
\usepackage{pythonhighlight}

\shorttitle{The CCHP: JAGB Method}
\shortauthors{Lee et al.}
\graphicspath{{./}{}}

\begin{document}

\title{The Chicago-Carnegie Hubble Program:\\ 
The JWST J-region Asymptotic Giant Branch (JAGB) Extragalactic Distance Scale
\footnote{Based on observations made with the NASA/ESA/CSA James Webb Space Telescope and obtained from the Mikulski Archive for Space Telescopes at the Space Telescope Science Institute, which is operated by the Association of Universities for Research in Astronomy, Inc., under NASA contract NAS 5-03127. These observations are associated with program 1995.}}

\author[0000-0002-5865-0220]{Abigail~J.~Lee}\affil{Department of Astronomy \& Astrophysics, University of Chicago, 5640 South Ellis Avenue, Chicago, IL 60637}\affiliation{Kavli Institute for Cosmological Physics, University of Chicago,  5640 South Ellis Avenue, Chicago, IL 60637}

\author[0000-0003-3431-9135]{Wendy~L.~Freedman}\affil{Department of Astronomy \& Astrophysics, University of Chicago, 5640 South Ellis Avenue, Chicago, IL 60637}\affiliation{Kavli Institute for Cosmological Physics, University of Chicago,  5640 South Ellis Avenue, Chicago, IL 60637}

\author[0000-0002-1576-1676]{Barry~F.~Madore}\affil{Observatories of the Carnegie Institution for Science 813 Santa Barbara St., Pasadena, CA~91101}\affil{Department of Astronomy \& Astrophysics, University of Chicago, 5640 South Ellis Avenue, Chicago, IL 60637}

\author{In~Sung~Jang}\affil{Department of Astronomy \& Astrophysics, University of Chicago, 5640 South Ellis Avenue, Chicago, IL 60637}\affiliation{Kavli Institute for Cosmological Physics, University of Chicago,  5640 South Ellis Avenue, Chicago, IL 60637}

\author[0000-0003-3339-8820]{Kayla~A.~Owens}\affil{Department of Astronomy \& Astrophysics, University of Chicago, 5640 South Ellis Avenue, Chicago, IL 60637}\affiliation{Kavli Institute for Cosmological Physics, University of Chicago,  5640 South Ellis Avenue, Chicago, IL 60637}

\author[0000-0001-9664-0560]{Taylor~J.~Hoyt}
\affil{Lawrence Berkeley National Laboratory, 1 Cyclotron Rd, Berkeley, CA, 94720}

\begin{abstract}

The J-region asymptotic giant branch (JAGB) method is a new standard candle based on the constant luminosities of carbon-rich AGB stars in the J band. The JAGB method is independent of the Cepheid and TRGB distance indicators. Therefore, we can leverage it to both cross-check Cepheid and TRGB distances for systematic errors and use it to measure an independent local $H_0$.
The JAGB method also boasts a number of advantages in measuring distances relative to the TRGB and Cepheids, several of which are especially amplified when combined with JWST's revolutionary resolving power.
First, JAGB stars are 1 mag brighter in the NIR than the TRGB, and can be discovered from single-epoch NIR photometry unlike Cepheids which require congruent optical imaging in at least 12 epochs. Thus, JAGB stars can be used to measure significantly farther distances than both the TRGB stars and Cepheids using the same amount of observing time. Dust extinction is also reduced in near-infrared observations and JAGB stars are ubiquitous in all galaxies with intermediate-age populations.
We present a novel algorithm that identifies the optimal location in a galaxy for applying the JAGB method, so as to minimize effects from crowding.
We then deploy this algorithm in JWST NIRCam imaging of seven SN Ia host galaxies to measure their JAGB distances, undertaking a completely blind analysis.
In our CCHP $H_0$ results paper \cite{freedman24}, we apply the JAGB distances measured in this paper to the Carnegie Supernova Program (CSP) SNe Ia sample, measuring a Hubble constant of $H_0=67.80\pm 2.17 ~\rm{(stat)}\pm 1.64 ~\rm{(sys)}~\rm{km ~s^{-1}~Mpc^{-1}}$.
\end{abstract}

\correspondingauthor{Abigail J. Lee}\email{abbyl@uchicago.edu}


\section{Introduction} \label{sec:intro}
The Hubble constant ($H_0$) determines the size and age of the Universe, and therefore is one of the most important parameters in modern cosmology. 
However, in the last 10 years, a $5\sigma$ tension has arisen between  measurements of the Hubble constant derived from the early universe (the CMB) and the local Universe (extragalactic distance ladders). This disagreement, denominated as the `Hubble tension,' points to potential cracks in the standard $\Lambda$CDM cosmology, since this model provides the basis for the early Universe CMB measurement. 
In 2020, the Planck collaboration measured the most precise $H_0$ to date derived from CMB temperature and polarization anisotropy maps of $H_0=67.4\pm0.5 ~\rm{km ~s^{-1}~Mpc^{-1}}$ \citep{2020A&A...641A...6P}. 
On the other hand, the local measurement of $H_0=73.0\pm1.0 ~\rm{km ~s^{-1}~Mpc^{-1}}$ instead determined at substantially lower redshifts, was derived from a type Ia supernovae distance ladder calibrated by Cepheid variable stars \citep{2022ApJ...934L...7R}, measured by the Supernova $H_0$ for the Equation of State (SH0ES) collaboration. 

One extremely intriguing path to reconciling the Hubble tension 
involves fundamentally altering the standard $\Lambda$CDM model to bring the two measurements into agreement.
However, despite the surge of proposed theories to reconcile the early and local Universe measurements (e.g., early dark energy, an evolving dark energy equation of state; \citealt{2021CQGra..38o3001D}), all fail to simultaneously solve the Hubble tension as well as explain observables that are already well-modeled by the standard  $\Lambda$CDM model (e.g., baryonic acoustic oscillations, Big Bang nucleosynthesis, model fits to the CMB anisotropy spectra).

Another proposed solution is the existence of potentially unaccounted-for systematic errors in the local Cepheid distance ladder. Whereas the first `new physics' solution has yet to provide a robust resolution to the Hubble tension, a recent local measurement of $H_0$ independent from Cepheids instead was shown to agree at the $2\sigma$ level with the CMB measurement, opening potential questions into the accuracy of the Cepheid measurement. This $H_0=69.8\pm1.7 ~\rm{km ~s^{-1}~Mpc^{-1}}$, measured by the Chicago-Carnegie Hubble Program (CCHP), was based on the tip of the red giant branch (TRGB) \citep{2019ApJ...882...34F, 2021ApJ...919...16F} calibration of the Carnegie Supernova Project (CSP) SN Ia sample. 

In \cite{2019ApJ...882...34F}, the CCHP compared TRGB and Cepheid distance moduli to 10 SN Ia host galaxies, finding a weighted averaged difference of $\rm{TRGB}-\rm{Cepheid}=+0.059$~mag, meaning the TRGB distances were measured to be systematically farther on average than the Cepheid distances.   
These significant differences indicate that investigating the potentially nefarious systematics of the local distance scale are still surely warranted before fully committing to the `new physics' approach. Now, whether the Cepheid or TRGB distance ladder is more accurate in measuring distances is the primary question. 
One potentially powerful avenue to answer this is through utilizing a `tie-breaker' distance ladder, i.e. comparing the TRGB and Cepheid SN Ia calibrator galaxy distances with distances measured from a third, independent yet equally accurate and precise distance indicator. However, until recently, such a pertinent distance indicator has yet to exist.
That is, until the CCHP recently developed an incredibly promising new standard candle based on the constant luminosities of carbon-rich asymptotic giant branch (AGB) stars in the J band (1.2 microns), called the J-region asymptotic giant branch (JAGB) method. 

In two seminal papers in 2020, \cite{2020ApJ...899...66M, 2020arXiv200510793F} first proposed the JAGB method as a standard candle when they observed the J-band magnitudes of carbon stars were conveniently constant with zero color dependence, and consistent from galaxy to galaxy. \cite{2020ApJ...899...66M, 2020arXiv200510793F} then calibrated the JAGB method zeropoint using detached-eclipsing binaries in the LMC and SMC and then subsequently measured distances to 14 nearby galaxies, finding these distances agreed with TRGB distances at the 3\% level. 
In the ensuing years, the CCHP has continued to test the JAGB method in preparation for a future $H_0$ measurement with JWST. For example, we have extensively shown the JAGB method is equally as precise and accurate at measuring distances as the TRGB and Cepheid Leavitt law in nearby galaxies using ground-based data \citep{2020arXiv201204536L, 2022ApJ...933..201L, fourstar}. The JAGB method has also been extensively tested by several other groups (e.g., \citealt{2020MNRAS.495.2858R, 2021arXiv210502120Z, 2021MNRAS.501..933P, 2023MNRAS.tmp..926P, 2024ApJ...966...20L}).

Now with the recent successful launch of JWST \citep{2023PASP..135d8001R, 2023arXiv230404869G} and the operational success of NIRCam \citep{2023PASP..135b8001R}, we have entered a new chapter of unprecedented precision and accuracy in studies of resolved stellar populations and the extragalactic distance scale. NIRCam's superb resolution and sensitivity in the near infrared far surpass HST's IR capabilities; JWST's NIRCam (FWHM = 0.04 arcsec) has a sampling resolution four times better than HST's WFC3/IR camera (FWHM = 0.15 arcsec). 
Indeed, we have measured comparably precise JAGB distances from: 
JWST for galaxies 20~Mpc away and
HST for galaxies in the Local Group (and ground-based imaging for galaxies 50~kpc away) \citep{2023arXiv230502453L, 2023arXiv231202282L}.

The CCHP was fortunately awarded a JWST cycle 1 program (JWST GO 1995; PIs: W. L. Freedman, B. F. Madore), aimed at reducing the current systematics in the Cepheid, TRGB, and JAGB distance scales to provide our most accurate measurement of $H_0$ to date. This program, which measures JAGB, TRGB, and Cepheid distances to 10 SN Ia host galaxies, is described in the JWST CCHP $H_0$ results paper \citep{freedman24}.
The descriptions of our TRGB and Cepheid measurements of $H_0$ are also found in companion papers: \cite{hoyt25} and Owens et al. (in preparation).

Inter-comparing distances between these three distance indicators is a powerful cross-check test for systematics. 
Each will be affected independently by crowding, reddening, and metallicity, because of the fundamental differences in their measurement techniques and astrophysics. For example: 
\begin{enumerate}
    \item \textit{Stellar Populations.} JAGB stars are an intermediate-age population (300 Myr-1 Gyr), Cepheids are a younger, metal-rich population ($<300$~Myr), and RGB stars are an older, metal-poor population ($>4$~Gyr). 
    \item \textit{Spatial Distribution.} JAGB stars are (ideally) photometered in the outer, low-reddening disks of galaxies where intermediate-age populations are still abundant yet crowding and reddening are reduced. Cepheids can be found in the more crowded star-forming regions of galaxies, and RGB stars are (also ideally) photometered in the sparse stellar halo.
    \item \textit{Underlying physics.} And finally, the astrophysical mechanism by which the three are standard candles are completely independent. JAGB stars form via the third dredge-up in TP-AGB stars, Cepheid P-L relations result from mechanical pulsation cycles in the atmospheres of these stars, and the TRGB marks the helium flash which ignites the beginning of core helium burning for low mass red giants. 
\end{enumerate}

The outline of this paper is as follows. In Section \ref{sec:data}, we describe the target galaxies and JWST data. In Section \ref{sec:jagb} we review the JAGB method, its history, calibration, and application to our targets, and then provide a summary of the uncertainties in the method.  Finally, in Section \ref{sec:conclusion}, we present a summary of this paper and discuss future prospects for improving the JAGB distance scale in the upcoming decade.

\section{Data}\label{sec:data}
Observations were taken as part of the CCHP's JWST cycle 1 program \#1995 (PIs: W. Freedman, B. Madore), for which we imaged 11 galaxies (10 SN Ia hosts and 1 geometric anchor) from November 2022 to January 2024 with JWST's NIRCam. A more detailed description of our observing program can be found in \cite{hoyt24}. NIRCam is JWST's primary imager and can simultaneously observe in two channels, a short-wavelength (SW) channel and a long-wavelength (LW) channel, via a beam splitting dichroic which reflects the short wavelengths and transmits the long wavelengths.
We designated the short-wavelength filter to be F115W (J band equivalent, $1.15$~microns) because the JAGB method must be applied as a standard candle in the J band, where the magnitudes of the JAGB stars are constant with color \citep{2020ApJ...899...66M}.
For the long-wavelength filter, 
 we originally chose the reddest NIRCam long-wavelength filter F444W ($4.42$~microns), to create the largest possible color baseline for separating our target carbon-rich AGB stars from contaminant populations like oxygen-rich AGB stars via their colors. 
 However, after inspecting the images of the first two galaxies in our program, NGC 7250 and NGC 4536, we elected to switch our LW filter to F356W (3.56~microns) to take advantage of its increased angular resolution. 
We found the (F115W$-$F356W) color was as effective as the (F115W$-$F444W) color at separating oxygen-rich and carbon-rich AGB stars. We kept F444W as the LW filter in the two nearest galaxies, M101 and NGC 4258, to provide a metallicity test for our Cepheid program.\footnote{ The F444W filter contains a CO bandhead that is sensitive to Cepheid metallicity \citep{2016MNRAS.459.1170S}, so could theoretically be used to test for a radial Cepheid metallicity effect in M101 and NGC 4258. } 
 
\begin{figure*}
\gridline{
\fig{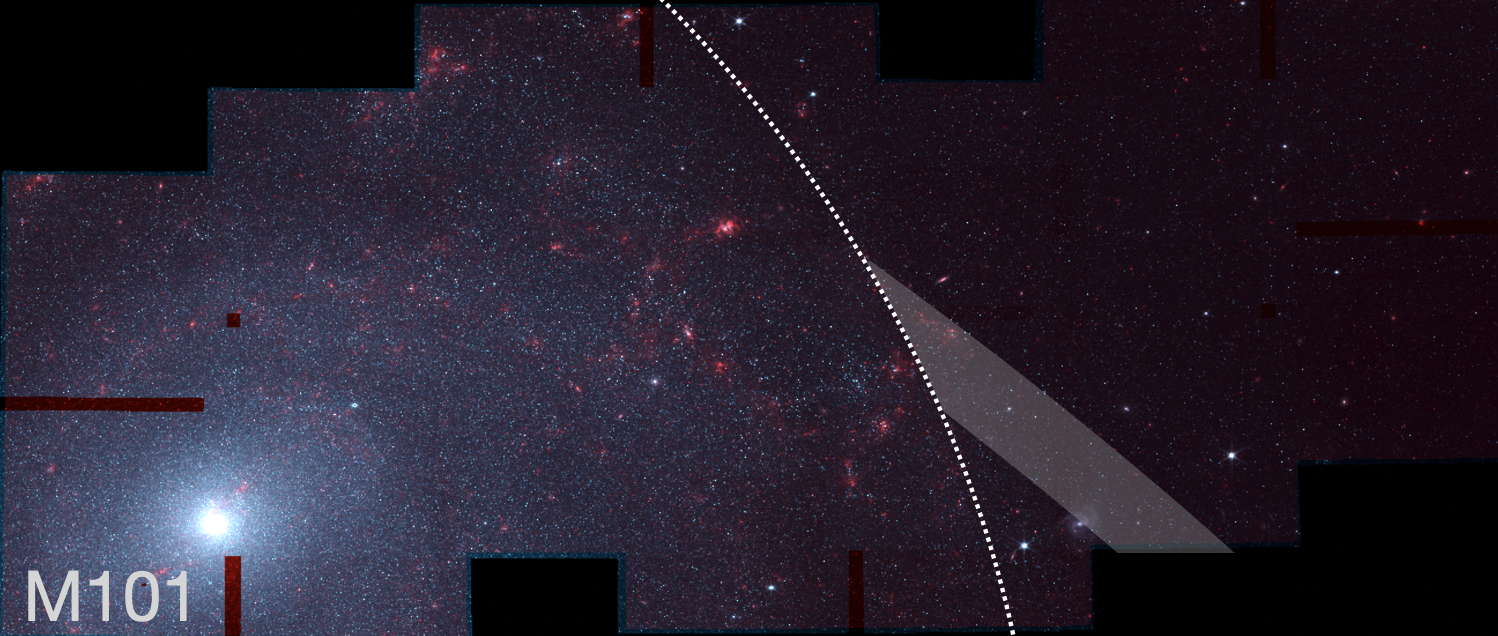}{0.34\textwidth}{}
\fig{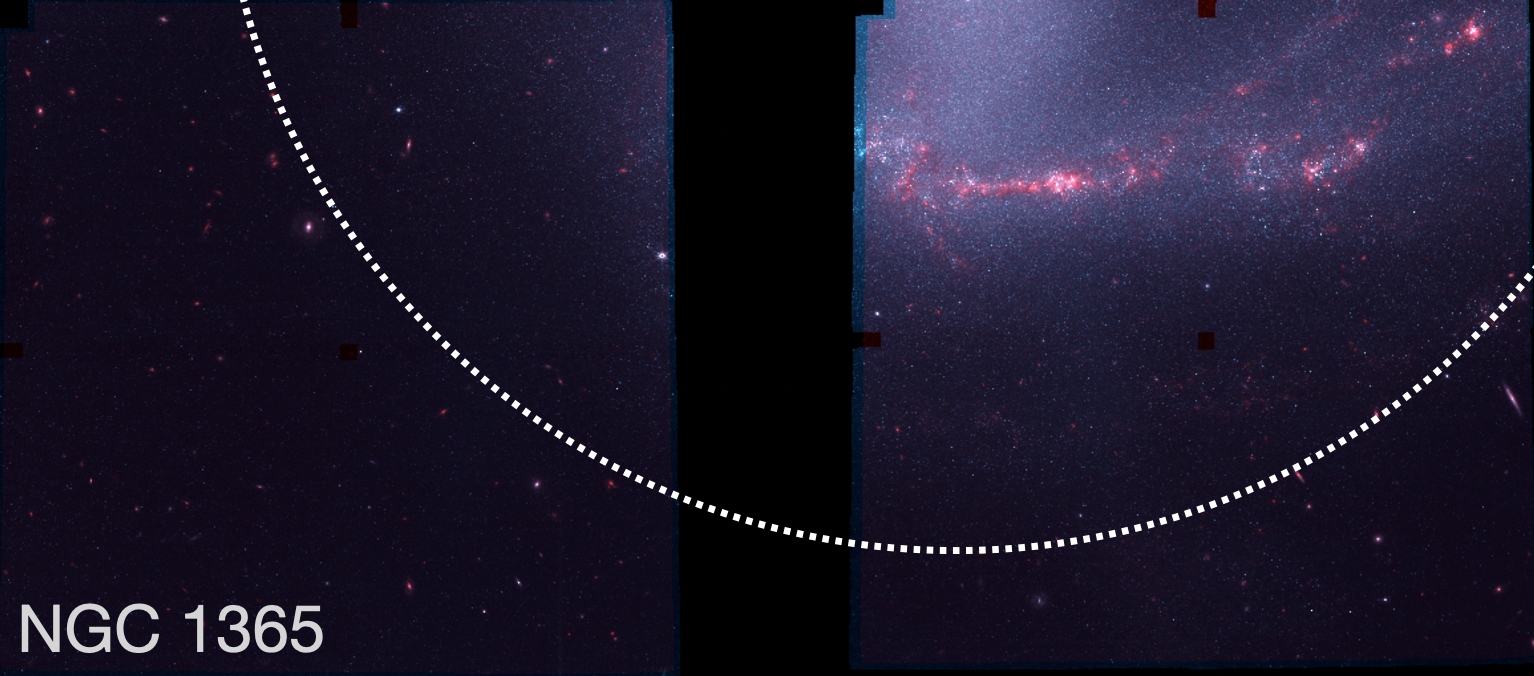}{0.33\textwidth}{} \fig{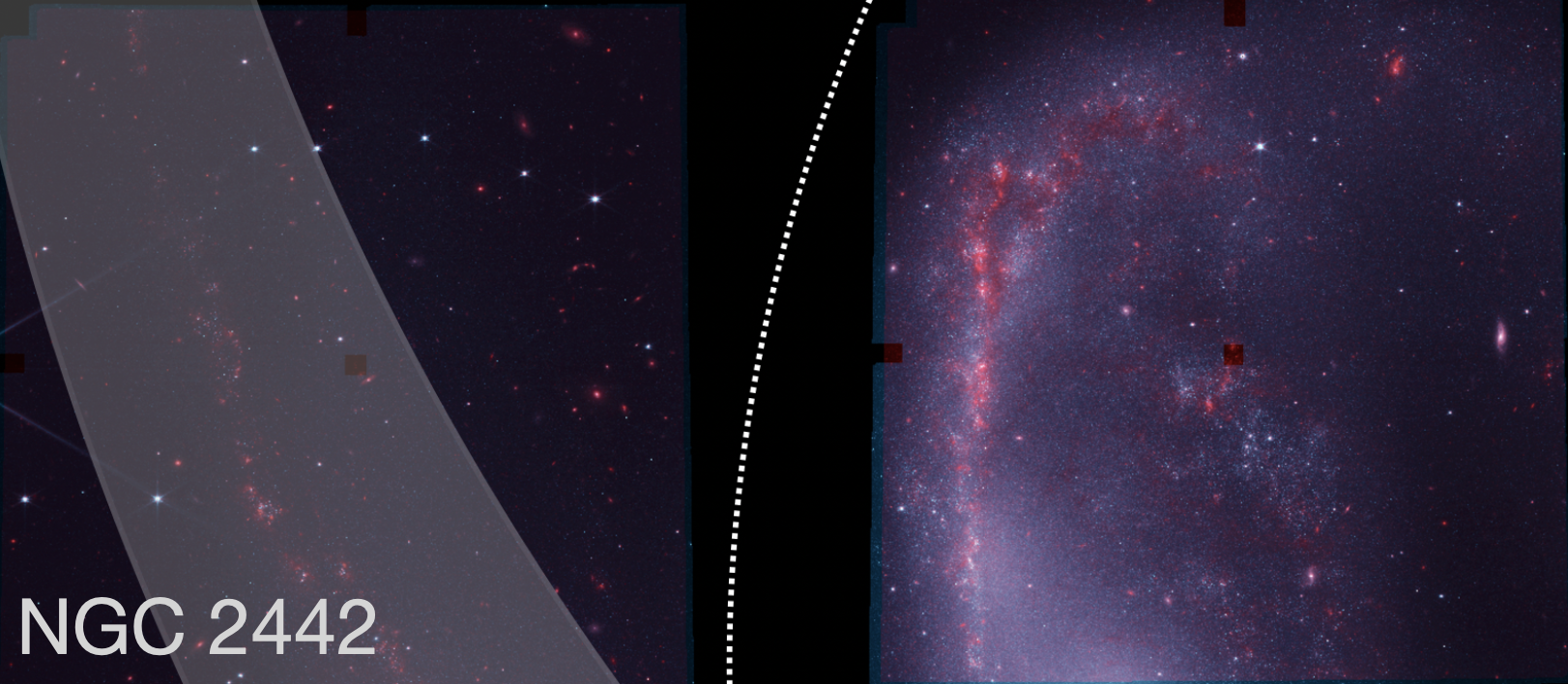}{0.33\textwidth}{}
}
\gridline{
\fig{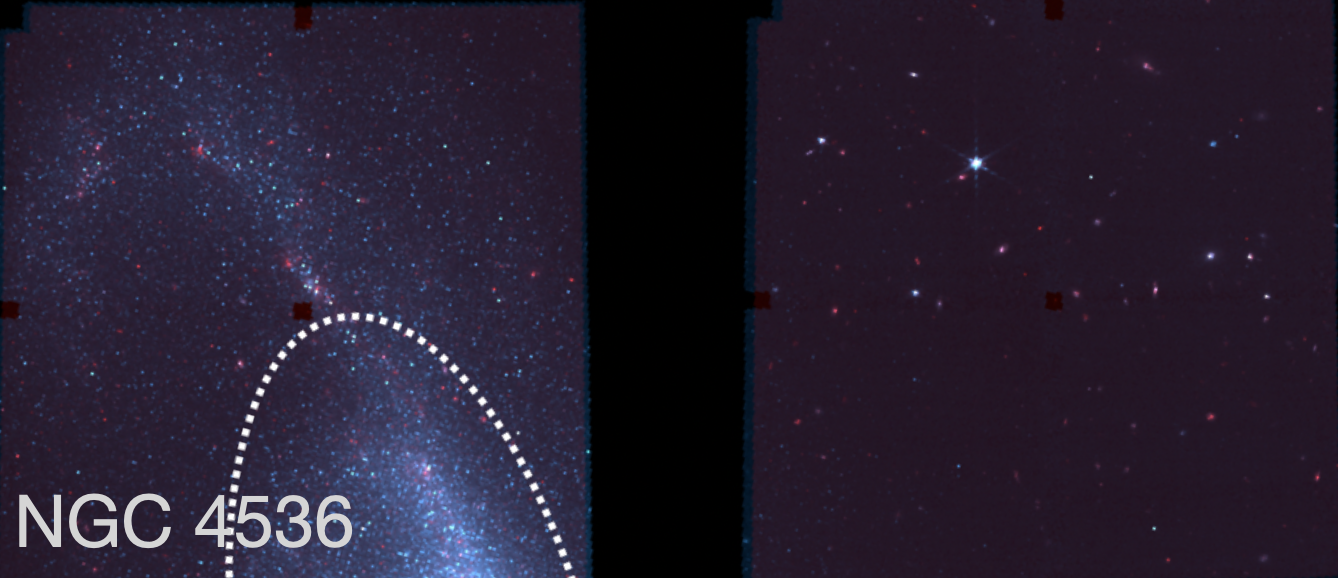}{0.33\textwidth}{}
\fig{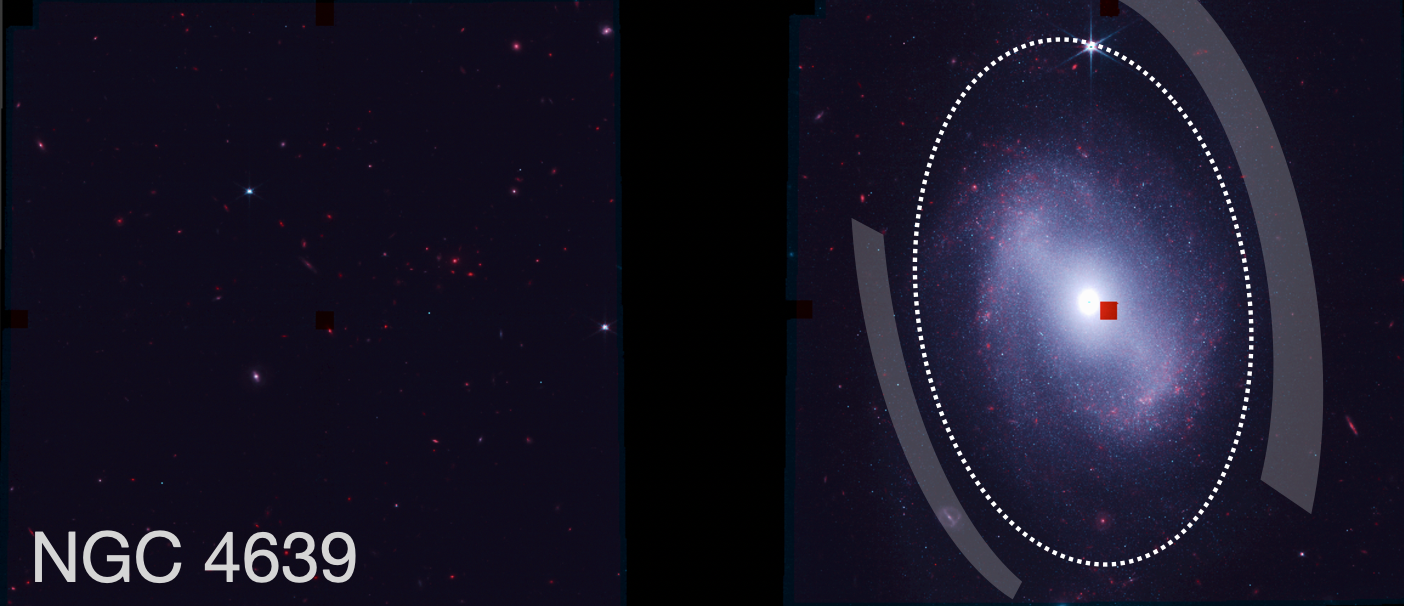}{0.33\textwidth}{}
\fig{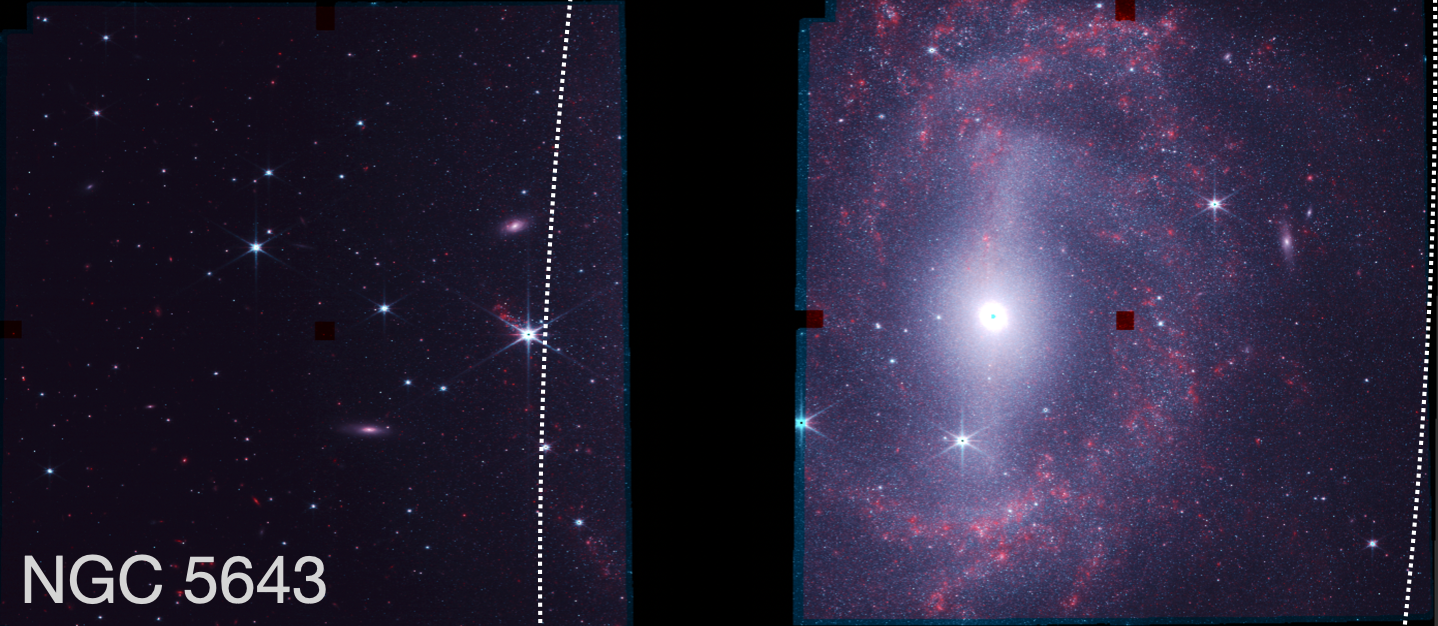}{0.33\textwidth}{}
}
\gridline{
\fig{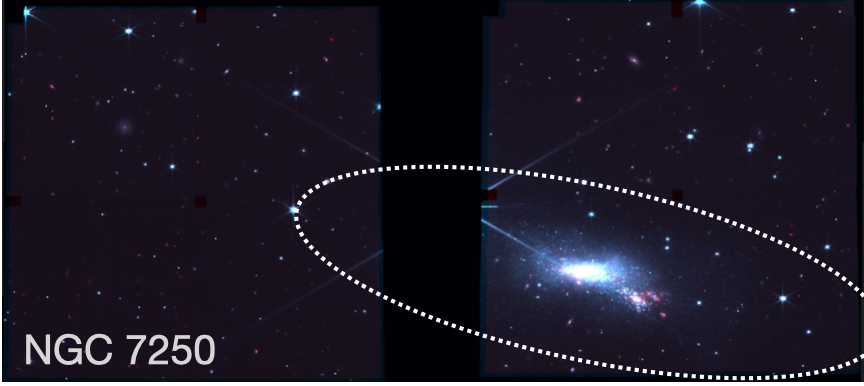}{0.33\textwidth}{}
}
\caption{NIRCam images of the seven SN Ia host galaxies studied in this work. The dotted line demarcates the `outer disk' and `inner disk'; only data outside of this line were used for the JAGB analysis. In NGC 2442, NGC 4639, and M101, the spiral arms were masked within the shaded grey regions.} \label{fig:images} 
\end{figure*}

 A montage of the seven SN Ia host galaxies studied in this work is shown in Figure \ref{fig:images}.

\subsection{Photometry}\label{subsec:phot}

The details of the CCHP's photometry procedure are described in Jang et al. (in prep). We briefly summarize them here.
First, we acquired all level 2b \texttt{*\_cal.fits} images from MAST.\footnote{Our images were processed from the following JWST pipeline software versions: \texttt{CAL\_VER = 1.11.4}, \texttt{CRDS\_VER = 11.17.2},  
\texttt{SDP\_VER = 2023\_2a}, and \texttt{CRDS\_CTX = jwst\_1149.pmap}.} 
Next, we extracted PSF photometry from the images using the NIRCam module of DOLPHOT v2.0 \citep{2000PASP..112.1383D, 2016ascl.soft08013D, 2023arXiv230104659W,2024arXiv240203504W}. We include the details of this procedure below. We also emphasize we performed this entire analysis blinded. That is, we performed the entire analysis without knowing the final measured distances. In practice, we added random photometric offsets to the photometry which were only removed after the analysis was finalized in Section \ref{subsubsec:spatial}. Our blinding procedure is described in detail in the CCHP $H_0$ results paper in \cite{freedman24}.

First, we performed initial photometry on the individual images to detect the bright stars. We used these stars to re-align all the F115W images, and then drizzled them to create an astrometric reference image for DOLPHOT. 
Within DOLPHOT, we utilized the \textit{warmstart} mode. Specifically, we extracted photometry solely from the SW F115W images, and then reduced  the LW images using the source list positions of the stars from the first run. We found this procedure subtracted stars more cleanly from the images than running DOLPHOT on the SW and LW images simultaneously. We note the SHoES team \citep{2023ApJ...956L..18R} and a JWST TRGB calibration program \citep{2024arXiv240603532N} independently also found the \textit{warmstart} mode resulted in the most cleanly subtracted residual images. All our photometry is on the Sirius-Vega calibration system.

The catalogs output from DOLPHOT were then culled of non-stellar sources (e.g., artifacts, cosmic rays, extended sources) using the quality-metric cuts chosen by the JWST Resolved Stellar Populations ERS team \citep{2023RNAAS...7...23W, 2023arXiv230104659W}. These quality-metric parameters prioritized sample purity over completeness, i.e. optimized for the removal of contaminants over retaining the largest number of stellar objects. We list these stellar quality cuts in Table \ref{tab:qualitycut}. The selection criteria were fulfilled in the long-wavelength and short-wavelength bands simultaneously. Here, we give brief descriptions of these DOLPHOT-returned parameters used to clean our photometric catalogs. The \textit{sharp} parameter represents how well the PSF model fit a star's flux, being zero for a perfectly-fit star, positive for objects where the flux is too concentrated (e.g., cosmic rays), and negative for objects were the flux is too spread out (e.g., extended objects like background galaxies). The \textit{crowd} parameter, which has units of mag, reports how much brighter the star would have been measured if nearby stars had not been fit simultaneously. A large \textit{crowd} value indicates the star was photometered in a crowded regions. The \textit{flag} parameter from DOLPHOT represents how well a star is recovered; the DOLPHOT manual recommends using values of 2 or less for precision photometry. Finally, the \textit{object type} parameter classifies objects based on their PSF fits; 1 denotes a `good star' and 2 denotes stars that are too faint for PSF determinations. Again, as recommended by the DOLPHOT manual, we kept \textit{object types} 1 and 2 in our catalogs. 

\begin{figure}
\centering
\includegraphics[width=\columnwidth]{"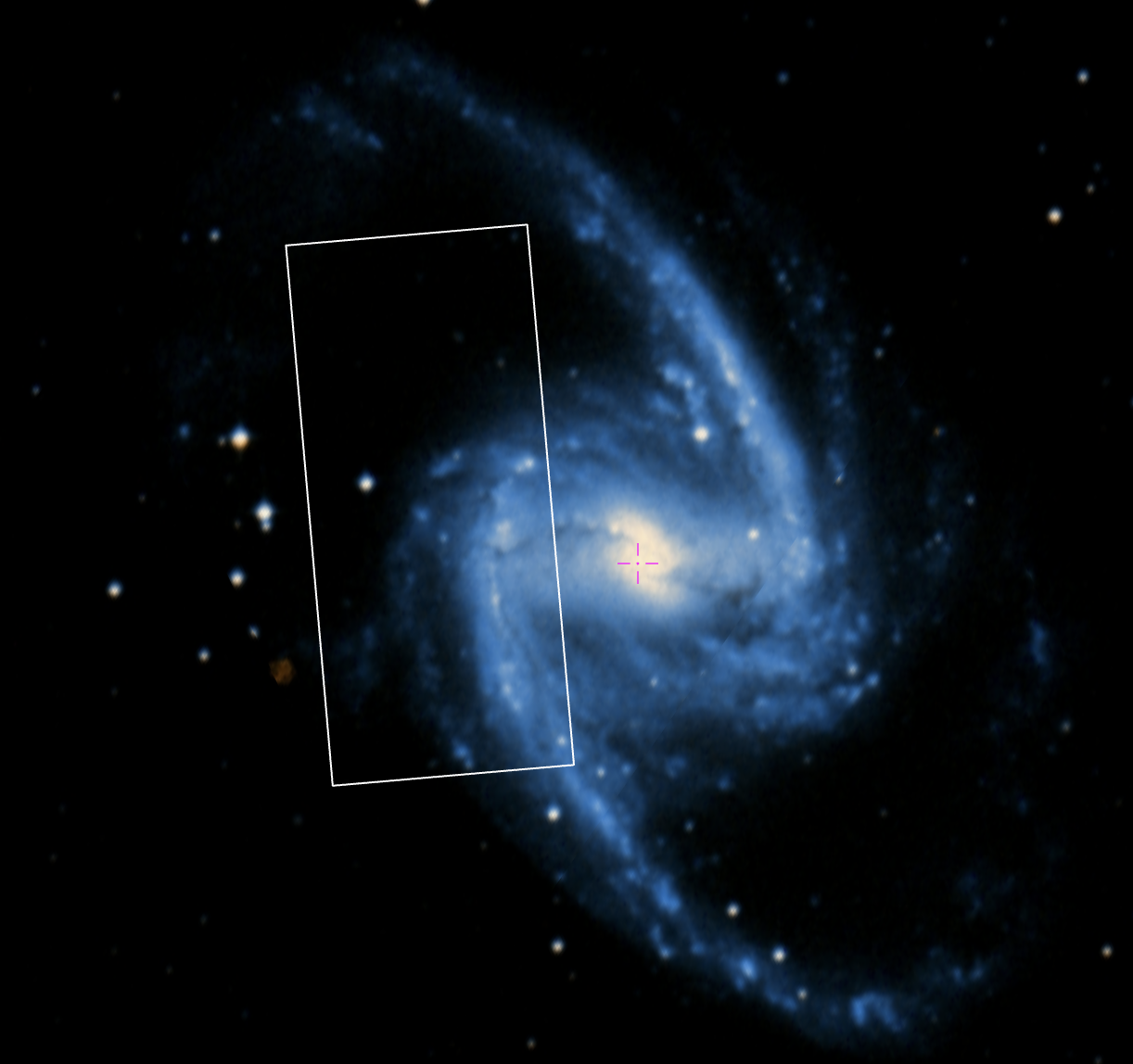"}
\caption{Our NIRCam pointing of NGC 1365, shown by the white rectangle, is parallel to the major axis of the galaxy. This hindered our ability to cleanly separate the JAGB stars in the inner disk and outer disk of this galaxy with an elliptical annulus calculated from the JAGB stars' deprojected galactocentric radii. We therefore opted to use a simple circular annulus for NGC 1365 to separate the inner and outer disk. }
\label{fig:point_n1365}
\end{figure}

Next, we calculated the deprojected galactocentric radius of each source in our catalogs via the host galaxy's central coordinate, position angle (P.A.), and inclination ($i$) values obtained from the Extragalactic Distance database \citep{2020ApJ...902..145K}. To convert from angular to physical radial distance, we utilized the SHoES Cepheid distances from \cite{2022ApJ...934L...7R}. 
These radial distances were eventually utilized to  seperate the JAGB stars into the `inner' and `outer' regions of each galaxy, shown by the ellipses in Figure \ref{fig:images}.
We note for NGC 1365, because our NIRCam pointing was parallel to the major axis of the galaxy, as shown in Figure \ref{fig:point_n1365}, an elliptical annulus (based on the deprojected galactocentric radial distances of the stars) would fail to cleanly separate the inner and outer disk of NGC 1365 like in our other galaxies.   
Therefore in this galaxy, the radial distances were calculated as the standard distance from the central coordinate instead of the deprojected distance (resulting in a circular instead of elliptical annulus).

\subsection{Galactic Foreground Extinction}
Finally, to correct for Galactic foreground extinction, we queried the \cite{1998ApJ...500..525S} full-sky Galactic $A_V$ dust map recalibrated by \cite{2011ApJ...737..103S}, from the online IRSA Galactic Dust Reddening and Extinction tool\footnote{\url{https://irsa.ipac.caltech.edu/applications/DUST/}}. 
Then, to convert from $A_V$ to $A_{F115W}$, we adopted $A_{F115W}/A_V=0.31943$, as computed by the PARSEC team in \cite{2019A&A...632A.105C}\footnote{\url{https://gitlab.com/cycyustc/ybc_tables/-/tree/master/rYBC/jwst_nircam_wide?ref_type=heads}} using the \cite{1989ApJ...345..245C} extinction law with $R_V=3.1$.

In Table \ref{tab:galaxysample}, we list the galaxies and their morphological types, NIRCam exposure times, foreground F115W extinctions, SN Ia names, P.A., and inclination values.

\begin{deluxetable}{cccccc}
\tablecaption{Quality-metric Criteria Used to Cull Our DOLPHOT Photometric Catalogs}\label{tab:qualitycut}
\tablehead{
\colhead{Band} & 
\colhead{S/N} & 
\colhead{$\rm{Sharp}^2$} & 
\colhead{Crowd} &
\colhead{Flag} & 
\colhead{Object Type}}
\startdata
F115W & $\ge4$ &$\le 0.01$ &$\le 0.5$ &$\le 2$ & $\le1$ \\
F356W &$\ge4$& $\le 0.01$&$\le 0.5$ &$\le 2$ & $\le1$\\
F444W &$\ge4$& $\le 0.01$&$\le 0.5$&$\le 2$ & $\le1$\\
\enddata
\tablecomments{Taken from \cite{2023arXiv230104659W}.}
\end{deluxetable}

\begin{deluxetable*}{ccccccc}
\tablecaption{JAGB Calibration Sample}\label{tab:galaxysample}
\tablehead{
\colhead{Galaxy} & 
\colhead{Type} & 
\colhead{Exposure time (s)} &
\colhead{$A_{F115W}$ (mag)\tablenotemark{a}} & 
\colhead{SN Ia Name}& 
\colhead{P.A. (deg)\tablenotemark{b}} & 
\colhead{$i$ (deg)\tablenotemark{b}}
}
\startdata
M101 &   SAB(rs)cd & 2802& 0.01  &SN 2011fe &  0.0 & 48 \\
NGC 1365 & SB(s)b & 3736 & 0.02  & SN 2012fr & \tablenotemark{\textasteriskcentered} & \tablenotemark{\textasteriskcentered}\\
NGC 2442 & SAB(s)bc pec & 2802 & 0.17  & 2015F & 23.2	& 81  \\
NGC 4536 &  SAB(rs)bc& 2802  & 0.02& SN 1981B & 118.5 & 69\\
NGC 4639 & SAB(rs)bc & 2802  & 0.02& 1990N  & 136.6 & 52 \\
NGC 5643 & 	SAB(rs)c & 2802& 0.14 &   2013aa, 2017cbv & 87.6 & 81\\
NGC 7250 & Sdm & 3769 &0.13 & 2013dy & 161.0 & 72\\
\hline
NGC 3972 & SA(s)bc & 3769& 0.01&  2011by & 117.2 & 78\\
NGC 4038 & SB(s)m pec & 2802 & 0.04 & 2007sr & 0.0 & 81\\
NGC 4424 & 	SB(s)a & 3769 & 0.02 & 2012cg & 95.3 & 69\\
\enddata
\tablerefs{(a) \citealt{1998ApJ...500..525S} (b)  Extragalactic Distance Database; \citealt{2020ApJ...902..145K}.  }
\tablenotetext{ \textasteriskcentered}{Because of the way our pointing was configured in NGC 1365, we calculated standard radial distances instead of deprojected radial distances, as explained in Section \ref{subsec:phot}.}
\tablecomments{The three galaxies below the line were discarded from the main JAGB calibration sample because their JAGB magnitudes never converged to a single value. This is described in detail in Section \ref{subsec:anom}.}
\end{deluxetable*}

\section{The JAGB Distance Scale}\label{sec:jagb}

\subsection{Description of JAGB Method's Theoretical Basis}

The AGB is the final nuclear-fusing stage of an intermediate-mass star's ($1-8M_{\odot}$) life. All AGB stars undergo alternating helium and hydrogen shell fusion in their interiors.
Because the energy from the helium shell fusion is too great to be transported through the star via radiation alone, convective cells form to compensate. 
These convective cells will also transport nuclear byproducts from the interior of the star onto the stellar surface. 
This process, called the \textit{third dredge-up} event, enriches the star's atmosphere with carbon created from the triple-$\alpha$ reaction in the helium shell, and forms molecules such as $\text{C}_{\text{2}}$ and CN.
The theoretical characterization of the third dredge-up event has recently improved significantly in the last decade due to the development of the \texttt{COLIBRI} stellar isochrones, the first models that have fully captured details in the TP-AGB evolution like convection, overshoot, hot bottom burning, mass loss, dredge-up, and pulsation \citep{2013MNRAS.434..488M, 2017ApJ...835...77M, 2019MNRAS.485.5666P, 2020MNRAS.498.3283P}. Excellent reviews on the third-dredge up event and formation of carbon stars can be found in \cite{2003agbs.conf.....H, 2017use..book.....L}.

For AGB stars with masses of $\approx 2-5M_{\odot}$, a sufficient number of dredge-up events will eventually cause the abundance of carbon to  exceed that of oxygen (C/O$>1$) on the stellar surface, and the star then transitions from an oxygen-rich to carbon-rich AGB star.  This mass range occurs because AGB stars less massive than $\sim 2M_{\odot}$ fail to evolve into carbon stars since they lose their entire stellar envelope after a few thermal pulses (as the mass of the envelope was so small to begin with), and therefore transition into planetary nebula before the conversion to carbon star can take place. Stars more massive than $\sim 5M_{\odot}$ are also thwarted from evolving into carbon stars because they undergo hot-bottom burning. Here, the carbon in the stars' interior is burned into nitrogen as it is transported throughout the star because the star is so massive and therefore extremely hot \citep{2003agbs.conf.....H, 2005ARA&A..43..435H,2013MNRAS.434..488M}. Therefore, only AGB stars with a narrow range of initial masses (and therefore luminosities) eventually evolve into carbon stars. Thus, the small range of NIR magnitudes observed for JAGB stars ($M_{JAGB} ~\pm~0.3$~mag) can be straightforwardly attributed to the astrophysics of AGB stars. Herein lies the foundational theoretical basis for carbon stars as standard candles.

\begin{figure}
\centering
\includegraphics[width=\columnwidth]{"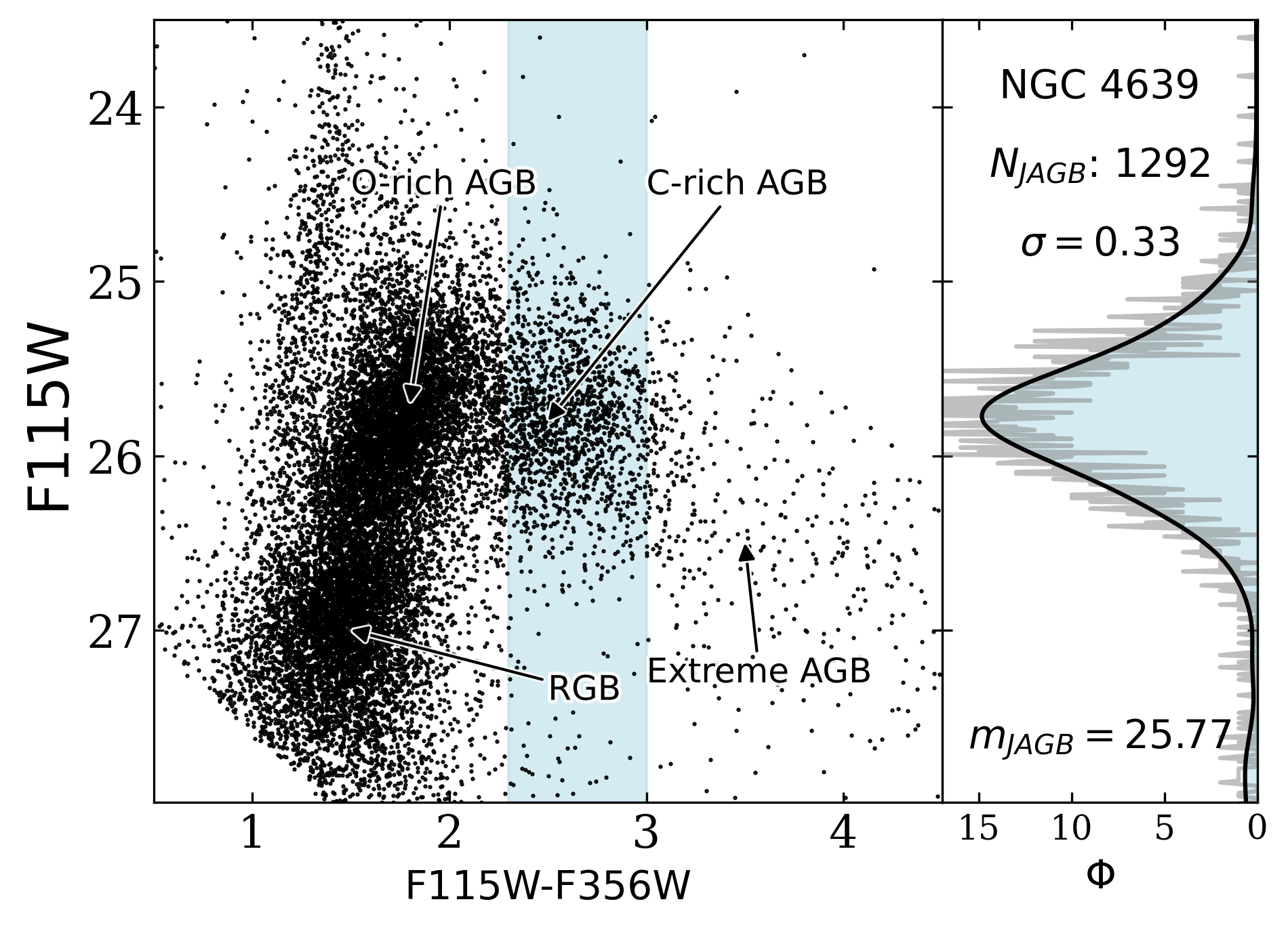"}
\caption{(Left panel) Color-magnitude diagram of the SN Ia host galaxy NGC 4639. The JAGB stars were selected within the light blue shaded region. (Right panel) GLOESS-smoothed JAGB star luminosity function in black overlaid on top of the binned luminosity function in grey. The number of JAGB stars within $\pm0.75$~mag of the mode is plotted in the upper right corner, as well as the dispersion for those stars about the mode. 
The measured JAGB magnitude is also shown in the bottom right. Different stellar populations are labeled.} 
\label{fig:rep_CMD}
\end{figure}

Carbon stars can also be easily photometrically distinguished from other stellar populations because the $\rm{C_2}$ and CN molecules in their atmospheres increase their opacity in typical photometric bandpasses. Therefore, carbon stars have cooler effective temperatures (and therefore much redder colors) relative to their oxygen-rich predecessors \citep{2003A&A...403..225M}. 
We can thus separate carbon stars from oxygen-rich AGB stars solely via their near-infrared colors. In Figure \ref{fig:rep_CMD}, we show a NIR CMD of one of the SN Ia host galaxies from our program, NGC 4639. The carbon stars can be cleanly delineated from other stellar populations via their colors alone. Furthermore, contamination from competing stellar populations is almost negligible, because JAGB stars are the brightest and reddest stellar populations in a given galaxy ($M_{JAGB}=-6.2$~mag in the ground-based J band). Only background galaxies lie in the same color-magnitude space as JAGB stars, but these can be straightforwardly eliminated using sharpness cuts (see Section \ref{subsec:phot}). 

\subsection{Historical Background to the JAGB method}

Carbon stars were proposed as distance indicators almost 50 years ago by \cite{1986ApJ...305..634C}, after they noted similarities in the I-band luminosity functions of carbon stars in the Local Group. 
15 years later, \cite{2000ApJ...542..804N, 2001ApJ...548..712W} first used carbon stars as standardizable candles in the near-infrared Ks band by calibrating the slope and intercept of the Ks magnitude vs. (J$-$Ks) color relation of carbon stars. 
They then applied this calibration to successfully map the back-to-front geometry of the LMC. They called this population of color-selected carbon stars the `J-region;' hence the origin of the \textit{J-region} AGB  name (therefore note the `J' in J-AGB does not stand for `J-type' carbon star or `J band'). 
Twenty years later, \cite{2020ApJ...899...66M, 2020arXiv200510793F} realized carbon stars have a constant magnitude in the J band. They used the \textit{mean} J-band magnitude of carbon stars to measure distances to 14 nearby galaxies. 
Next, \cite{2022ApJ...933..201L} first suggested using the \textit{mode} of the JAGB LF instead of the mean, because the mode is more robust to fainter contaminant populations in the carbon star luminosity function such as background galaxies.
Using the modal magnitude, \cite{fourstar} then measured and compared JAGB and TRGB distances to 11 galaxies, finding an inter-method scatter of $\pm0.07$~mag.
This confirmed the mode of the JAGB LF was an accurate standard candle relative to the TRGB at the 3\% level.

\subsection{Advantages and Disadvantages of the JAGB Method}
Below, we enumerate the many advantages of the JAGB method for measuring distances.

\begin{enumerate}

    \item JAGB stars are easily identified solely from their near-infrared photometric colors, as the brightest population of reddest stars in a galaxy.
    \item Measuring JAGB distances requires only one epoch of observations, unlike measuring Cepheid distances which requires at least a dozen epochs to extract the Cepheid  periods, amplitudes, and mean magnitudes.
    \item Utilizing near-infrared observations reduces effects of dust extinction. For comparison, extinction in the optical I band is 2 times larger than extinction in the J band \citep{1989ApJ...345..245C,2005ApJ...619..931I}. Reddening in optical observations may therefore introduce larger systematics in distances measured from the Cepheid P-L relation and I-band TRGB.
    \item In the J band, JAGB stars ($M_J=-6.2$~mag) are about 1~mag brighter than the TRGB ($M_J\approx-5.1$~mag) and about the same brightness as a 25-day Cepheid. However, because Cepheids need to first be discovered in optical wavelengths from their amplitudes, and 12 phase points are needed to measure their periods, less total observing time is required for the JAGB method than Cepheids to measure comparable distances.
    \item JAGB stars can be used to measure distances to all galaxies with intermediate-age populations and therefore the method can be applied to a wide range of galaxy types, unlike Cepheids which can only be found in late-type spiral galaxies with low inclinations.
    \item The JAGB method is capable of delivering incredibly statistically precise distances. The observed dispersion about the modal JAGB magnitude is $\pm0.3$~mag. The error on the JAGB magnitude therefore decreases as $0.3/\sqrt{N}$~mag, where $N$ is the number of JAGB stars in the galaxy. Therefore, a sample of 500 JAGB stars delivers a JAGB magnitude with a corresponding error on the mode of 0.01~mag. For reference, spiral galaxies typically contain thousands of JAGB stars. 
    
    In Table \ref{tab:galaxydistances}, we list 
     the total number of JAGB stars in the outer disk of each galaxy. We emphasize that with at least $\sim 1000$ JAGB stars contributing to the final measured distance, the statistical precision of the JAGB method is unchallenged by both the Cepheids and the TRGB (with typically $\sim100$ Cepheid stars and RGB stars contributing to their measurements). 

    \item \cite{fourstar} measured JAGB and TRGB distance to 11 galaxies from the same imaging. The residuals obtained from subtracting the distance moduli from the two methods yielded an rms scatter of $\sigma_{\rm{JAGB-TRGB}} = \pm0.07$~mag. Therefore, all systematics in the JAGB method and TRGB method (e.g., crowding, differential reddening, star formation histories) must have been contained within these $\pm0.07$ mag bounds for this sample of galaxies, because the JAGB and TRGB distance indicators are drawn from entirely distinct stellar populations and are thus affected by these systematics independently. This small scatter suggests that each of these two methods can individually provide distances that are statistically good to 2\% or better.  Note a similar test by \cite{2021arXiv210502120Z} for a JAGB-Cepheid distance comparison delivered a similar $\pm0.09$~mag scatter. In conclusion, any systematics in the JAGB method resulting from different SFHs, internal reddenings, and metallicities have been shown to be constrained at the 2-3\% level or less.
\end{enumerate}

A possible limitation of the JAGB method is that theoretically, metallicity is predicted to influence the JAGB stars' magnitudes. 
Theory predicts the minimum mass for carbon star formation is larger at higher metallicity, because the third dredge-up event's efficiency decreases with increasing metallicity \citep{2020MNRAS.498.3283P}. Furthermore, carbon stars in metal-rich environments contain more oxygen in their atmospheres, which will preferentially bind with the dredged-up carbon to form CO instead of carbon molecules such as CN or $\rm{C_2}$, thereby also hindering the carbon-enhancement of these stars' atmospheres. Thus, the mode of the JAGB LF metallicity is theorized to be brighter in metal-rich environments. \citep{2020MNRAS.498.3283P}

Empirically, however, definite consensus has yet to be reached on whether metallicity has a significant effect on the shape or mode of the JAGB LF. 
\cite{2020arXiv200510793F} compared the JAGB magnitude to the [Fe/H] metallicity of 12 host galaxies, finding a statistically insignificant correlation of $-0.03\pm0.04 ~\rm{mag~dex^{-1}}$.
On the other hand, \cite{2021MNRAS.501..933P} speculated metallicity may affect the skew of the JAGB LF. They found the LFs of the two higher-metallicity galaxies in their sample, the LMC and NGC 6822, exhibited more skew than the two lower-metallicity galaxies in their sample, the SMC and IC 1613. 
However, they also noted that definitively constraining the effect of metallicity on the JAGB LF  would require homogeneous [Fe/H] parameters for each galaxy in their sample.
In a follow up paper, \cite{2023MNRAS.tmp..926P} expanded their sample from four to six galaxies, adding NGC 3109 and WLM. They found the JAGB LFs in NGC 3109 and WLM to be symmetric, like the JAGB LFs in the SMC and IC 1613. 

We now test for a metallicity effect in their sample by utilizing the recent homogeneously-analyzed C/M star ratios provided by
\cite{2022Univ....8..465R} as a metallicity probe of the AGB stars, where a higher C/M ratio indicates a lower metallicity.\footnote{The ratio of C-type to M-type AGB stars is the most direct probe of AGB star metallicities, because fewer C-type AGB stars are expected to form in metal-rich environments. } 
The C/M ratios of five of six of these galaxies in the \cite{2023MNRAS.tmp..926P} sample were measured by \cite{2022Univ....8..465R} and are listed here in ascending order (and therefore in descending order of metallicity):  LMC (highest metallicity), SMC, IC 1613, NGC 6822, WLM (lowest metallicity). Here, we see that the skew in the JAGB LFs of LMC and NGC 6822, and likewise symmetry in the JAGB LFs of WLM, IC 1613, and the SMC cannot be straightforwardly explained by a metallicity effect on the magnitudes of the JAGB stars. Furthermore, recent tests by \cite{fourstar} indicate any systematic errors incurred due to metallicity are constrained at least at the 3\% level, by finding excellent agreement between JAGB and TRGB distances to galaxies with a wide range of metallicities. \cite{2023arXiv230502453L} also directly tested for a JAGB metallicity dependence in M31, by comparing the JAGB magnitude to the average [M/H] metallicity in different spatial regions of M31's disk, finding zero effect. We are currently undertaking the same test in the lower-metallicity galaxy M33 (Lee, in preparation), to continue to empirically test for and constrain a metallicity effect on the JAGB method. 

The second limitation of the JAGB method is that JAGB distances may incur significant systematic errors when measured in the high-surface-brightness regions (e.g., inner disks) of galaxies.
For example, the JAGB magnitude is systematically brighter in the inner regions of the galaxies studied in this paper (see Section \ref{subsubsec:spatial}), likely due to crowding effects. 
However, this effect can be mitigated by solely applying the JAGB method in the outer disks and halos of galaxies, where crowding effects are minimized, as shown by \cite{2022ApJ...933..201L, 2023arXiv230502453L, 2023arXiv231202282L}. 
In Section \ref{subsubsec:spatial}, we describe in detail our methodology for selecting the suitable `outer disk' region of a galaxy for measuring JAGB distances.

\subsection{Measuring JAGB distances to SN Ia Host Galaxies}

Here, we summarize the CCHP's procedure for measuring JAGB distances.
First, we color selected the JAGB stars. For galaxies with F444W data, JAGB stars were selected as having colors between $2.4<(F115W-F444W)<3.2$~mag. 
For galaxies with F356W data, JAGB stars were selected as having colors between $2.3<(F115W-F356W)<3.0$~mag. 
Then, we binned the F115W magnitudes of the JAGB stars using bins of 0.01~mag. Next, we smoothed the binned luminosity function using a nonparametric interpolation technique: the Gaussian-windowed, Locally Weighted Scatterplot Smoothing (GLOESS) algorithm \citep{loess_ref, loader, 2004AJ....128.2239P}. The GLOESS smoothing technique is effective at suppressing false (noise-induced) edges, and has been used in several astrophysical contexts like smoothing variable-star light curves (e.g., \citealt{2004AJ....128.2239P}) and for smoothing the RGB LF for measuring the TRGB (e.g., \citealt{2019ApJ...882...34F}). The only user input is the smoothing parameter $\sigma_s$.
For all galaxies, we used a smoothing parameter of $\sigma_s=0.25$~mag.  In Section \ref{subsubsec:statunc}, we describe how we incorporated an uncertainty due to our choice of smoothing parameter. The peak location (mode) of the smoothed luminosity function then marks the JAGB magnitude. 

In the follow section, we describe our algorithm for selecting the suitable region of a galaxy for the JAGB measurement.

\subsubsection{Choice of Spatial Selection for the JAGB method}\label{subsubsec:spatial}

The JAGB method is optimally applied in the outer disks and halos of galaxies where plentiful numbers of carbon stars exist \citep{2003agbs.conf.....H}, yet where systematic effects from crowding, blending, and reddening are also minimized. These systematics can manifest themselves through the shape of the JAGB LF. 
For example, the JAGB LF exhibits a distinct peak location and Gaussian shape in low-reddening, uncrowded regions of a galaxy. 
On the other hand, the JAGB LF lacks a clear peak and/or can appear asymmetric in reddened, crowded regions. This phenomenon was first observed in the galaxy M33 by \cite{2022ApJ...933..201L}, who split the photometry of M33 into four concentric radially-separated regions, finding the JAGB LFs in the outer two regions (outer disk and halo region) had symmetric Gaussian shapes with modes that agreed to within 0.01~mag. On the other hand, the JAGB LFs in the two inner regions were asymmetric with modes that varied by up to 0.7~mag compared with the outer regions. The photometry in the inner two regions were then discarded for the final JAGB distance measurement.
\cite{2023arXiv230502453L} has also shown the dispersions of the JAGB LFs in the lowest-reddening regions of M31's disk were on average 0.1~mag smaller relative to the dispersions of the JAGB LFs measured in the highest-reddening regions of M31's disk.

In \cite{2023arXiv231202282L}, we began developing a methodology for systematically selecting the optimal `outer disk' regions of galaxies for JAGB measurements, to minimize systematic effects resulting from crowding and reddening.  We used data from the first three galaxies in our JWST CCHP sample: NGC 7250, NGC 4536, and NGC 
3972 to develop this methodology. We first split the photometry into eight radially-separated regions, and measured the JAGB magnitude in each bin.
In all three galaxies, we observed the same pattern: the JAGB magnitude was brightest in the inner regions and then grew fainter as a function of radial distance before eventually converging to a constant magnitude in the outer regions. 
The outer regions that agreed in the mode to within 0.05~mag (2\% in distance) with the outermost eighth bin were then aggregated to create the final `outer disk' JAGB sample.\footnote{Note in this paper, we used $m_{JAGB}$ instead of $\Delta m_{JAGB}$ (the change from a fiducial mode) which was used in \cite{2023arXiv231202282L} to preserve the blinding of the photometry in that paper. We also note \cite{2023arXiv231202282L} compared the JAGB magnitude versus the average sky parameter returned from DOLPHOT instead just the radial distance used in this study. The choice of x-axis (e.g., average sky value vs. radial distance) is irrelevant to the final choice of radial cut. Ultimately, we decided using radial distance instead of average sky parameter as the x-axis choice was a clearer visual representation of the `convergence' pattern.} 

\begin{figure*}
\gridline{\fig{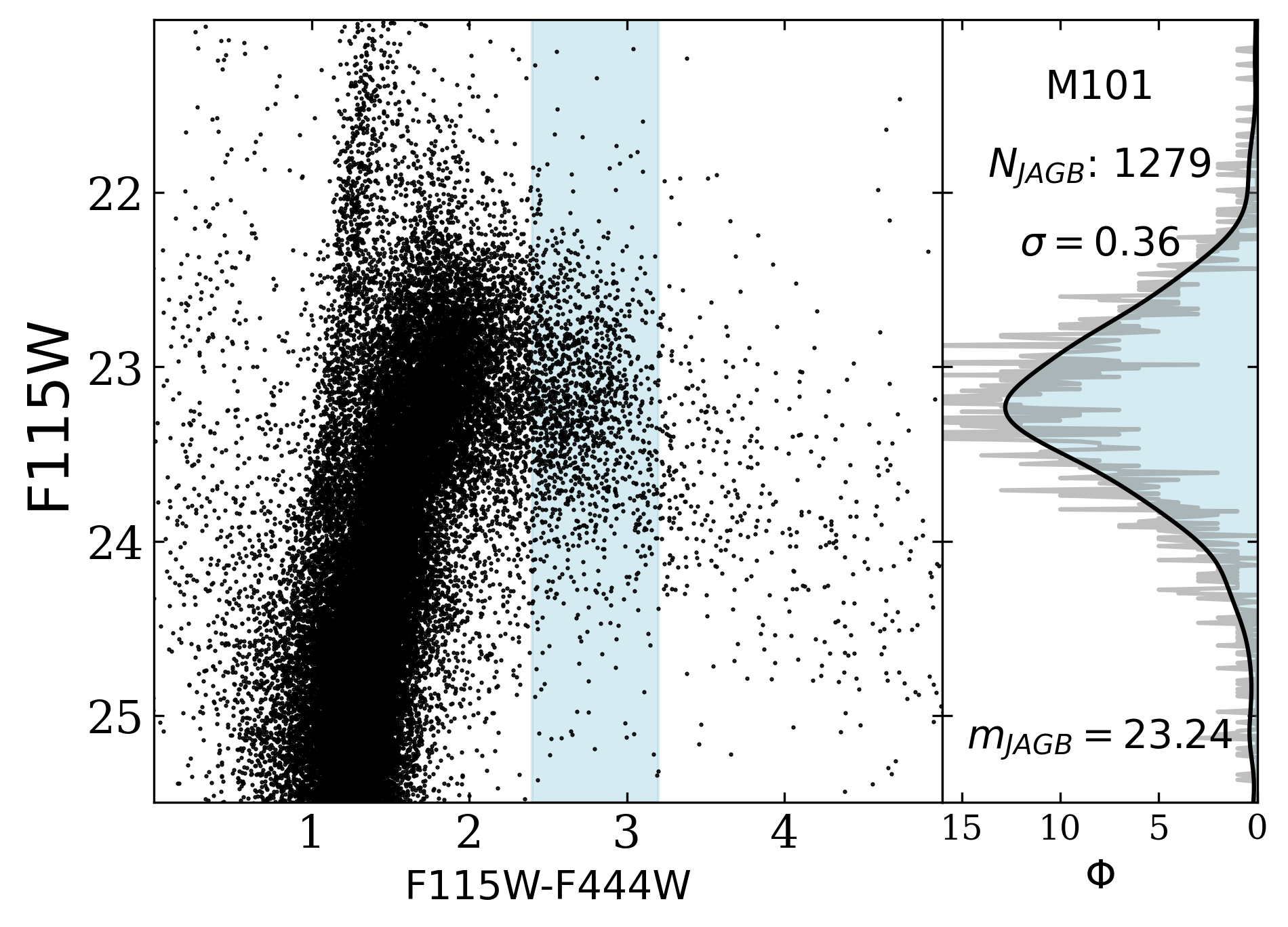}{0.5\textwidth}{}
          \fig{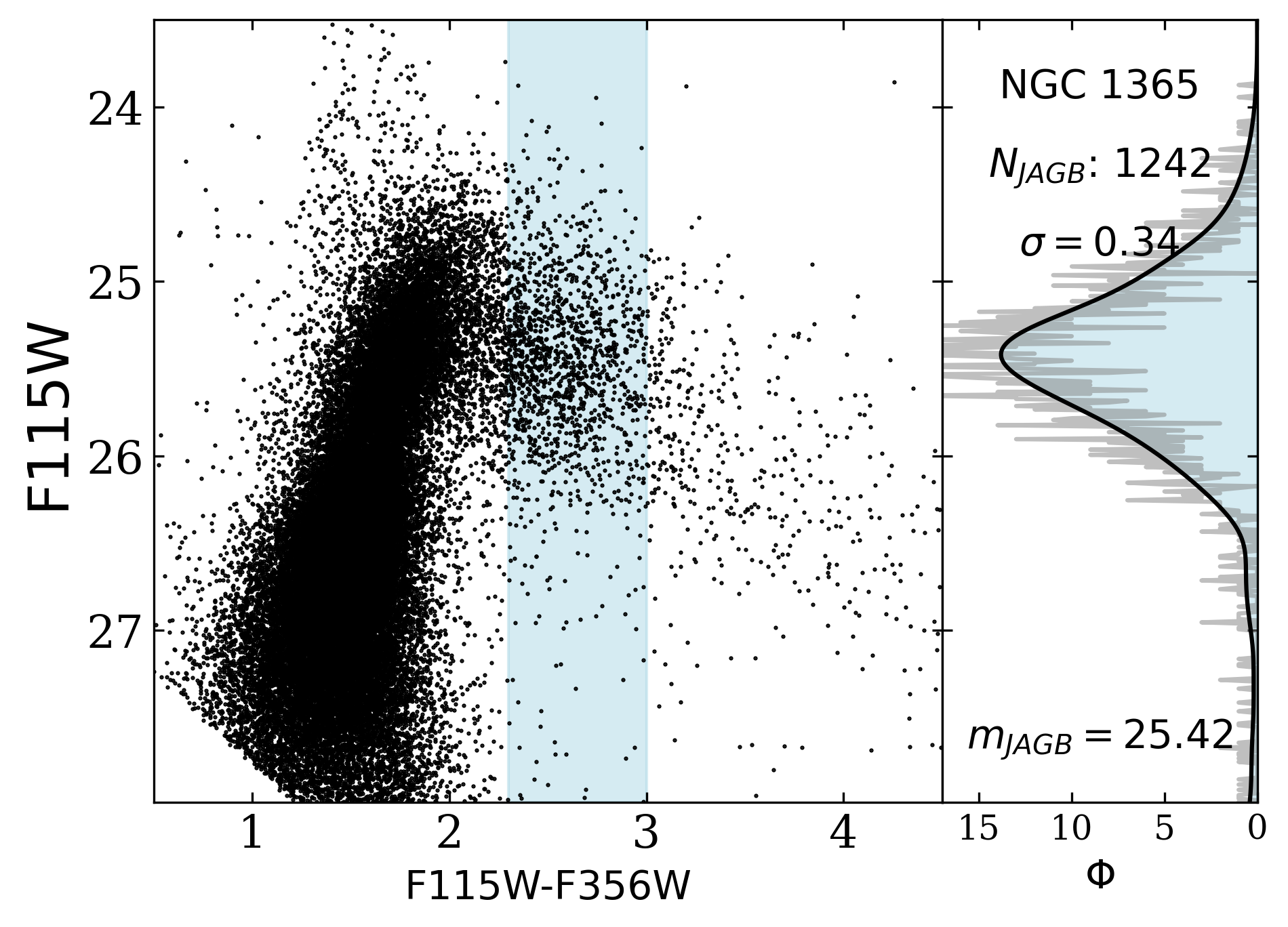}{0.5\textwidth}{}
          }
\gridline{\fig{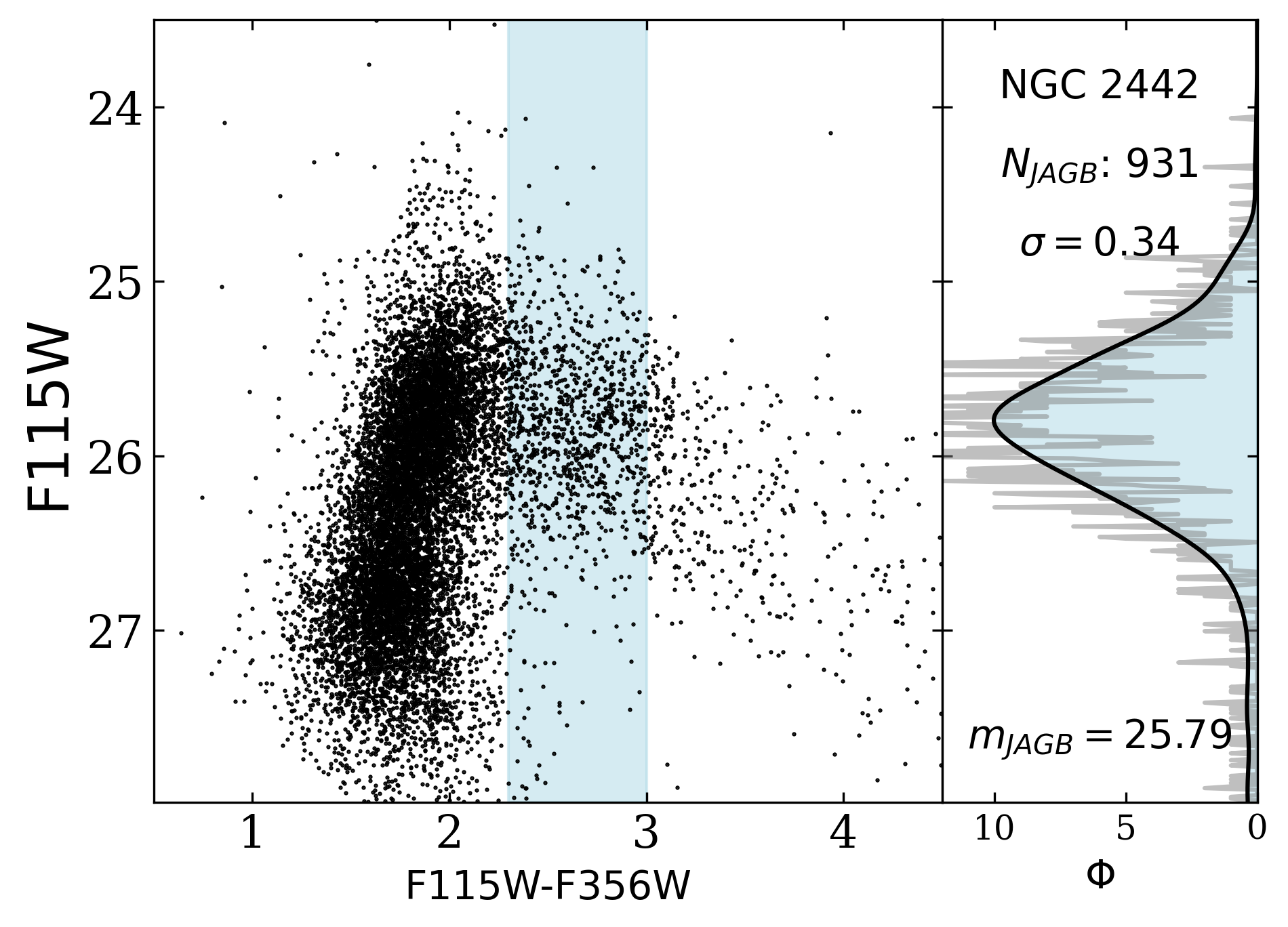}{0.5\textwidth}{}
          \fig{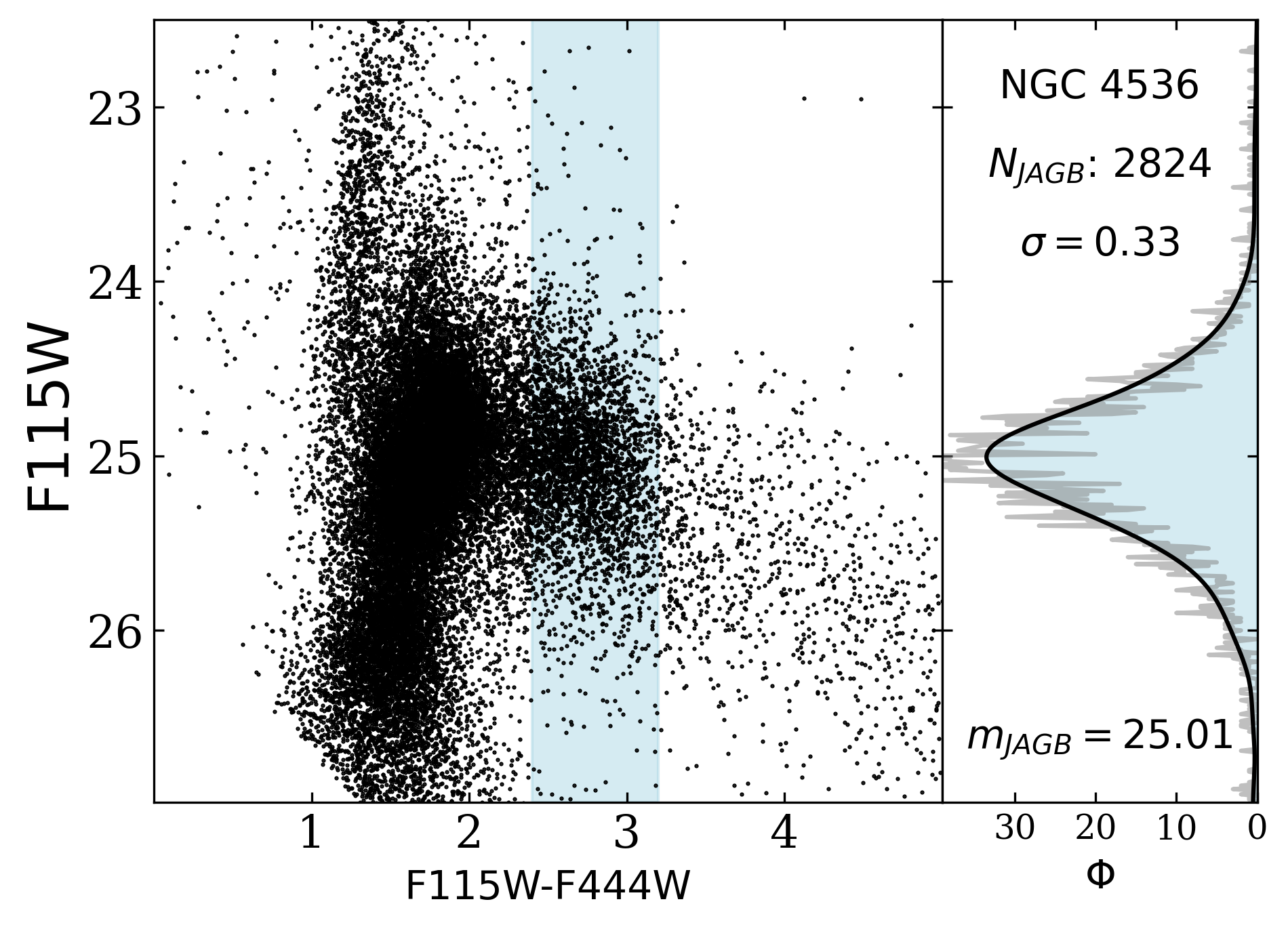}{0.5\textwidth}{}
          }
\gridline{\fig{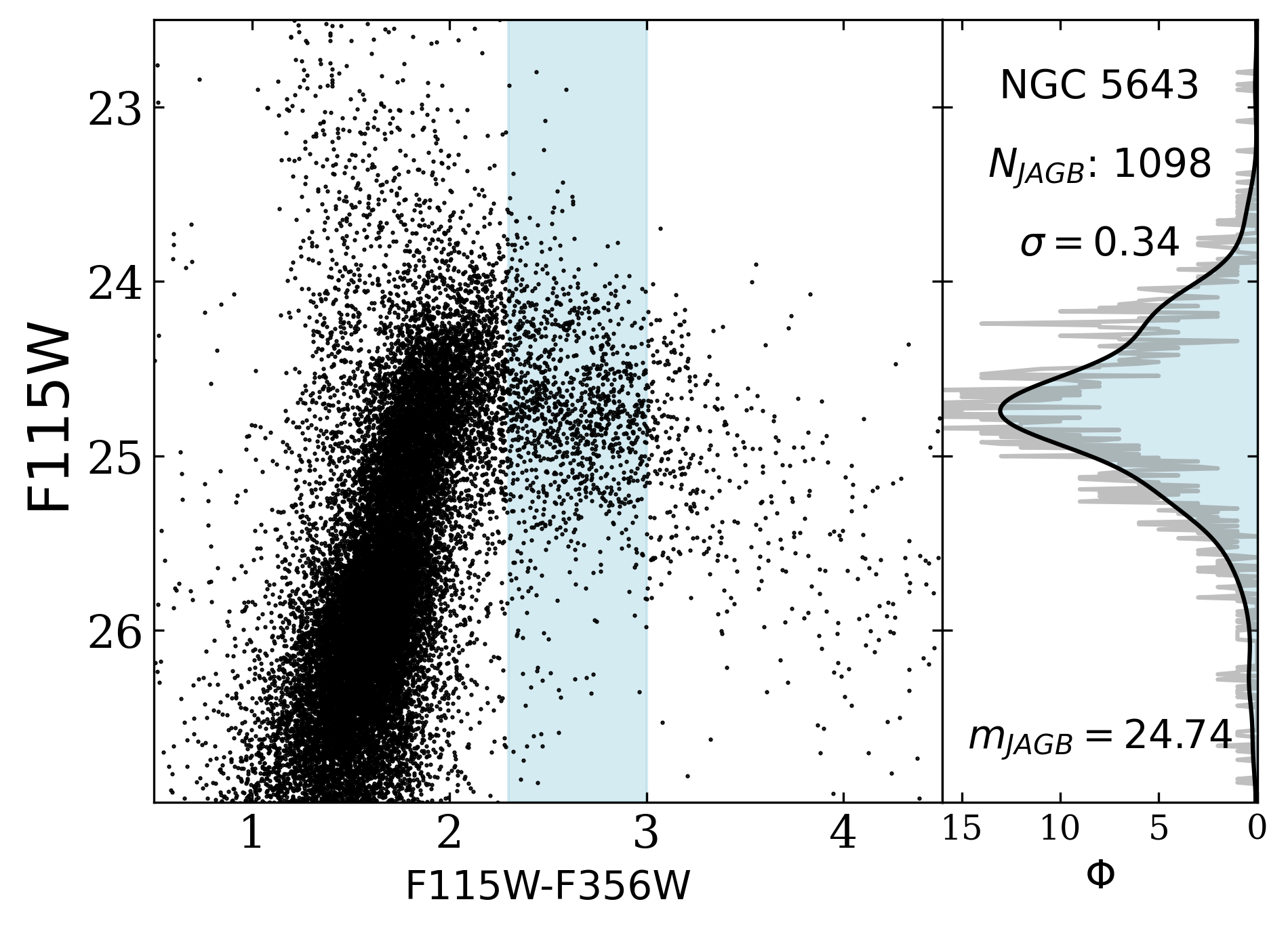}{0.5\textwidth}{}
          \fig{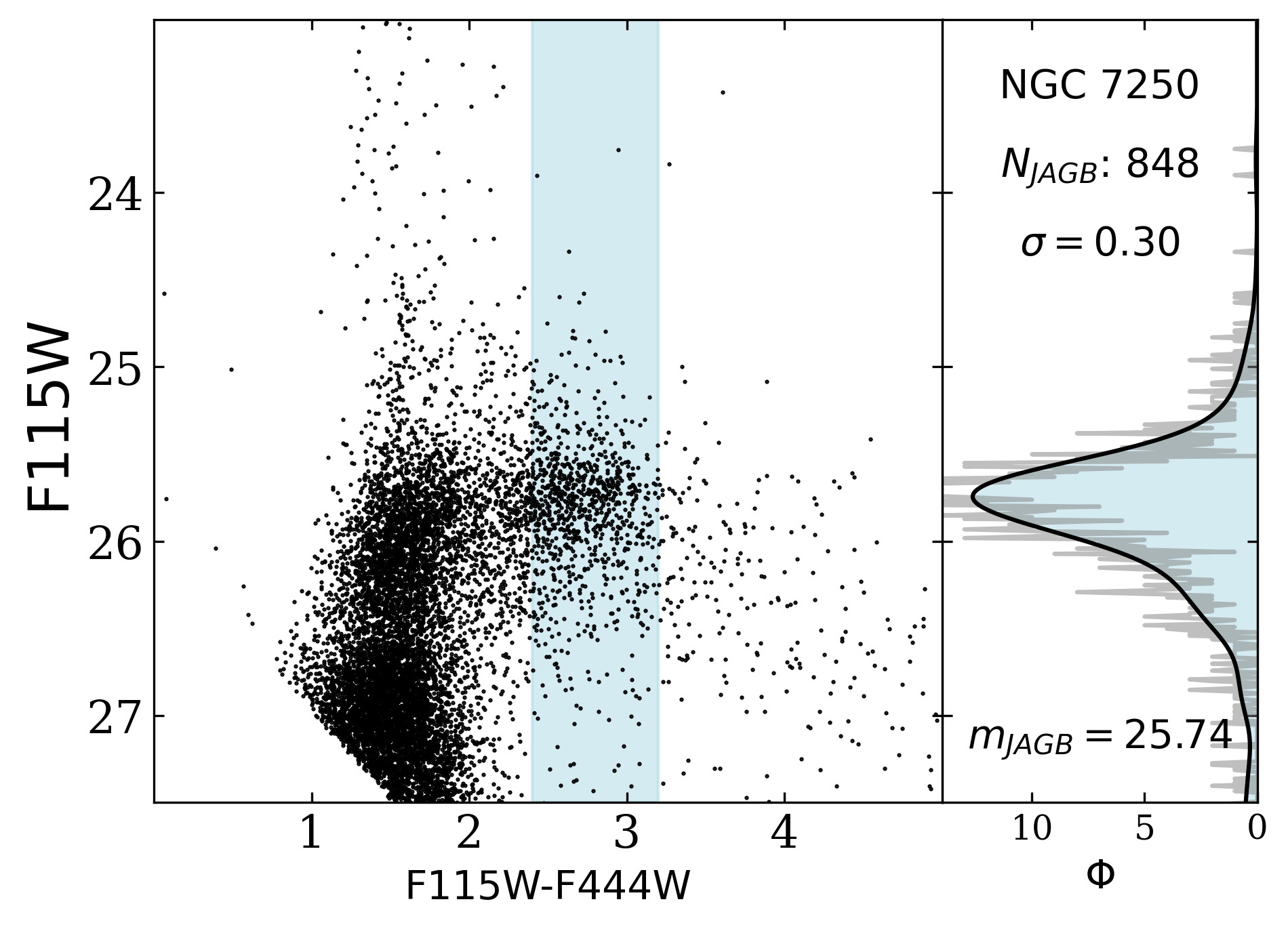}{0.5\textwidth}{}
          }          
\caption{(Left panels) Color-magnitude diagrams for six of the SN Ia host galaxies (see Figure \ref{fig:rep_CMD} for NGC 4639). The JAGB stars were selected within the light blue shaded regions. (Right panels) GLOESS-smoothed JAGB star luminosity functions in black overlaid on top of the binned luminosity function in grey. The number of JAGB stars within $\pm0.75$~mag of the mode is plotted in the upper right corner, as well as the dispersion for those stars about the mode. The y-axis range is 4.5~mag for all galaxies.
The measured JAGB magnitude for each galaxy is also shown in the bottom righthand corner of each plot.}\label{fig:CMD}
\end{figure*}

Now, with our full sample of galaxies available from JWST, we present our finalized methodology for systematically selecting the outer disk via  \textit{convergence plots}. 
The full algorithm is summarized below:

\begin{enumerate}
    \item The JAGB stars were first selected by their colors, and ordered by their semi-major axis distance.
    \item We measured the modal magnitude $m_{JAGB}$ of the JAGB LF constructed from the first 500 JAGB stars, i.e., the 500 innermost JAGB stars.\footnote{A sample of 500 JAGB stars can deliver a distance that is statistically good to a precision of 1\%.} This data point represents the first bin of the convergence plot shown in Figure \ref{fig:settledown_main}.
    \item We then measured $m_{JAGB}$ for the 50th to 550th innermost JAGB stars. Therefore, 450 of the JAGB stars in the second bin are in common with the first bin measured in step 2. This data point represents the second bin of the convergence plot.  We continued to iterate through the entire JAGB sample until reaching the outermost 500 JAGB stars. The result of this procedure is shown in the top panels of Figure \ref{fig:settledown_main}, where we plot the measured JAGB mode vs. the average radial distance for each bin in every galaxy.
    \item Next, we numerically determined $dm/dr$, the smoothed derivative of $m_{JAGB}$ as a function of radial distance.\footnote{The code used to calculate $dm/dr$ can be found at \url{https://github.com/abiglee7/CCHP-JAGB/blob/main/jagb_der.py}.}We smoothed $dm/dr$ according to the same criteria for each galaxy, where the smoothing interval was 1/6 the number of $m_{JAGB}$ data points.
    
    \item The radial distance at which $dm/dr$ equals zero for the first time then denotes
    the radial cut to be used for that galaxy.  
    This first zero-crossing of the derivative signals $m_{JAGB}$ has stabilized in magnitude and ceased growing fainter for the first time.

     Furthermore, to prevent selecting a radial cut where the derivative equaled zero due to random fluctuations, we imposed an additional criteria. We calculated the maximum $m_{JAGB}$, minimum $m_{JAGB}$, median $m_{JAGB}$, and standard deviation $\sigma$ of the $m_{JAGB}$ data points outside of the selected radial cut. The maximum and minimum $m_{JAGB}$ were then required to be greater and less than the median$\pm 2\sigma$, respectively. The smoothing interval was increased until this criterion was fulfilled. This criterion affected three galaxies, M101, NGC 2442, and NGC 4258, where the first zero-crossing of the smoothed derivative was clearly due to a random fluctuation and not where the derivative actually converged.
     
    \item All the data outside of this radial cut were then aggregated to create the final `outer disk' catalog, which is plotted in the CMD for each galaxy in Figures \ref{fig:rep_CMD}, \ref{fig:CMD}, and \ref{fig:n4258_cmd}.
    \item If $m_{JAGB}$ never converged or the galaxy lacked a sufficient number of carbon stars outside of the selected radial cut ($<500$~stars), that galaxy was discarded from the SN Ia calibrating sample. The three galaxies fulfilling these criteria are discussed in Section \ref{subsec:anom}.
\end{enumerate}

The final selected radial cuts are shown overlaid on NIRCam color images of the galaxies in our sample in Figure \ref{fig:images}.

 This algorithm is improved over the preliminary algorithm presented in \cite{2023arXiv231202282L} in two main ways. First, whereas \cite{2023arXiv231202282L} split the photometry into eight independent bins for each galaxy, we now measure the JAGB magnitude in overlapping bins, where each bin has 500 stars total and 50 stars in common with its neighboring bins (therefore a galaxy with more total JAGB stars will have more radial bins). 
This has allowed us to more easily observe micro-changes in the convergence plots (for example, a spiral arm  with significant crowding is now clearly observed in these plots). Furthermore,  we now avoid having to arbitrarily choose the number of spatial bins. Second, the ideal radial cut is now calculated via the first zero-crossing of the first derivative of the convergence plot, $dm/dr$. 
This change now allows for small-scale fluctuations in $m_{JAGB}$ past the adopted radial cut. 
For example, in Figure \ref{fig:settledown_main}, in NGC 4536 and NGC 4258,  $m_{JAGB}$ has clearly converged past the radial cut, yet still varies at the 0.05~mag level.  Our preliminary algorithm presented in  \cite{2023arXiv231202282L} would have output a radial cut significantly farther out into the disk because of these small-scale fluctuations,  even though $m_{JAGB}$ had already clearly converged at smaller radial distance, thus
leading to a significantly smaller sample of JAGB stars. In Section \ref{subsubsec:statunc}, we describe how we adopted a statistical error accounting for the `noisiness' of the convergence plot. 

We also noticed the presence of a spiral arm in the convergence plots sometimes caused additional noisiness. This caused $m_{JAGB}$ to converge at a radial distance with too few JAGB stars or $dm/dr$ to never equal zero in four galaxies: M101, NGC 2442, NGC 4258, and NGC 4639.
We masked these spiral arms using increasingly larger masks until $m_{JAGB}$ successfully converged.  The functional forms of the masking criteria are detailed in Appendix \ref{sec:mask}.
The convergence plots without the spiral arms masked are shown in grey in Figure \ref{fig:settledown_main}. We emphasize we performed this procedure during the blinded stage of our analysis. We only masked the spiral arms so that $m_{JAGB}$ successfully converged, not to change the final measured distance. 
Leaving the spiral arms unmasked while using the newly adopted 
radial cuts yielded almost a negligible change in $H_0$ of 0.3\% (larger).
This is also further discussed in Appendix \ref{sec:mask}.

We observed the same pattern in all our convergence plots where the mode was generally measured to be brightest in the inner, high-surface brightness regions of a galaxy. 
In Appendix \ref{sec:inner}, we show CMDs for the `inner region' of each galaxy, which were produced using the stars inside the chosen radial cut. In our final sample of seven calibrating galaxies, we found the average difference between the measured JAGB magnitude in the inner vs. outer regions to be $\rm{<JAGB_{inner} - JAGB_{outer}}> = -0.21$~mag, meaning the JAGB magnitude was on average 0.21 mag brighter in the inner disks of these galaxies. Thus, we continue to caution the JAGB magnitude can be significantly biased when measured in the inner disks of galaxies.
This pattern may result from crowding, the effect of which will be larger in the higher surface brightness regions in galaxies.


\begin{figure*}
\gridline{\fig{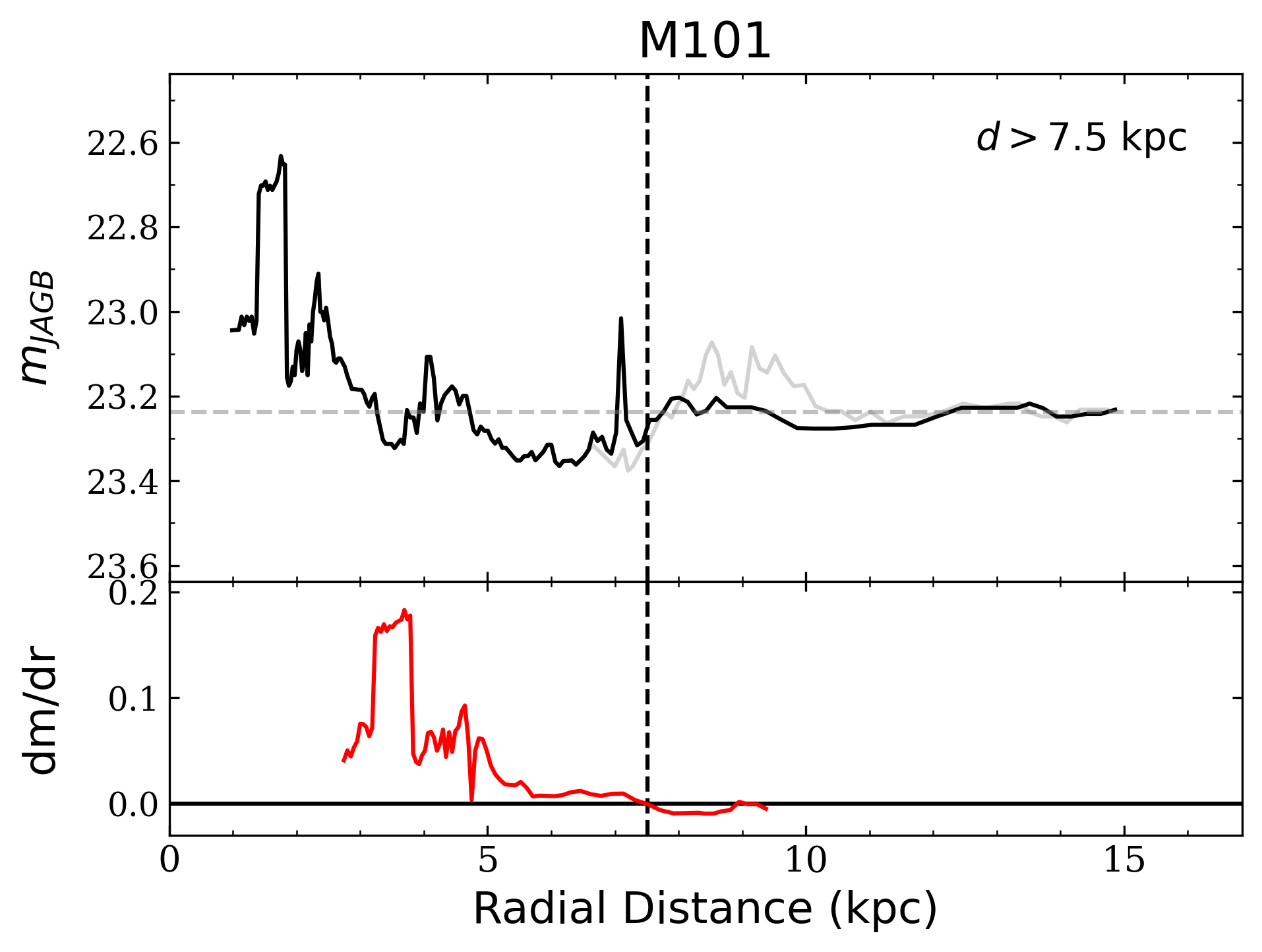}{0.5\textwidth}{}
          \fig{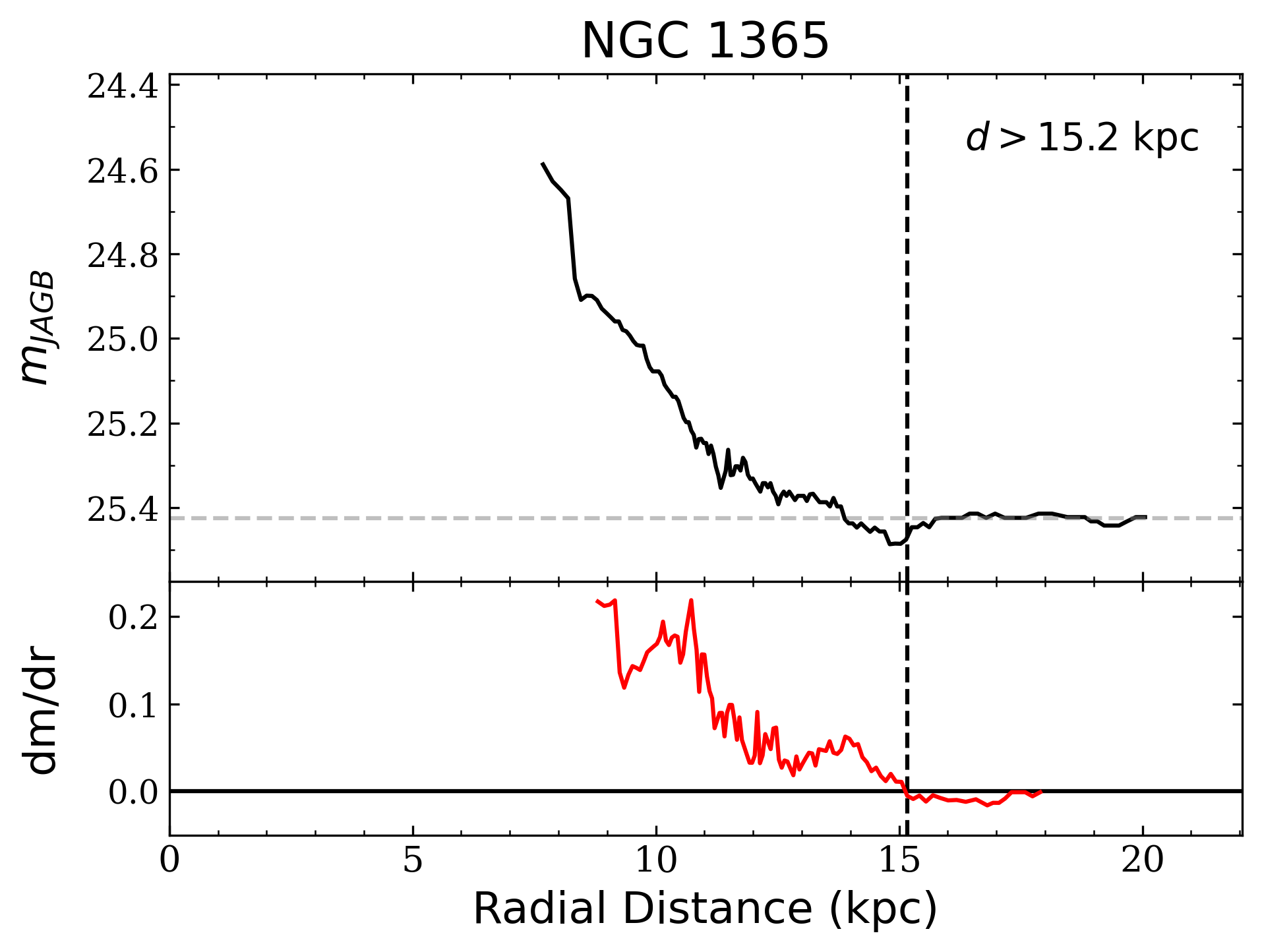}{0.5\textwidth}{}}
\gridline{\fig{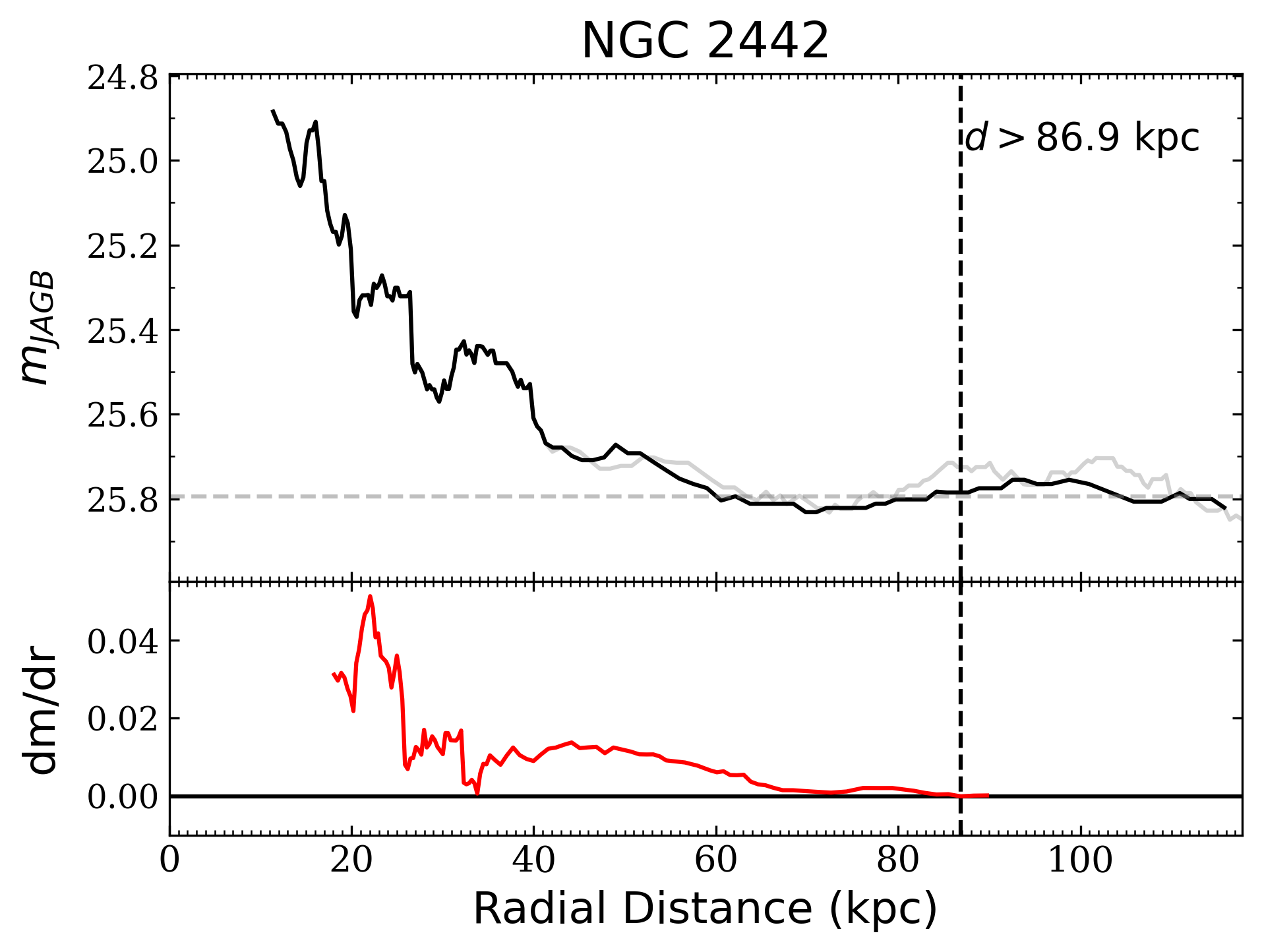}{0.5\textwidth}{}
          \fig{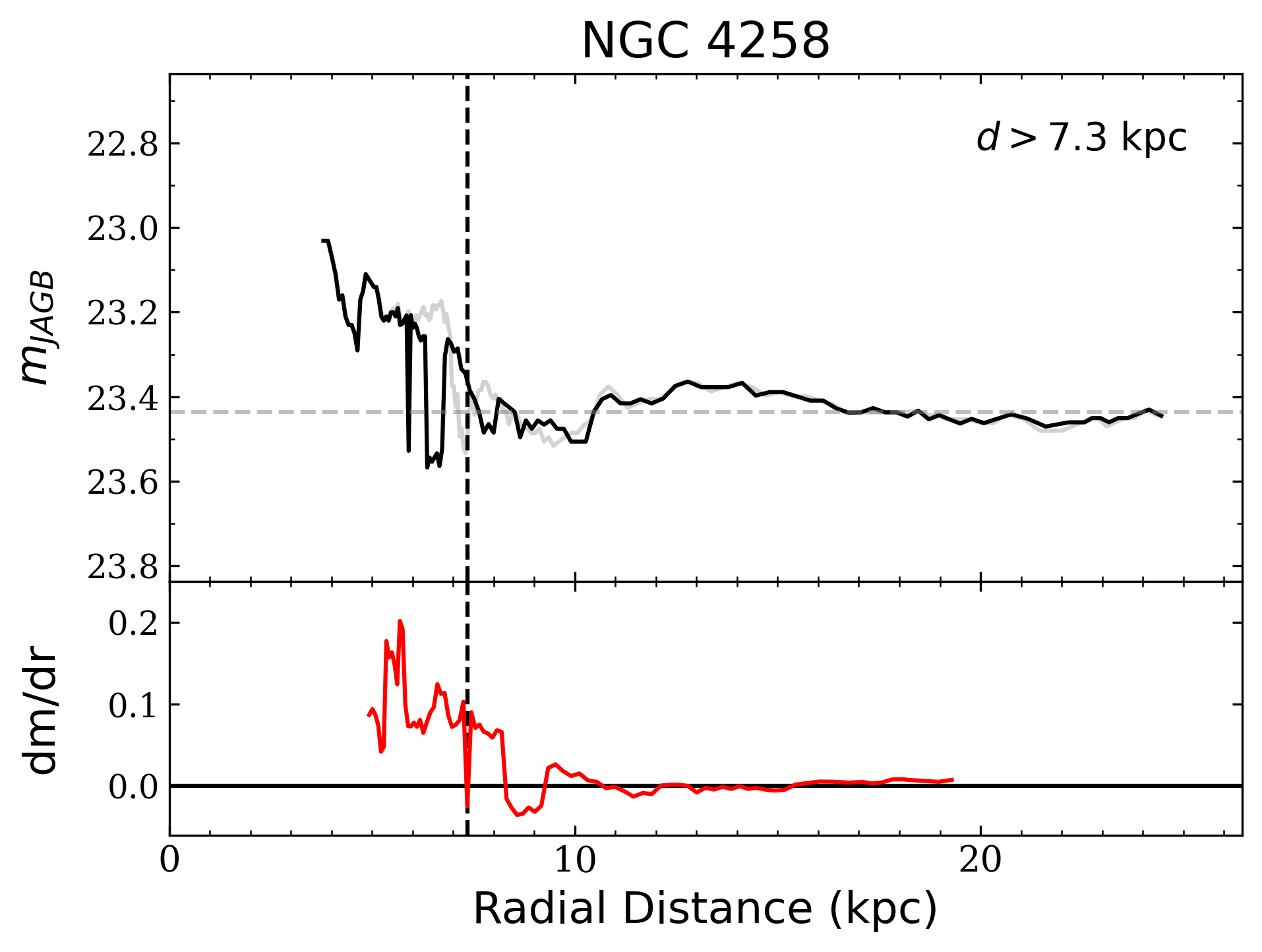}{0.5\textwidth}{}
          }
\caption{(Top panels) The JAGB magnitude as a function of radial distance in our seven SN Ia host galaxies. Each data point represents the mode measured from a JAGB LF composed of 500 JAGB stars. Every data point contains 50 JAGB stars in common with its neighboring data points.  The magnitude y-range is 1.2~mag for every plot.  In M101, NGC 2442, NGC 4258, and NGC 4639, the grey line represents the measured $m_{JAGB}$ as a function of radial distance with the spiral arms unmasked.  
(Bottom panels) The derivative of the JAGB magnitude as a function of radial distance. 
The first zero crossing of the derivative, i.e. where the JAGB magnitude `converges' to a stable magnitude for the first time, marks the chosen radial cut in that galaxy.
This is denoted by the dotted black line, where all the stars to the right of this line were aggregated to create the final `outer disk' JAGB star sample for that galaxy.
The physical radial distance cut is shown in the upper righthand corner of each plot. 
}\label{fig:settledown_main}
\end{figure*}

\begin{figure*}\figurenum{4}
\gridline{\fig{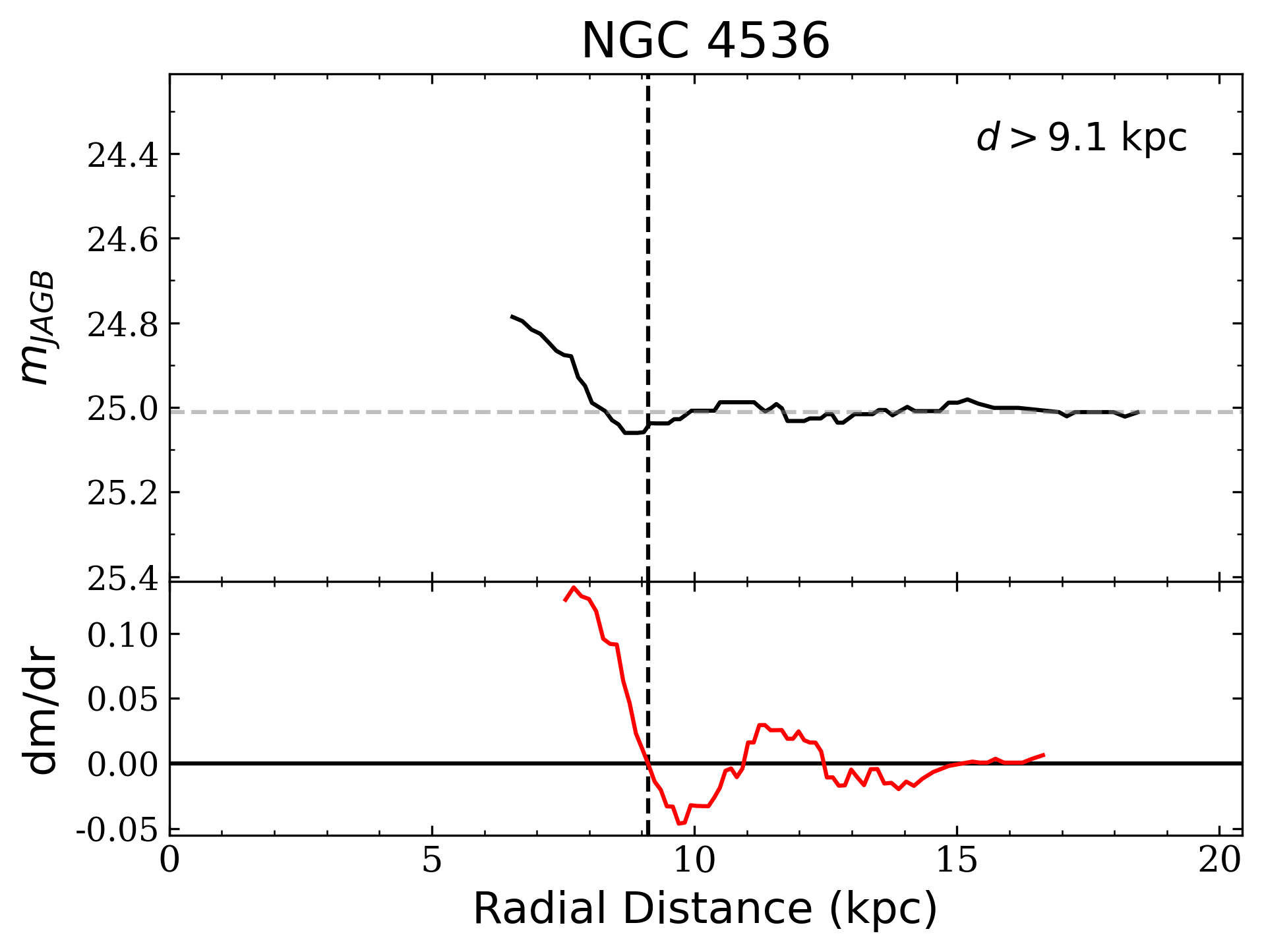}{0.5\textwidth}{} 
          \fig{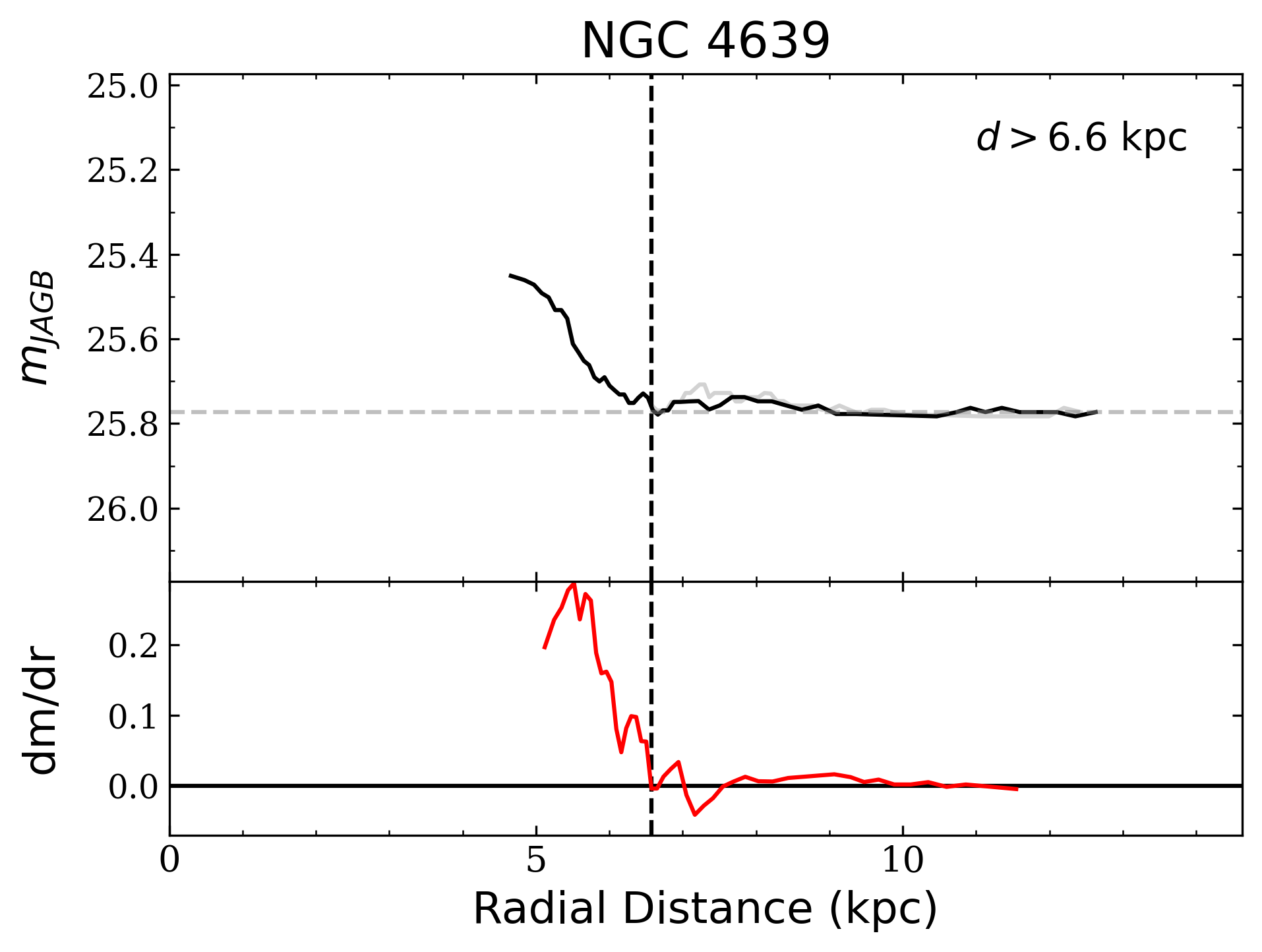}{0.5\textwidth}{}
          }
\gridline{\fig{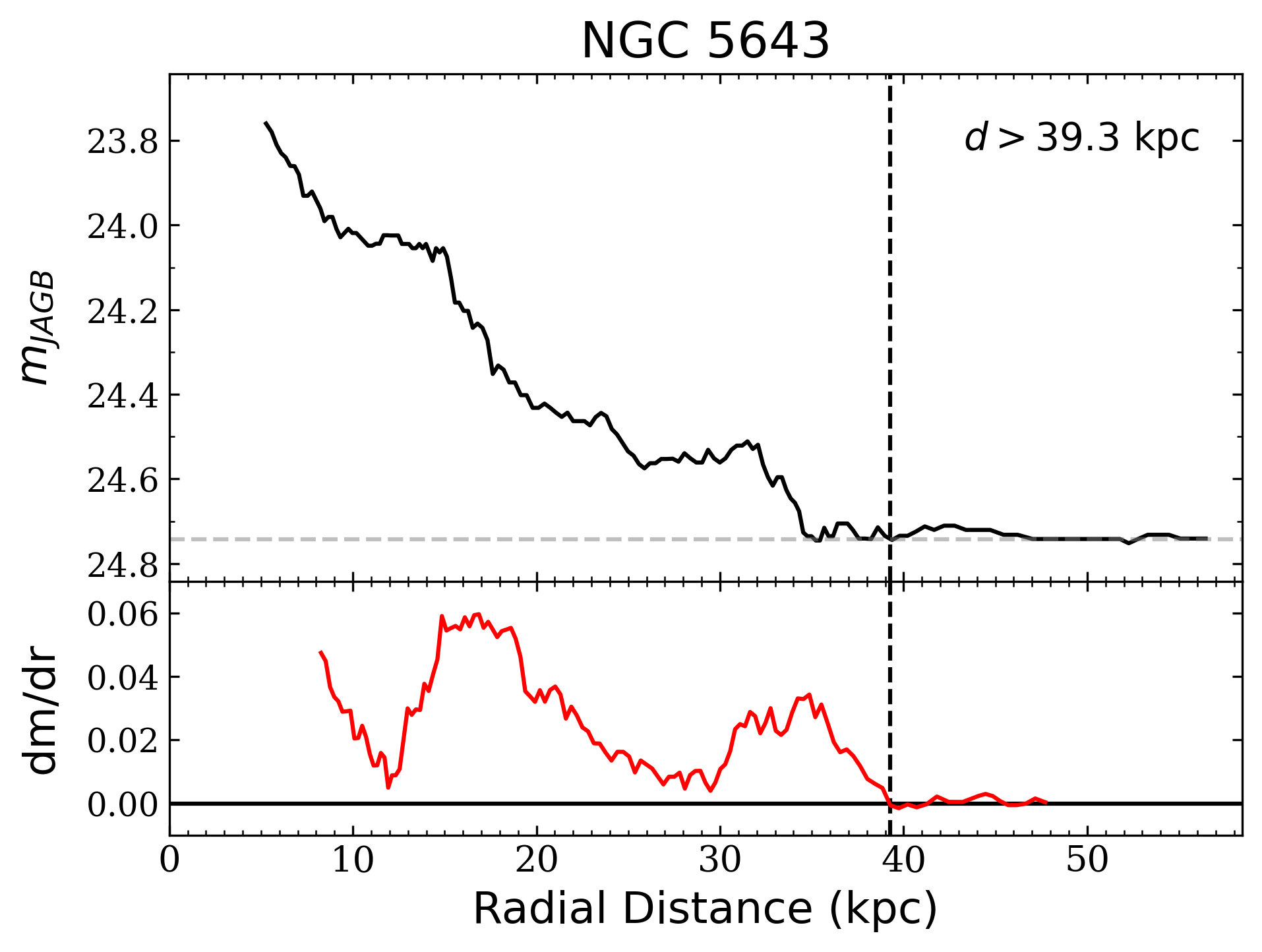}{0.5\textwidth}{}
          \fig{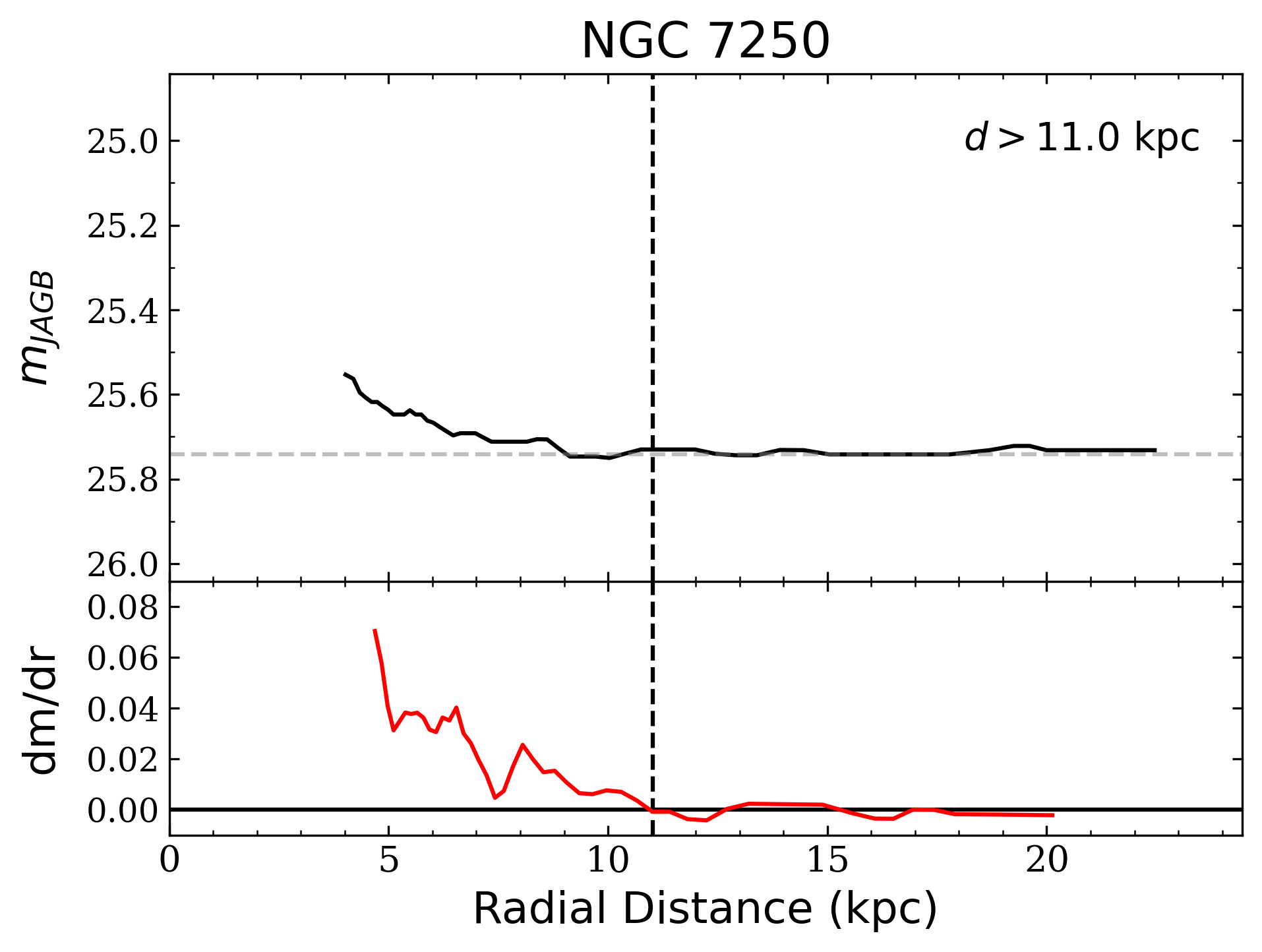}{0.5\textwidth}{}}
\caption{(cont.)}
\end{figure*}

\subsection{Galaxies in which the JAGB magnitude failed to converge}\label{subsec:anom}

\begin{figure*}
\gridline{\fig{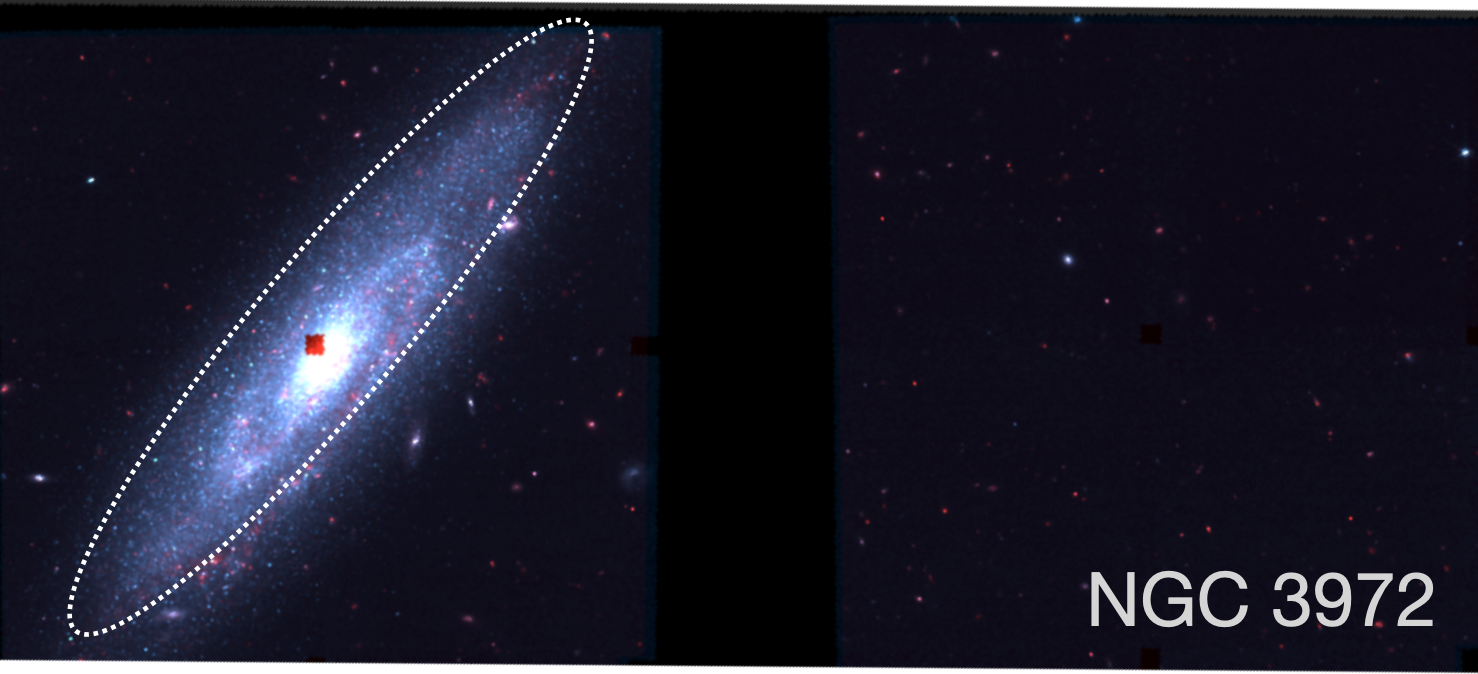}{0.33\textwidth}{} 
          \fig{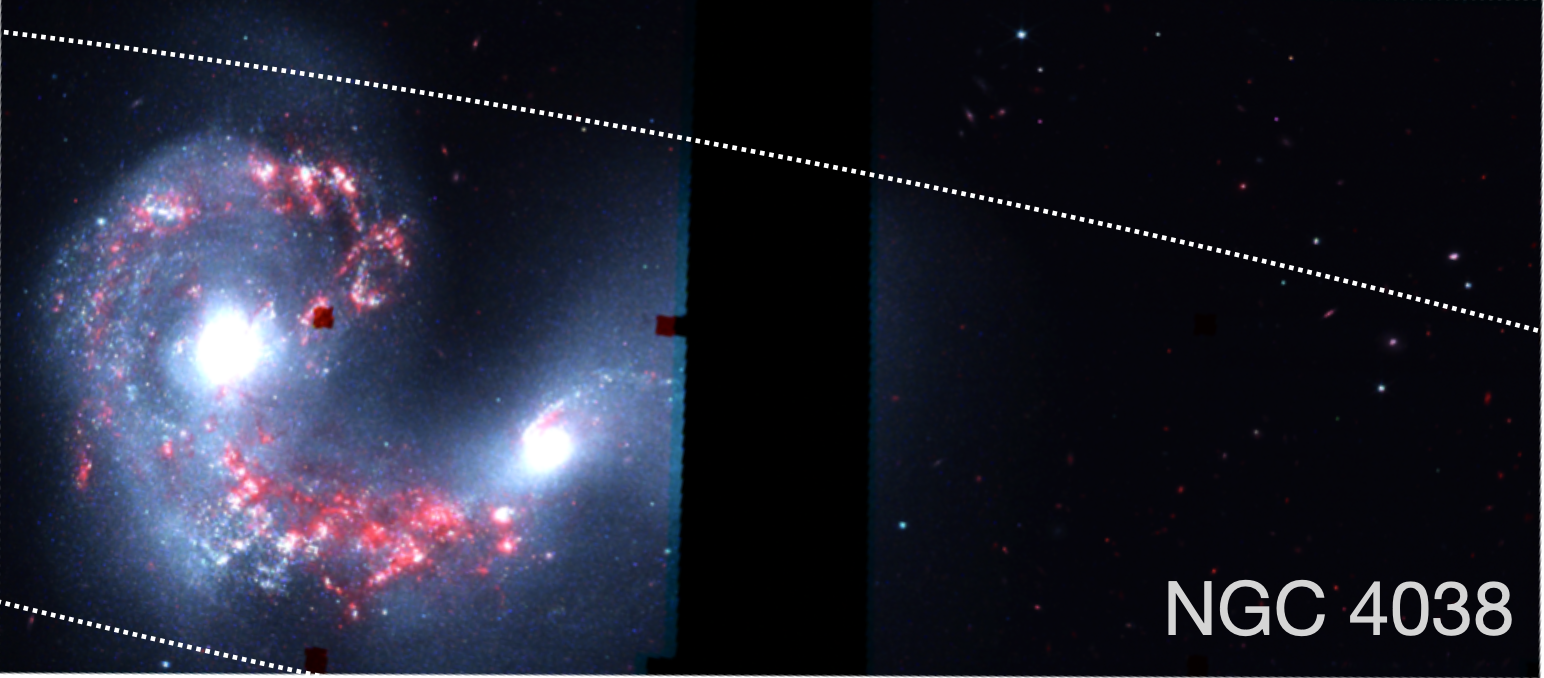}{0.34\textwidth}{} 
          \fig{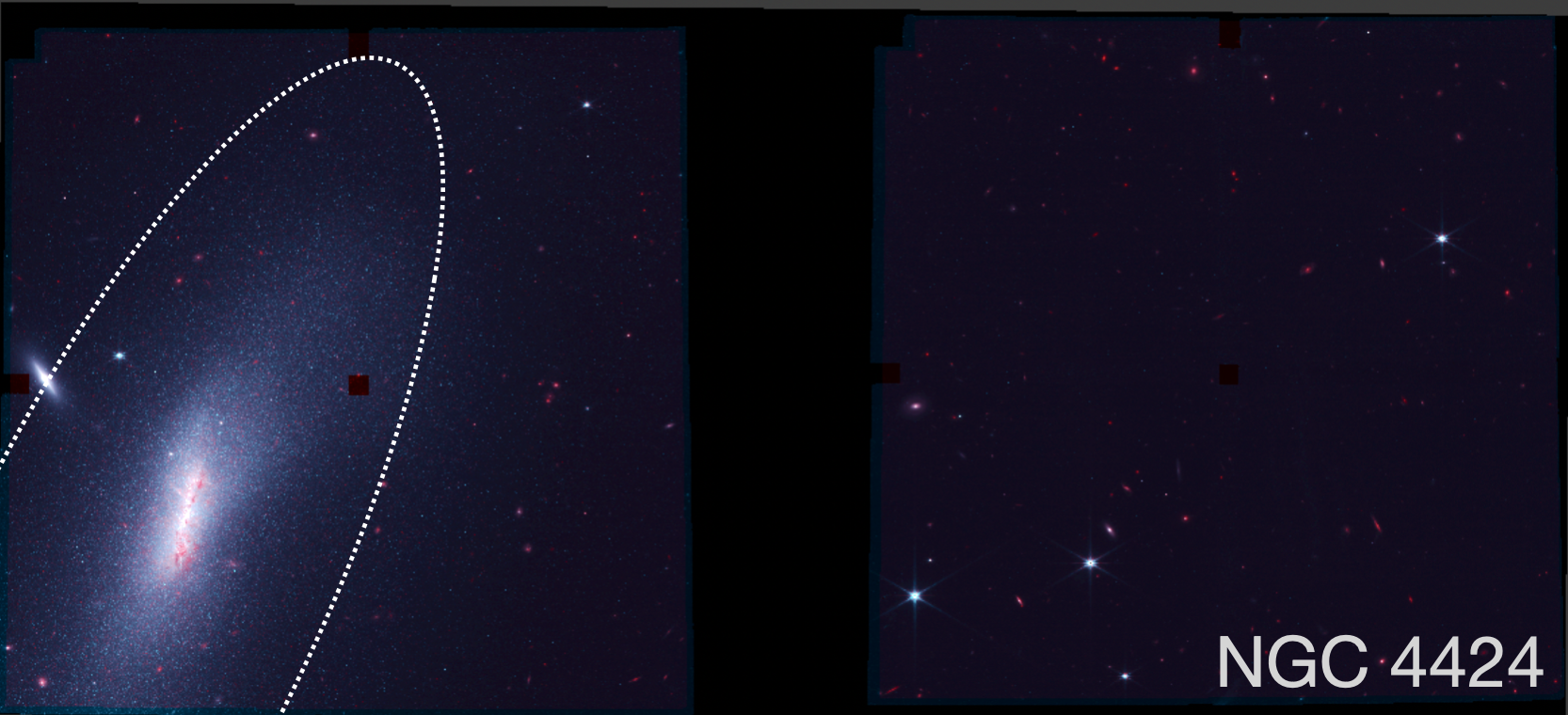}{0.33\textwidth}{}
          }
\gridline{\fig{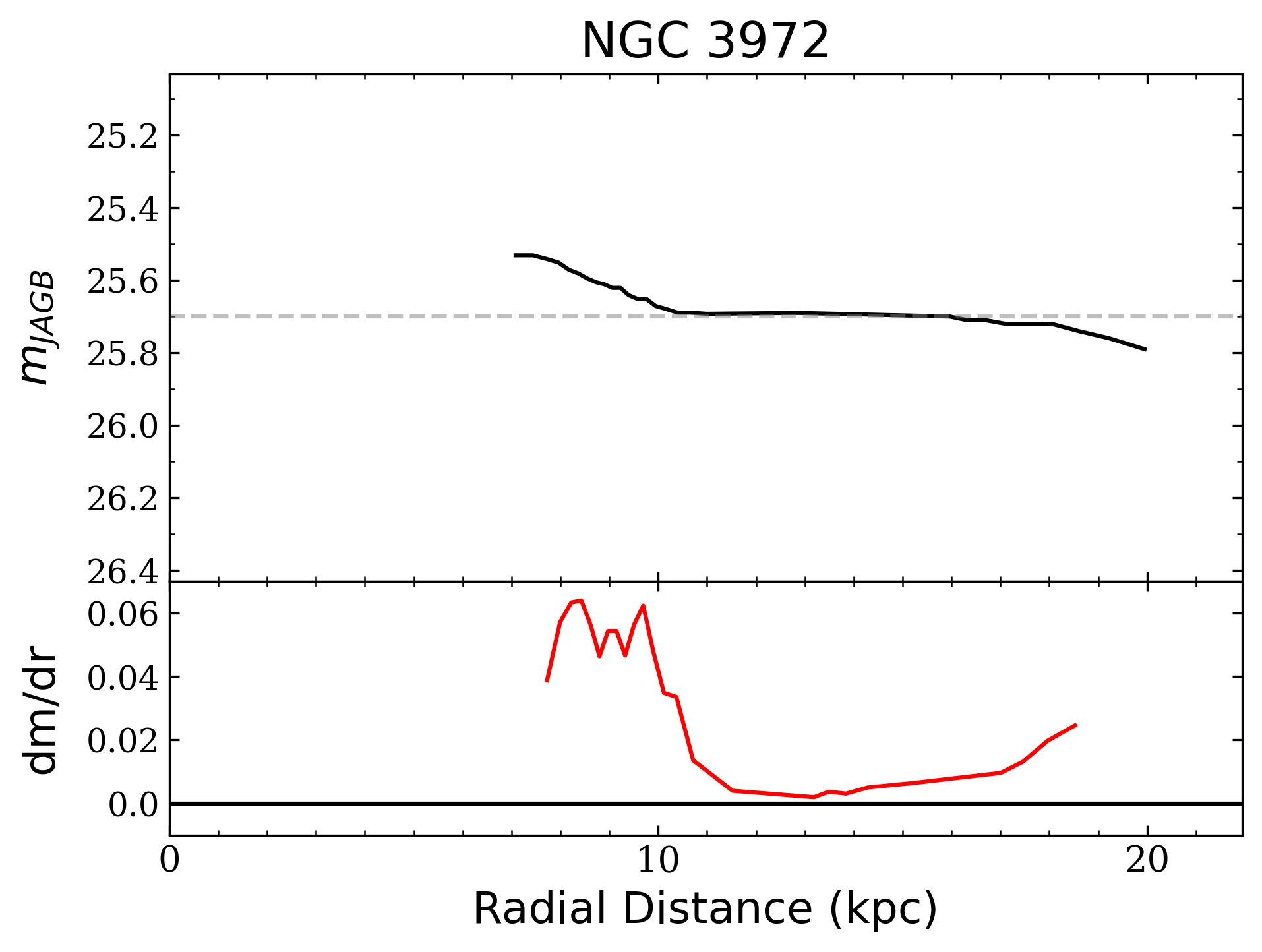}{0.33\textwidth}{} 
          \fig{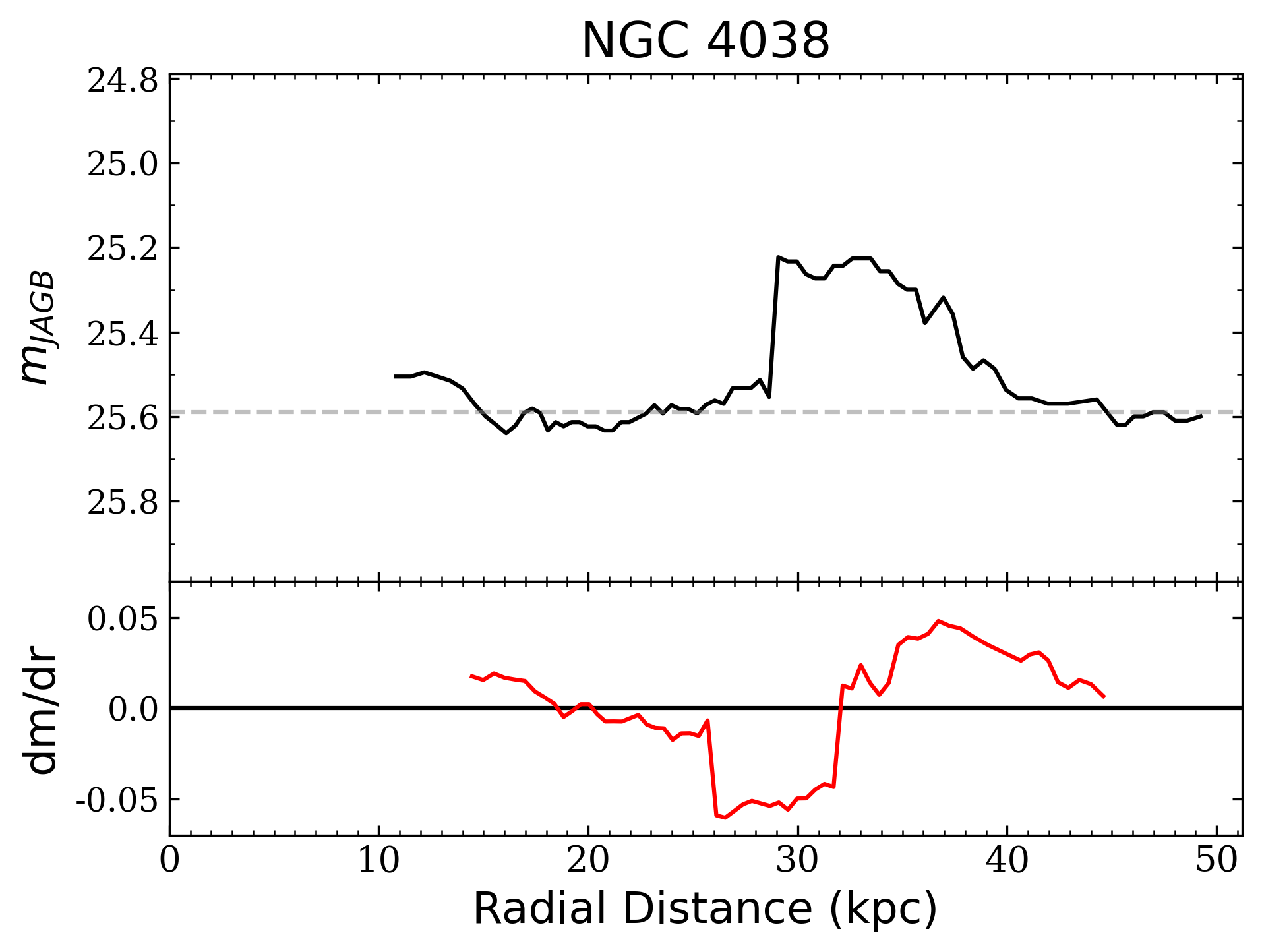}{0.33\textwidth}{} 
          \fig{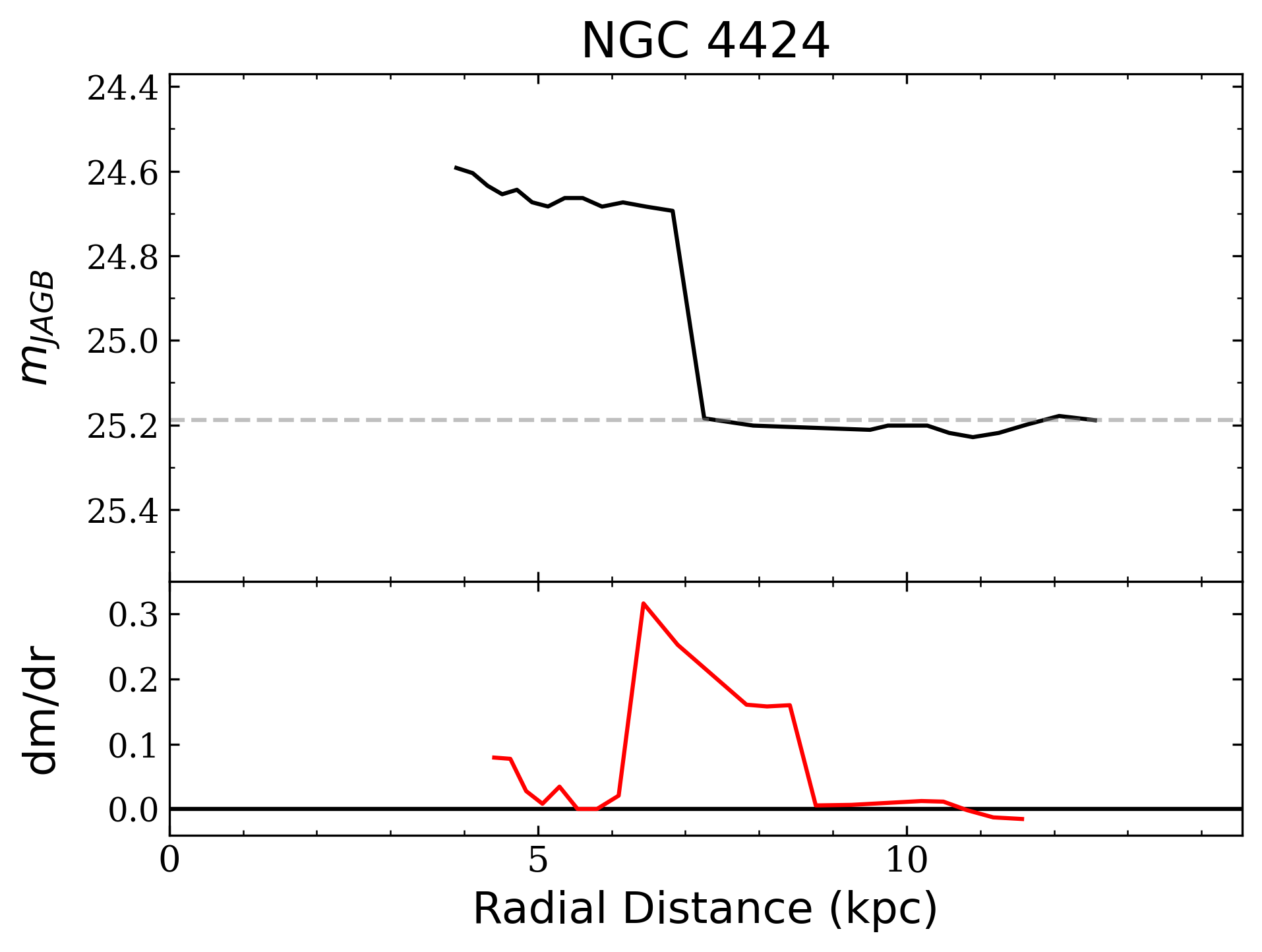}{0.33\textwidth}{}
          }
\gridline{\fig{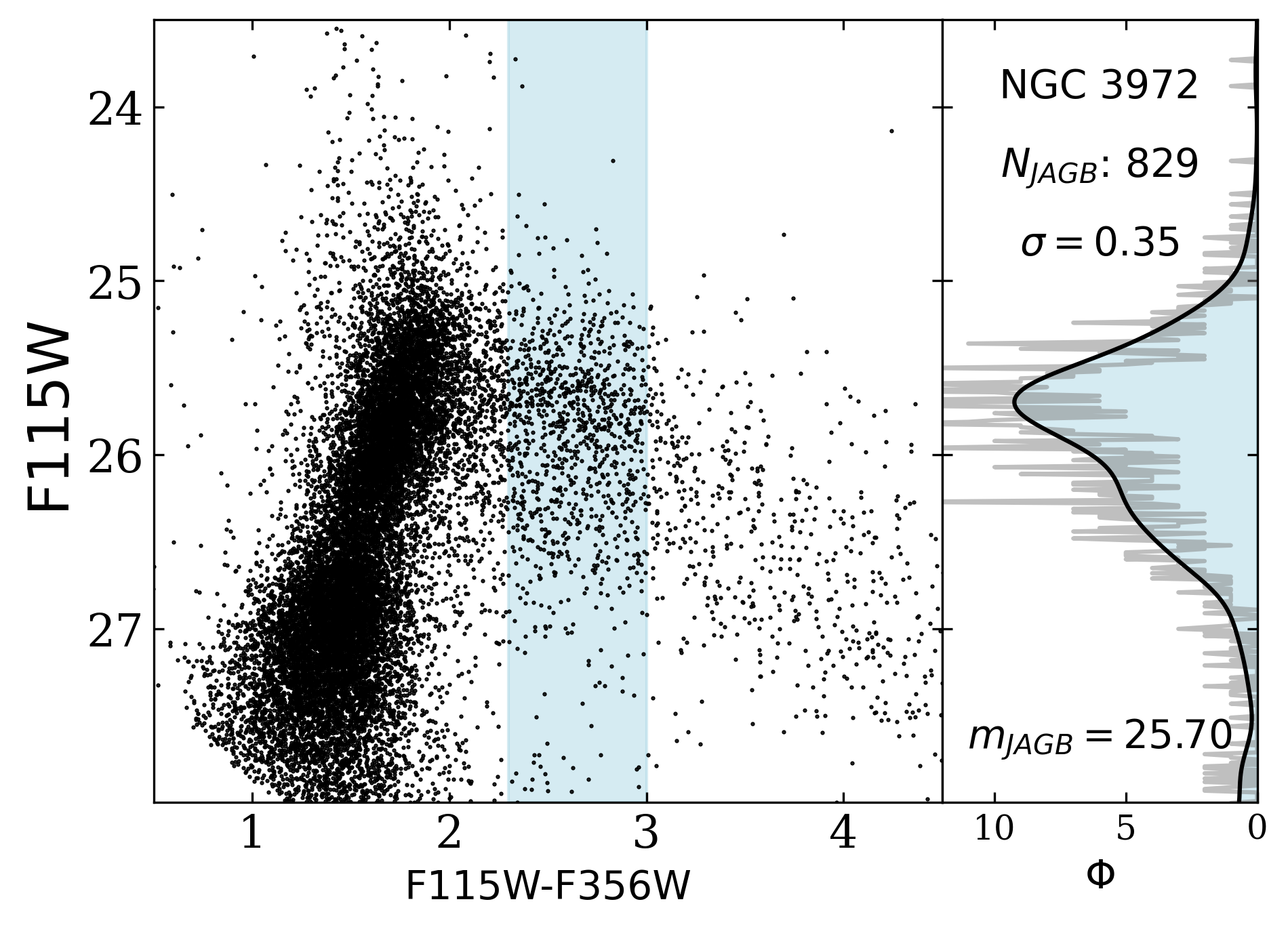}{0.33\textwidth}{} 
          \fig{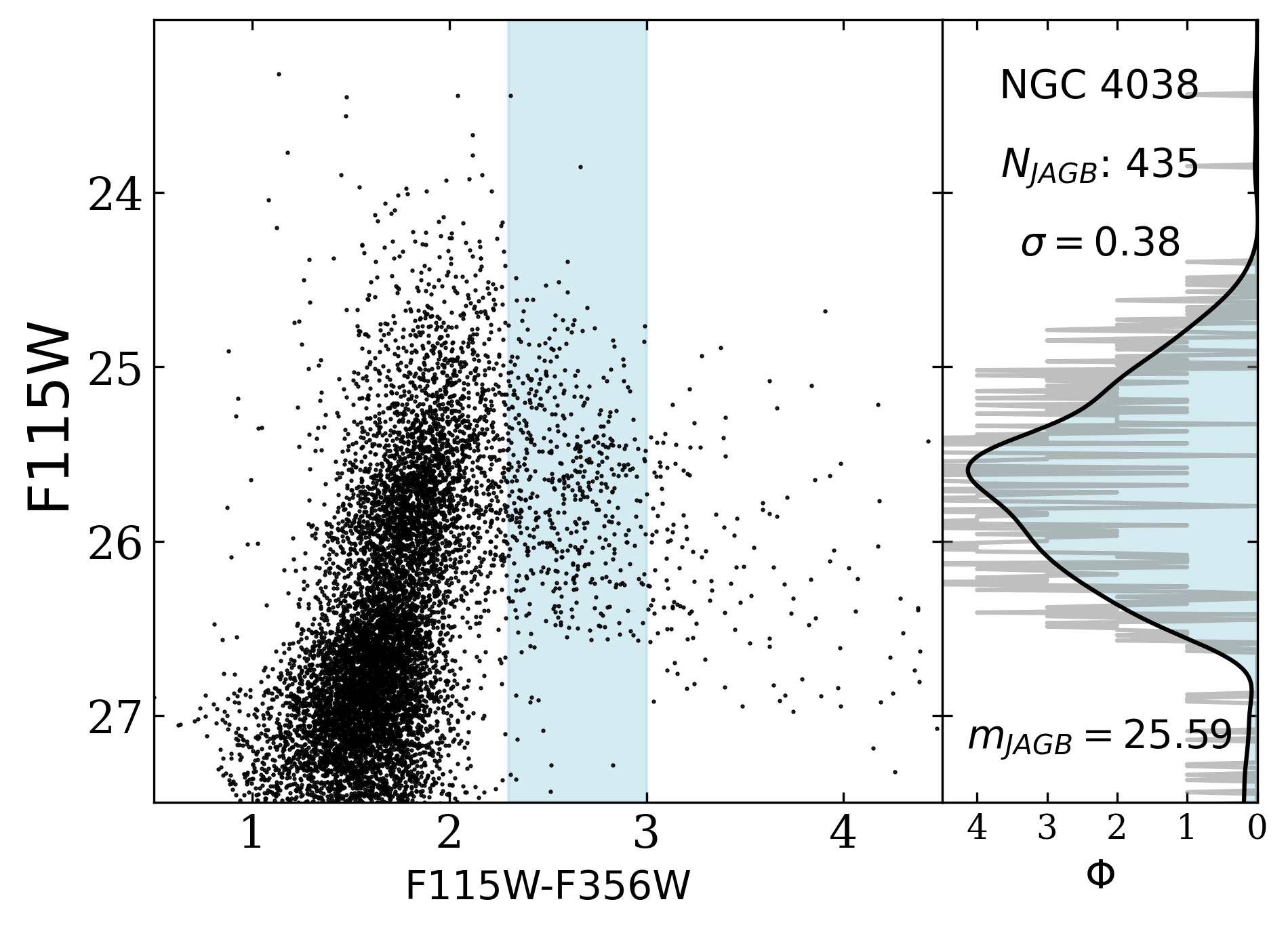}{0.33\textwidth}{}
          \fig{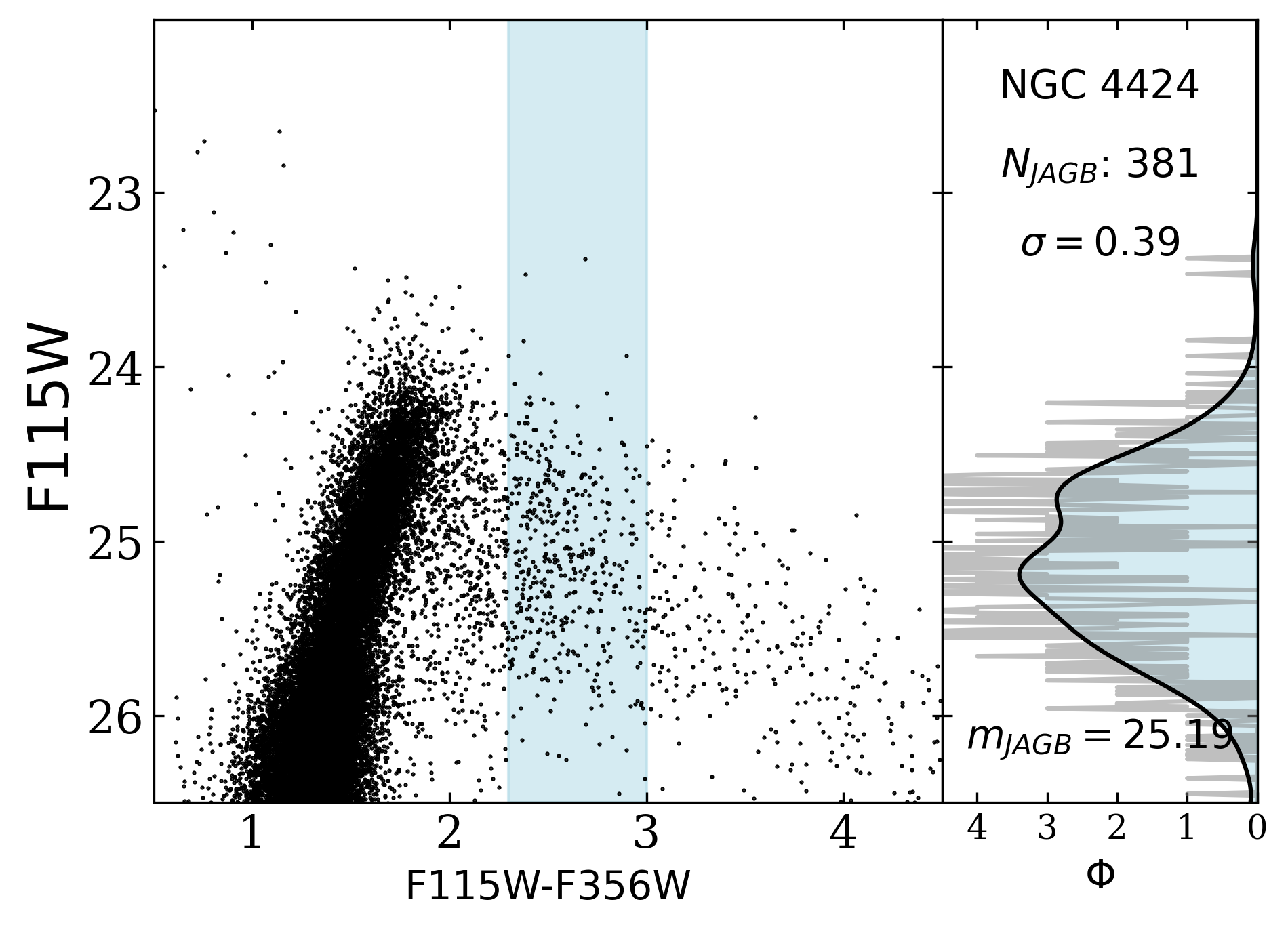}{0.33\textwidth}{}
          } 
\caption{
(Top panels) NIRCam images of the three SN Ia host galaxies in our program in which the JAGB magnitude failed to converge (see Figure \ref{fig:images} for a full description).
(Middle panels) Convergence tests of the JAGB magnitude (see Figure \ref{fig:settledown_main} for a full description).
The JAGB magnitude failed to converge at a radial distance where sufficient numbers of JAGB stars still remained. 
We therefore visually selected the following radial cuts to create the CMDs in the bottom panel for NGC 3972, NGC 4038, and NGC 4424, respectively: 9.0 kpc, 42.0 kpc, 7.3 kpc. 
(Bottom panels) Color-magnitude diagrams (see Figure \ref{fig:CMD} for a full description).
}\label{fig:appendix0}
\end{figure*}

The JAGB magnitude failed to converge at a radial cut outside of which sufficient numbers of JAGB stars remained ($>500$ JAGB stars) for three galaxies in our sample, NGC 3972, NGC 4038, and NGC 4424.
However, all three galaxies' failures to provide converged JAGB magnitudes appear to be straightforwardly explained by astrophysical reasons such as the presence of a merger or high levels of crowding. We describe these effects in detail in the following subsections. We note that NGC 4424 and NGC 4038 have also been shown to be challenging targets for measuring Cepheid and TRGB distances, which we also review in their respective subsections.
NIRCam color images, convergence plots, and CMDs of the three galaxies are shown in Figure \ref{fig:appendix0}.

We also emphasize that because we performed this analysis blinded, we discarded these three galaxies from our SN Ia calibrating sample before knowing their measured distances. It was at this point, after finalizing the analysis and SN Ia calibrating sample, that we removed the random photometric offset from our catalogs to reveal each galaxy's measured JAGB distance.

\subsubsection{NGC 4424: A ram pressure stripped galaxy without recent star formation}

NGC 4424 is a barred spiral galaxy at a distance of $\sim15$~Mpc. The CCHP TRGB distance to NGC 4424 was measured to be $\mu_0=31.05\pm0.06$~mag \citep{2018ApJ...861..104H} (derived using the updated CCHP calibration from \citealt{2021ApJ...919...16F}), and the SHoES Cepheid distance was measured to be $\mu_0=30.86\pm0.13$~mag \citep{2022ApJ...934L...7R}. These distances disagree at the 9\% level.

As shown in Figure \ref{fig:appendix0}, the JAGB magnitude in NGC 4424 converged at a radial distance of $\sim7.5$~kpc. However, at this cut, only 365 JAGB stars remained, which is an insufficient number of stars for a precise 1\% JAGB distance. 
NGC 4424 contains so few JAGB stars because its recent star formation  was quenched due to ram-pressure stripping from a galaxy merger $\lesssim 500$~Myr ago \citep{2018A&A...620A.164B}. This prevented the formation of  intermediate-aged and younger stellar populations like JAGB stars and Cepheids, respectively. 

Accordingly, we also note the distance to NGC 4424 has proven to be difficult to measure via Cepheids.
\cite{2016ApJ...826...56R} first measured a Cepheid distance to NGC 4424 of $\mu_0=31.08\pm0.29$~mag, using only 3 Cepheids total in their P-L relation (note the P-L relation with the second-least number of Cepheids had 13). 
 \cite{2022ApJ...934L...7R} increased their sample to 9 Cepheids and re-measured a distance of $\mu_0=30.86\pm0.13$~mag, which was 10\% different in distance from their 2016 measurement. Clearly, NGC 4424 is a challenging target for measuring Cepheid and JAGB distances, due to its almost nonexistent young and intermediate-aged populations.

\subsubsection{NGC 4038: A galaxy currently merging with NGC 4039}

NGC 4038 is a barred spiral galaxy at a distance of $\sim21$~Mpc.
The CCHP TRGB distance to NGC 4038 was measured to be $\mu_0=31.65\pm0.05$~mag \citep{2015ApJ...807..133J} (derived using the updated CCHP calibration from \citealt{2021ApJ...919...16F}), which agrees well with the SHoES Cepheid distance of $\mu_0=31.62\pm0.12$~mag \citep{2022ApJ...934L...7R}.

NGC 4038 is currently merging with NGC 4039, as shown in Figure \ref{fig:appendix0}, where NGC 4038 is the left galaxy in the image and NGC 4039 is the right galaxy. This interacting double is known as the Antennae Galaxies. Because of this merger, the outer disk of NGC 4038 overlaps significantly with the inner disk of NGC 4039. 
Consequently, without a clean `outer disk' region, the JAGB magnitude never cleanly converges, as shown in Figure \ref{fig:appendix0}.  
Visually choosing a radial cut of 45~kpc based on Figure \ref{fig:appendix0} yielded a sample of only 279 JAGB stars, which is an insufficient number for a precise JAGB measurement.  
We note that although NGC 4038 technically `passed' our algorithm's criteria when the first derivative equals zero at $d\approx 20$~kpc,  the JAGB magnitude clearly increases again by $\sim0.3$~mag due to the presence of NGC 4039. Therefore, due to the exceptional circumstances of NGC 4038's merger interactions, we chose to discard NGC 4038 from our SN Ia calibration sample.

We note we also tried  (1) using circular ellipses centered on NGC 4038 (as we did with NGC 1365 in Section \ref{subsec:phot}), (2) analyzing photometry from only the right NIRCam module in Figure \ref{fig:appendix0}, (3) calculating the deprojected galactocentric radial distances of each star using varying inclination values, and (4) masking NGC 4039. Unfortunately, none of these combinations led to a clear convergence pattern of the JAGB magnitude with radial distance.

NGC 4038 has also proven a challenging target for Cepheid distances. \cite{2016ApJ...826...56R} first measured a distance to NGC 4038 of $\mu_0=31.29\pm0.11$~mag, and then re-measured this distance in \cite{2022ApJ...934L...7R} to be $\mu_0=31.62\pm0.12$~mag. These two distances differ by 16\%. This change resulted because \cite{2022ApJ...934L...7R} included ultra-long period Cepheids ($P>100$~days) that  \cite{2016ApJ...826...56R} originally excluded. Ultra-long period Cepheids are often found in unusual galaxies that are simultaneously star-forming and late-type 
yet metal-poor like NGC 4038, and can make distance determinations via Cepheids especially challenging because their P-L relation suffers from larger uncertainties than classical Cepheids P-L relations \citep{2009ApJ...695..874B, 2012Ap&SS.341..143F, 2017ApJ...836...74J}.

In conclusion, NGC 4038's unusual properties and current merger with NGC 4038 make it a difficult target for measuring distances for the Cepheids and the JAGB method.

\subsubsection{NGC 3972: A highly inclined galaxy lacking abundant intermediate-age populations}

NGC 3972 is a spiral galaxy at a distance of $\sim21$~Mpc. The Cepheid SHoES distance to NGC 3972 was measured to be $\mu_0=31.64\pm0.09$~mag \citep{2022ApJ...934L...7R}. 

We note NGC 3972 passed our initial convergence test in our first JWST CCHP exploratory paper \citep{2023arXiv231202282L}. In that paper, the JAGB magnitude converged to within 0.05~mag in the outer disk past 7.6~kpc. However, in this paper, the JAGB magnitude never converged, continually decreasing in total by 0.25~mag past 7.6~kpc. 
This can be explained by two changes  in our data processing; we are now using DOLPHOT v2.0 instead of the beta version. Second, the quality-metric cuts used to select stars were standardized for all galaxies in this paper, whereas we visually adjusted the quality-metric cuts to optimize the CMD for each individual galaxy in \cite{2023arXiv231202282L}.

Unlike NGC 4038 and NGC 4424, the reason for the lack of convergence in NGC 3972 is less clear; however, it is likely that
  NGC 3972's large inclination made it difficult to photometer the stars due to crowding effects.
This theory is supported by NGC 3972's surprisingly small sample of JAGB stars, which runs counter to expectations for a late-type star-forming galaxy.
With only $1519$~JAGB stars, NGC 3972 contained the second-smallest intermediate-age population in our sample, after NGC 4424 which contained 1397 JAGB stars (note the galaxy with the third largest intermediate-age population was NGC 7250 with 2917 JAGB stars, and the average number of JAGB stars in a given galaxy was 5000 for the full SN Ia sample). Therefore, galaxies with large inclinations may be difficult targets for JAGB measurements.

\subsubsection{Distances to NGC 3972, NGC 4038, NGC 4424}\label{subsubsec:alternateH0}

Although NGC 3972, NGC 4038, and NGC 4424 never technically converged according to our algorithm, we still visually selected  `outer disk` radial cuts to provide distances to these galaxies. 
Then, we  re-measured $H_0$ while including these galaxies in the SN Ia calibration.
The distances to these three galaxies and their uncertainties are presented in Tables \ref{tab:errors} and \ref{tab:galaxydistances}, although they were unused for our `primary' Hubble constant measurement. As described in Section \ref{subsec:h0} and in \cite{freedman24}, we re-measured $H_0$ by including NGC 4038, NGC 3972, and NGC 4424 to quantify the effect of their exclusion in our primary Hubble constant measurement. 

While the JAGB magnitude in NGC 3972 never converged, as shown in Figure \ref{fig:appendix0}, we selected a radial cut of 9.0~kpc.  In NGC 4038, while the JAGB magnitude technically converged at $d\approx20$~kpc, the presence of NGC 4039 from $28<d<38$~kpc made this galaxy pair an extraordinary exception to our algorithm. Nevertheless, we selected a radial cut of 42.0~kpc. Finally, in NGC 4424, we selected a radial cut of 7.3~kpc.

\begin{deluxetable*}{ccccccc}
\tablecaption{$m_{JAGB}$ Error Budget}\label{tab:errors}
\tablehead{
\colhead{Galaxy} & 
\colhead{Error on the mode} &
\colhead{Choice of $\sigma_s$ error} &
\colhead{Convergence error} &
\colhead{Aperture correction error} & 
\colhead{Extinction error} \\
\colhead{} &
\colhead{(stat)} &
\colhead{(stat)} &
\colhead{(stat)} &
\colhead{(sys)} &
\colhead{(sys)} 
}
\startdata
M101 & 0.01  & 0.03 &  0.004 & 0.03 &   0.01  \\
NGC 1365& 0.01  & 0.02  & 0.002 &  0.03 & 0.01   \\
NGC 2442& 0.01  &  0.00 & 0.005 &  0.03 & 0.03  \\
NGC 4536& 0.01  & 0.01 & 0.002  & 0.03&   0.01  \\
NGC 4639 & 0.01  & 0.02& 0.003 & 0.03&   0.01  \\
NGC 5643& 0.01  & 0.00 & 0.003 &0.03 &   0.02   \\
NGC 7250 & 0.01  & 0.02 & 0.002 &  0.03 & 0.02 \\
\hline
NGC 3972 &  0.01 & 0.03 & 0.018 &0.03 & 0.01  \\
NGC 4038 & 0.02 & 0.06 & 0.006 & 0.03 & 0.02 \\
NGC 4424 & 0.02 & 0.03 & 0.006 & 0.03 & 0.01
\enddata
\end{deluxetable*}

\begin{deluxetable}{cccccc}
\tablecaption{JAGB Distances}\label{tab:galaxydistances}
\tablehead{
\colhead{Galaxy} & 
\colhead{$\mu_0$} & 
\colhead{$\sigma_{stat}$} & 
\colhead{$\sigma_{sys}$} & 
\colhead{$d$} &
\colhead{No. JAGB stars}\\
\colhead{} & 
\colhead{(mag)} & 
\colhead{(mag)} & 
\colhead{(mag)} & 
\colhead{(kpc)} & 
\colhead{in Outer Disk}
}
\startdata
M101 & 29.21 & 0.03 & 0.03  & 7.0 & 1279 \\
NGC 1365 & 31.38 & 0.02 & 0.03 & 19.0 & 1242 \\
NGC 2442  & 31.60  & 0.01 & 0.04 & 21.0 & 931 \\
NGC 4536 & 30.97 & 0.01 & 0.03 & 15.7 & 2824\\
NGC 4639 & 31.73 & 0.02 & 0.03  & 22.1 & 1292\\
NGC 5643 & 30.58  & 0.01 & 0.04 & 13.1 & 1098\\
NGC 7250 & 31.59 & 0.02 & 0.04 & 20.9 & 848\\
\hline 
NGC 3972 & 31.67  & 0.04 & 0.03 & 21.7 & 829\\
NGC 4038 &  31.53 & 0.06 & 0.04 & 20.3 & 435\\
NGC 4424 & 31.15 &0.04 & 0.03 & 17.1 & 381\\
\enddata
\tablecomments{Note these distance uncertainties do not include the uncertainties from the JAGB zero-point listed in Table \ref{tab:zero}.}.
\end{deluxetable}

\subsection{Summary of Uncertainties}\label{subsec:unc}

In this section, we describe the known potential sources of uncertainty, which are also listed in Table \ref{tab:errors}. In Table \ref{tab:galaxydistances}, we list the final measured distance moduli and their total systematic and statistical uncertainties. 

\subsubsection{Statistical Uncertainties}\label{subsubsec:statunc}

\begin{enumerate}
    \item \textit{Error on the mode}. The dispersion of the JAGB LF encompasses uncertainties due to the JAGB stars' photometric errors, random errors due to the intrinsic variability of AGB stars, and differential extinction within the galaxy \citep{2001ApJ...548..712W}. The error on the mode, or the dispersion divided by the square root of the number of JAGB stars with magnitudes within $\pm0.75$~mag of the mode $m_{JAGB}$, therefore accounts for these uncertainties. 
    In all galaxies this error was measured to be 0.01~mag.

    This uncertainty calculation assumes the distribution of JAGB stars is Gaussian. To test whether this is a valid assumption for calculating the statistical uncertainty, we sub-divided the final distribution of $\sim2800$ JAGB stars in NGC 4536, the galaxy with the largest number of JAGB stars, into five radially-separated bins, each with $\sim570$~JAGB stars. Because the error on the mode for this total sample was measured to be 0.006~mag, the expected uncertainty in each sub-region should be $\sqrt{n/k} \times 0.006$, where $n$ is 2800 and $k$ is the number of JAGB stars in each subregion. We then measured the error on the mode in each region, finding: {0.015, 0.014, 0.014, 0.013, 0.013}~mag in each subregion, compared to the expected error of 0.014~mag. This test validated the scatter is close to Gaussian and that our method for using the error on the mode as a statistical uncertainty is applicable and valid. 
    \item \textit{Choice of smoothing parameter $\sigma_s$}. The choice of smoothing parameter $\sigma_s$ may affect the measured mode of the JAGB star LF. To test for any such statistical effects, we re-smoothed the JAGB LF in each galaxy using different smoothing parameters, iterating through \{0.15, 0.20, 0.25, 0.30, 0.35, 0.40\}~mag and then re-measured the mode. Then, we defined  the statistical error due to the choice of smoothing parameter as the maximum difference between the fiducial mode (measured with a smoothing parameter of $\sigma_s=0.25$~mag for all galaxies) and any of the measured modes. This approach has been previously adopted by the CCHP (e.g., \citealt{2023arXiv231202282L, fourstar}). 
    
    We note the mode measured from an increasingly smoothed luminosity function will eventually converge to the mean. Therefore, this `smoothing parameter uncertainty'  encapsulates any systematic differences incurred from the choice of JAGB statistic (i.e., mode vs. mean vs. median). We demonstrate in Appendix \ref{sec:mean_median} that if we use the mean instead of the mode as the chosen JAGB statistic, the distance moduli were measured to be $0.030$~mag brighter on average. This systematic offset is fully encapsulated within the smoothing parameter uncertainty of our zeropoint NGC 4258 (0.04 mag). 
    \item \textit{Convergence error}. We adopted a statistical error due to the fluctuations about the final converged JAGB magnitude in Figure \ref{fig:settledown_main}. This uncertainty was derived from the dispersion about all measured $m_{JAGB}$ outside of the radial cut, divided by the square root of the number of bins.  For example, the convergence plot outside of the radial cut in NGC 7250 barely fluctuates; NGC 7250 has a `convergence error' of 0.002 mag. On the other hand, for a galaxy like NGC 2442 which has noisier fluctuations, the measured convergence error was measured to be 0.005~mag. For NGC 3972, since the JAGB magnitude never converged, we calculated the $\sigma$ as half the full range of fluctuations in $m_{JAGB}$ past 9~kpc.

\end{enumerate}

\subsubsection{Systematic Uncertainties}

\begin{enumerate}

    \item \textit{Aperture corrections}. 
    We conservatively adopted a 0.02~mag uncertainty for the aperture correction. A more detailed explanation on how we arrived at this uncertainty can be found in our photometry overview paper (Jang et al., in prep). 
    \item \textit{Foreground extinction.} 
    \cite{1998ApJ...500..525S} cites an uncertainty of 16\% on the foreground extinction values from their dust map, which we adopted for our galaxies with large ($A_{F115W}>0.1$~mag) reddening values.  For our galaxies with small reddening values ($A_{F115W}<0.02$~mag), we adopted half the reddening value as its uncertainty. This approach has been adopted by the CCHP in all our distance scale papers (first established in \citealt{2018ApJ...861..104H}).
\end{enumerate}

\subsection{Adopted Zero-point Calibration in NGC 4258}

\begin{figure*}\figurenum{7}
\gridline{\fig{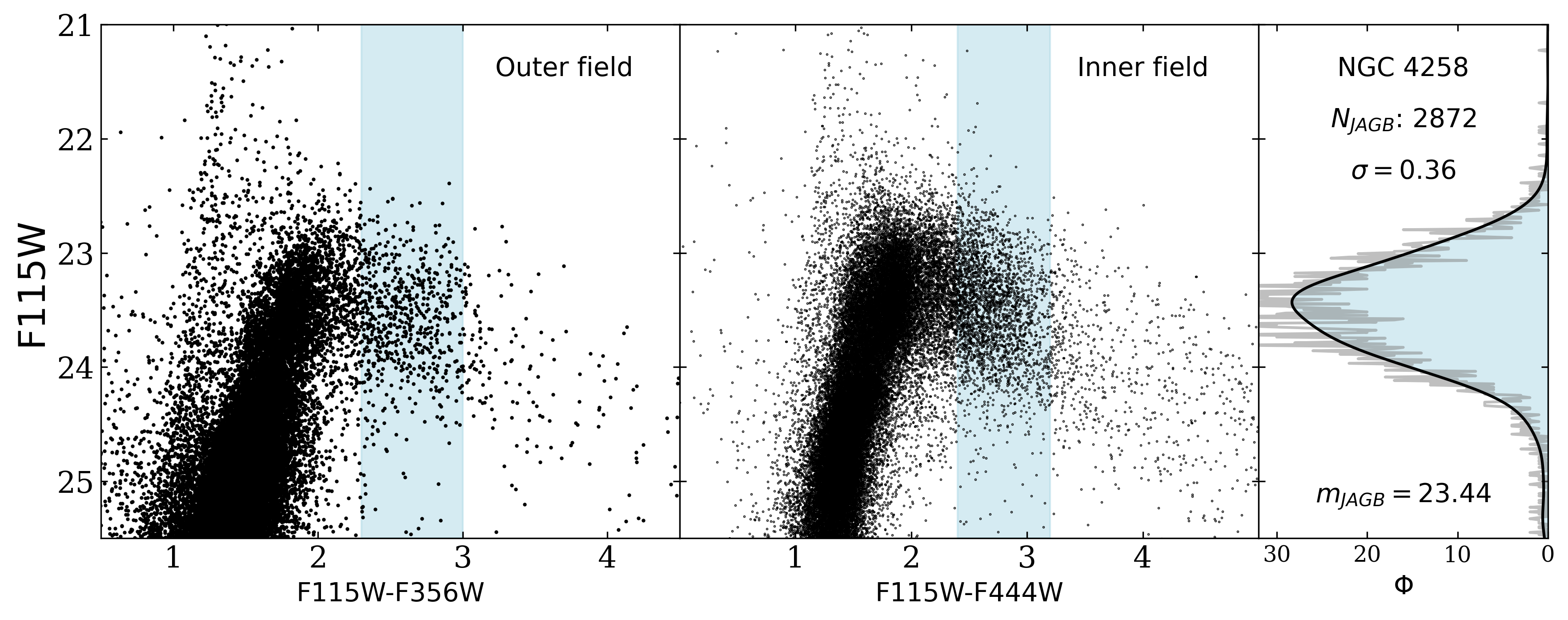}{\textwidth}{}
          {}
          }   
\caption{
(Left and middle panels) Color-magnitude diagrams for the two fields in our geometric anchor, NGC 4258. The JAGB stars were selected within the light blue shaded regions. (Right panel) The JAGB stars in the inner and outer fields were combined to make this aggregate JAGB LF. The number of JAGB stars within $\pm0.75$~mag of the mode is plotted in the upper right corner, as well as the dispersion for those stars about the mode.
The measured JAGB magnitude is also shown in the bottom right.
}\label{fig:n4258_cmd}
\end{figure*}

\begin{figure}
\centering
\includegraphics[width=\columnwidth]{"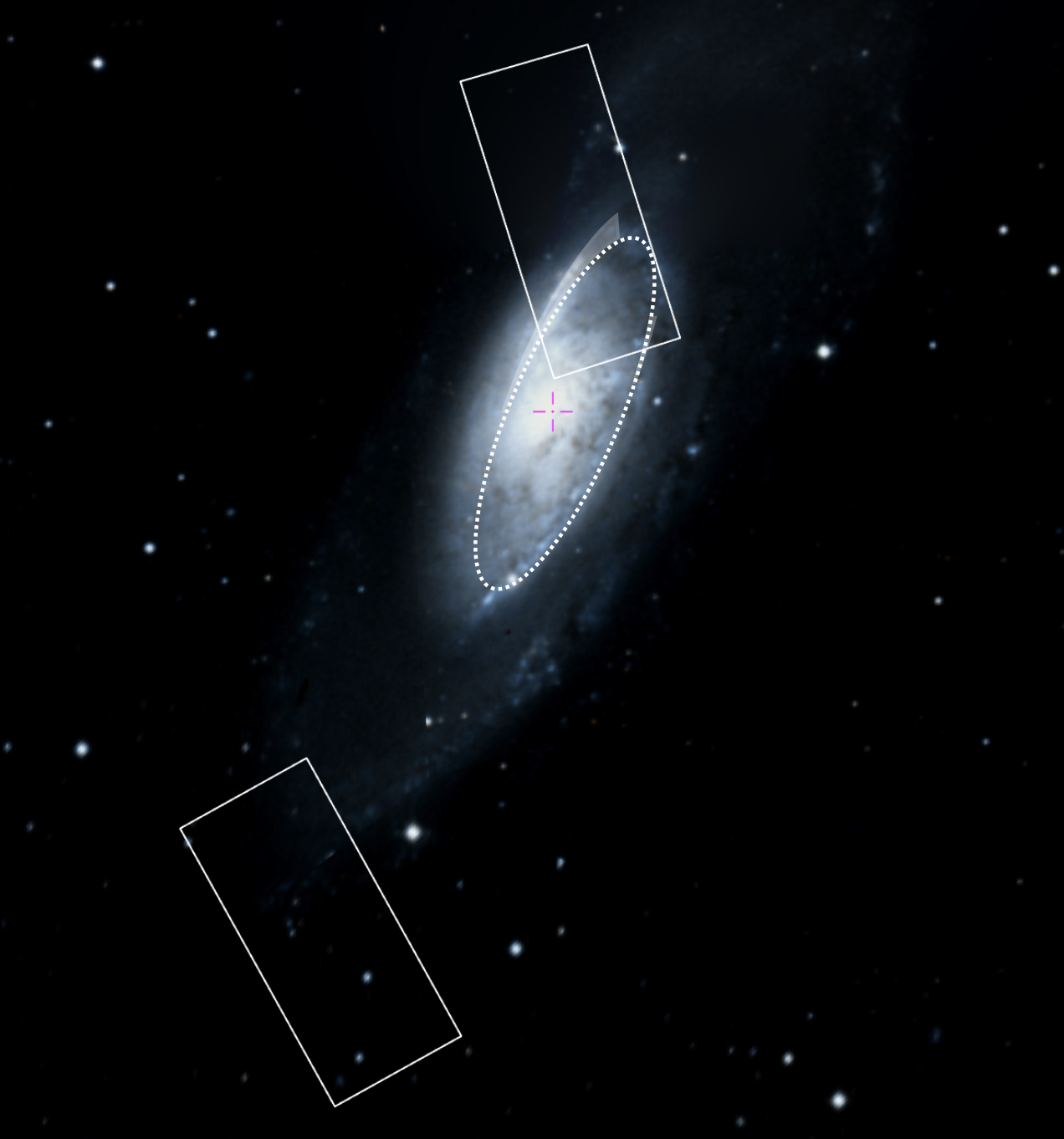"}\figurenum{8}
\caption{Our two NIRCam pointings of NGC 4258: the `inner disk field' in the upper right and the `outer disk field' in the bottom left. We only utilized data in the `outer disk,' outside of the dotted white ellipse. We also masked a spiral arm in the inner field within the shaded grey region. 
}
\label{fig:n4258}
\end{figure}

\begin{deluxetable}{cccc}
\tablecaption{Uncertainty in the JAGB Zero-point in NGC 4258}\label{tab:zero}
\tablehead{
\colhead{Source} & 
\colhead{Value (mag)} & 
\colhead{$\sigma_{stat}$ (mag) } & 
\colhead{$\sigma_{sys}$ (mag)}}
\startdata
$m_{JAGB}$ & 23.44 & 0.01 & \nodata \\
Aperture correction & \nodata  & \nodata& 0.02 \\
Choice of $\sigma_s$ & \nodata &  0.04& \nodata\\
Convergence error & \nodata & 0.004 & \nodata\\
NIRCam ZP & \nodata & \nodata & 0.02 \\
$A_{F115W}$ & 0.02 & \nodata & 0.01 \\
$\mu_0$ & 29.40 & 0.02 & 0.02 \\
\hline
$M_{JAGB}$ & $-5.98$ & 0.05 & 0.04
\enddata
\end{deluxetable}

The nearby water-megamaser galaxy NGC 4258 (P.A. = 150.0$^{\circ}$, $i=73^{\circ}$, $d=7.6$~Mpc) anchors our extragalactic distance scale via its 1.5\% geometric distance measured by \cite{2019ApJ...886L..27R}.
By mapping the proper motions of the water masers rotating around the black hole in the center of NGC 4258,
the positions, velocities, and accelerations of the masers were modeled to give the most precise physical distance to NGC 4258 to date: $\mu_0=29.40\pm~0.02$ (stat) $\pm~0.02$ (sys)~mag.

We now use the \cite{2019ApJ...886L..27R} distance to calibrate the JAGB method's absolute magnitude in NGC 4258.
First, we observed NGC 4258 in two NIRCam pointings, which covered the southeastern outer disk and northwestern inner disk (total exposure time in both fields = 2802 seconds), as shown in Figure \ref{fig:n4258}. 
We ran our convergence test on the combined JAGB star sample from the outer and inner disk fields. 
We also masked out the spiral arm in the inner disk field, as shown in Figure \ref{fig:settledown_main}, to eliminate its contribution to the noisiness of the convergence pattern.
The CMDs from both the inner and outer fields as well as their combined JAGB LF are shown in Figure \ref{fig:n4258_cmd}.

The JAGB magnitude in the combined inner and outer disk JAGB sample was measured to be $m_{JAGB}=23.44$~mag.
We corrected this for a Milky Way foreground extinction of $A_{F115W}=0.02$~mag \citep{1998ApJ...500..525S}. This yielded a JAGB zeropoint of $M_{JAGB}= -5.98\pm0.05$ (stat)~$\pm~0.04$ (sys)~mag. 
The sources of uncertainty on this zeropoint are summarized in Table \ref{tab:zero}.  In addition, we adopted a 1\% systematic error (0.02~mag) for the uncertainty on the NIRCam photometric zero-point. This uncertainty came from STScI, which currently quotes the NIRCam absolute flux uncertainties at $\lesssim1\%$.\footnote{\url{https://jwst-docs.stsci.edu/jwst-calibration-status/nircam-calibration-status/nircam-imaging-calibration-status}} 

We also measured the JAGB magnitude in the inner and outer fields separately, after adopting the radial cut of $d>7.3$~kpc. In the inner field, we measured a JAGB magnitude of $m_{JAGB}=23.41\pm0.04$ (stat)~mag, with $\sim2300$ JAGB stars contributing to the measurement. In the outer field, we measured a JAGB magnitude of $m_{JAGB}=23.51\pm0.06$ (stat)~mag, with $\sim600$ JAGB stars contributing to the measurement. The statistical uncertainty was calculated by adding in quadrature the  the error on the mode and  the smoothing parameter uncertainty. These two measurements agree to within $1.2\sigma$.

\subsection{Measurement of the Hubble constant}\label{subsec:h0}
Details of the JAGB calibration of SNe Ia, as well as comparisons between the JAGB distances measured in this paper with TRGB distances, can be found in the CCHP $H_0$ results paper \cite{freedman24}. We  give a brief summary here. 

\cite{freedman24} applied the JAGB distances measured in this paper to SNe Ia in the Carnegie Supernova Program (CSP; \citealt{2024ApJ...970...72U}). In our sample of seven galaxies, there are eight SN Ia calibrators (NGC 5643 contains two SN Ia). These distances were used as inputs to the MCMC analysis described in \cite{freedman24} for the CSP sample. 
The CSP JAGB $H_0$ was determined to be $H_0=67.80\pm 2.17 ~\rm{(stat)}\pm 1.64 ~\rm{(sys)}~\rm{km ~s^{-1}~Mpc^{-1}}$.  

In \cite{freedman24}, we also measured a Hubble constant by including NGC 3972, NGC 4038, and NGC 4424 as calibrators, as discussed in Section \ref{subsubsec:alternateH0}. Including these three additional SN Ia delivered a Hubble constant of $H_0=68.75\pm 2.23 ~\rm{(stat)}\pm 1.65 ~\rm{(sys)}~\rm{km ~s^{-1}~Mpc^{-1}}$.

\section{Summary \& Future Outlook}\label{sec:conclusion}
We have established a new SNe Ia distance scale calibrated by JAGB stars, that is parallel and independent to the TRGB and Cepheid distance scales. 
We developed a new methodology for selecting the `outer disk' for JAGB measurements, by determining when the JAGB magnitude converges in the outer regions of a galaxy. We caution JAGB measurements in the `inner disk' regions of galaxies will likely be significantly affected by crowding. For example, we found the average difference between the JAGB magnitudes measured in the inner disks and the outer disks in our sample of seven galaxies to be: $\rm{<JAGB_{inner} - JAGB_{outer}}> = -0.21$~mag, meaning the JAGB magnitude was on average 0.21 mag brighter in the inner disks of these galaxies.

We used JWST imaging to measure JAGB distances to seven SN Ia host galaxies, performing our analysis completely blinded. We determined a value of the Hubble constant of $H_0=67.80\pm 2.17 ~\rm{(stat)}\pm 1.64 ~\rm{(sys)}~\rm{km ~s^{-1}~Mpc^{-1}}$ \citep{freedman24}. This value is in excellent agreement with the CMB value measured by Planck of $H_0=67.4\pm0.5 ~\rm{km ~s^{-1}~Mpc^{-1}}$ \citep{2020A&A...641A...6P}.
The JAGB distances measured in this paper can cross-check the distances to the same galaxies measured via Cepheids and the TRGB (using the same JWST imaging), respectively presented in \cite{hoyt25} and Owens et al. (in preparation). We present these comparisons in our CCHP JWST $H_0$ results paper, \cite{freedman24}.

In the next few years, the JAGB method will continue to improve, aided by several upcoming studies and developments from new telescopes. We list four of them below.

\begin{enumerate}
    \item \textit{Upcoming NIR Facilities.} The upcoming Euclid mission and Nancy Grace Roman Telescope have large field-of-views which will be able to resolve stellar populations well out into the stellar halo of all nearby galaxies, offering plentiful opportunities to continue to test and characterize the JAGB method. Specifically, we can continue to apply and improve our `convergence algorithm' in nearby galaxies for which we will have significantly wider areal coverage.
    \item \textit{JWST.} The JWST will continue to provide imaging of SN Ia host galaxies. We will continue to measure their JAGB distances to increase the number of Type Ia supernovae calibrated by the JAGB method. As discussed in \cite{freedman24}, more JAGB distances to SN hosts are needed to provide a more representative sample of SNe Ia. Therefore, additional JWST data of farther SN Ia host galaxies, particularly at distances beyond 40~Mpc, will be imperative in definitively ruling out systematic effects in the local distance scale measurement of $H_0$. 
    \item \textit{JAGB Metallicity Tests.} \cite{2023arXiv230502453L} demonstrated that the JAGB method has zero metallicity or age dependence in the disk of the galaxy M31, by directly comparing the shape of the JAGB star LF to maps of metallicity and age. We plan to extend this study to more metal-poor galaxies, like M33.
    \item \textit{Gaia Parallaxes.} The release of Gaia DR4 in 2026 is also expected to improve the systematic error on the parallax zero-point, which has hindered previous 1\%  zeropoint calibrations of the Cepheid, TRGB, and JAGB distance scales. With \textit{Gaia}, we will add the Milky Way as a geometric anchor, thereby decreasing the JAGB zeropoint statistical and systematic uncertainties. 
\end{enumerate}

\begin{acknowledgments}
We thank Andy Dolphin, Dan Weisz, and the JWST Resolved Stellar Populations ERS team for developing the NIRCam module of DOLPHOT and for continually helping us troubleshoot issues. We thank Siyang Li and Adam Riess for  discussions which improved this paper. 
A.J.L. thanks Myung Gyoon Lee for useful discussions about NGC 4038/NGC 4039, and for hosting her at Seoul National University in Summer 2023, during which part of the data for this project were analyzed. Finally, we thank the referee for their constructive comments and suggested which improved this work.

A.J.L. was supported by the Future Investigators in NASA Earth and Space Science and Technology (FINESST) award number 80NSSC22K1602 during the completion of this work. We also thank the University of Chicago and the Observatories of the Carnegie Institution for their support of our long-term research into the calibration and determination of the expansion rate of the universe.

This research has made use of NASA's Astrophysics Data System Bibliographic Services. This research has made use of the NASA/IPAC infrared Science Archive (IRSA), which is operated by the Jet Propulsion Laboratory, California Institute of Technology, under contract with the National Aeronautics and Space Administration.

All the {\it JWST} imaging used in this paper can be found in MAST: \dataset[10.17909/ecf8-2z68]{http://dx.doi.org/10.17909/ecf8-2z68}. The photometry of the JAGB stars for every galaxy can be found at doi:\href{https://zenodo.org/records/14502265}{10.5281/zenodo.14502265}.

\end{acknowledgments}

\facilities{JWST (NIRCam)}
\software{Astropy \citep{2013A&A...558A..33A, 2018AJ....156..123A,2022ApJ...935..167A}, Dolphot \citep{2000PASP..112.1383D, 2016ascl.soft08013D}, Matplotlib \citep{2007CSE.....9...90H}, NumPy \citep{2020Natur.585..357H}, Pandas \citep{pandas}, scipy \citep{2020NatMe..17..261V}}

\appendix
\restartappendixnumbering

\section{Masking spiral arms}\label{sec:mask}
The following functions were used to mask the spiral arms in M101, NGC 2442, NGC 4258, and NGC 4639, where $y$ is the declination, $x$ is the right ascension, and $d$ is the radial distance: 
\begin{itemize}
    \item M101: $y<2.17x - 402.55$; $y>2.17x - 402.58$
    \item NGC 2442: $y>-1.59 x^2 + 364.14x -20829.12$; 
    $y<-1.59 x^2 + 364.14 x -20829.10$
    \item NGC 4258: $y<-15.00x^2 + 5541.08x -511678.75$;
    $y>-15.00x^2 + 5541.08x -511678.77$; $x>184.71$
    \item NGC 4639: (spiral arm 1) $d>7.0$; $d<9.0$;$y<13.27$; $y>13.24$;  $x<190.72$. (spiral arm 2) $d>7.5$; $d<10$;$y<13.28$; $y>13.25$;  $x>190.72$. 
\end{itemize}

If we chose to keep the spiral arms unmasked, the JAGB magnitudes in M101, NGC 2442, NGC 4258, and NGC 4639 would have been instead respectively measured to be: 23.21 (0.03 mag brighter), 25.79 (0.01 mag brighter), 23.44 (same magnitude), and 25.76 (0.01 mag brighter). This resulted in distance moduli that were on average 0.007 mag brighter for the seven SN Ia host galaxies, or a 0.3\% larger $H_0$.

\section{`Inner Disk' Color-magnitude diagrams}\label{sec:inner}
\begin{figure*}
\gridline{\fig{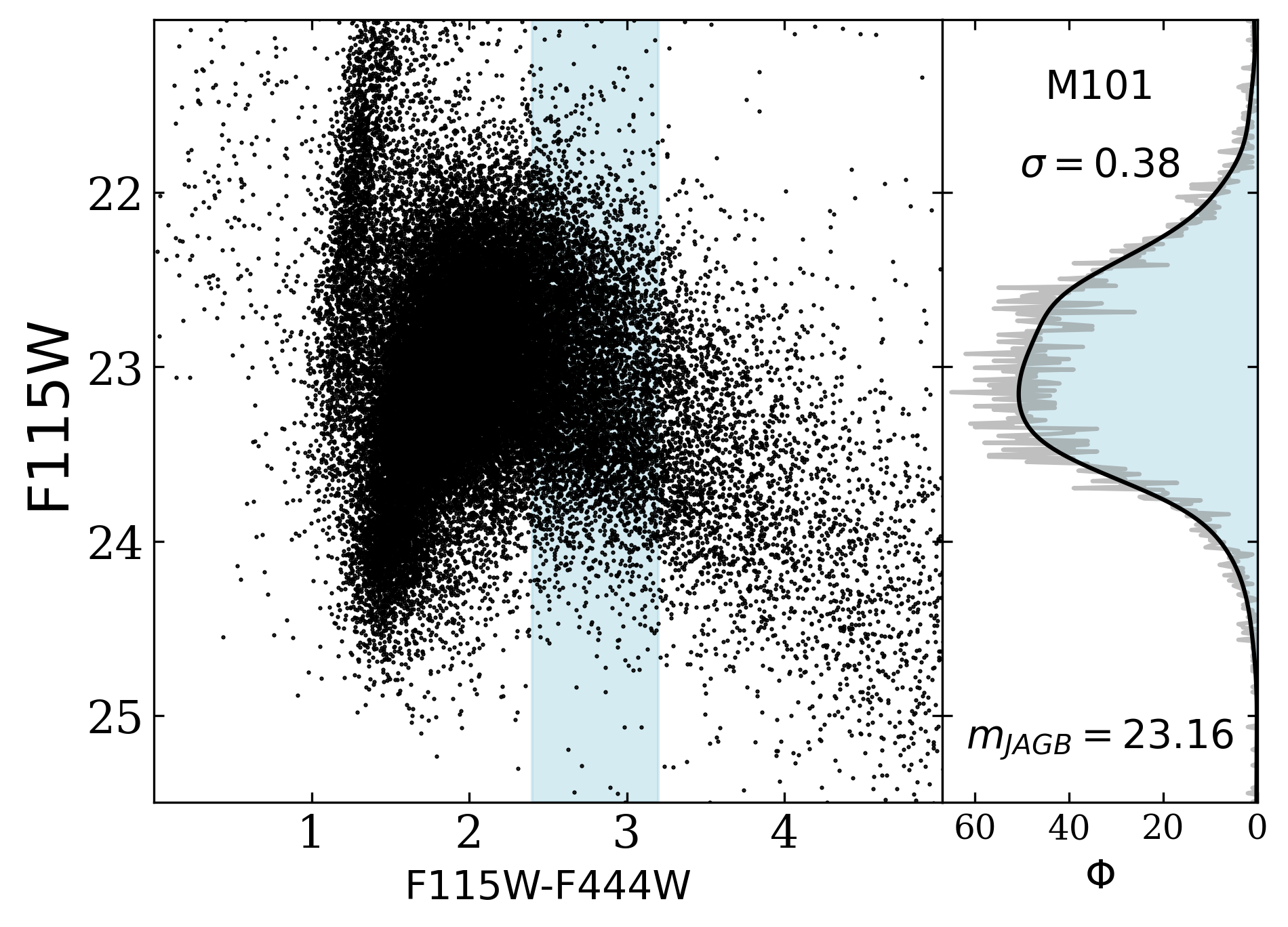}{0.333\textwidth}{}
          \fig{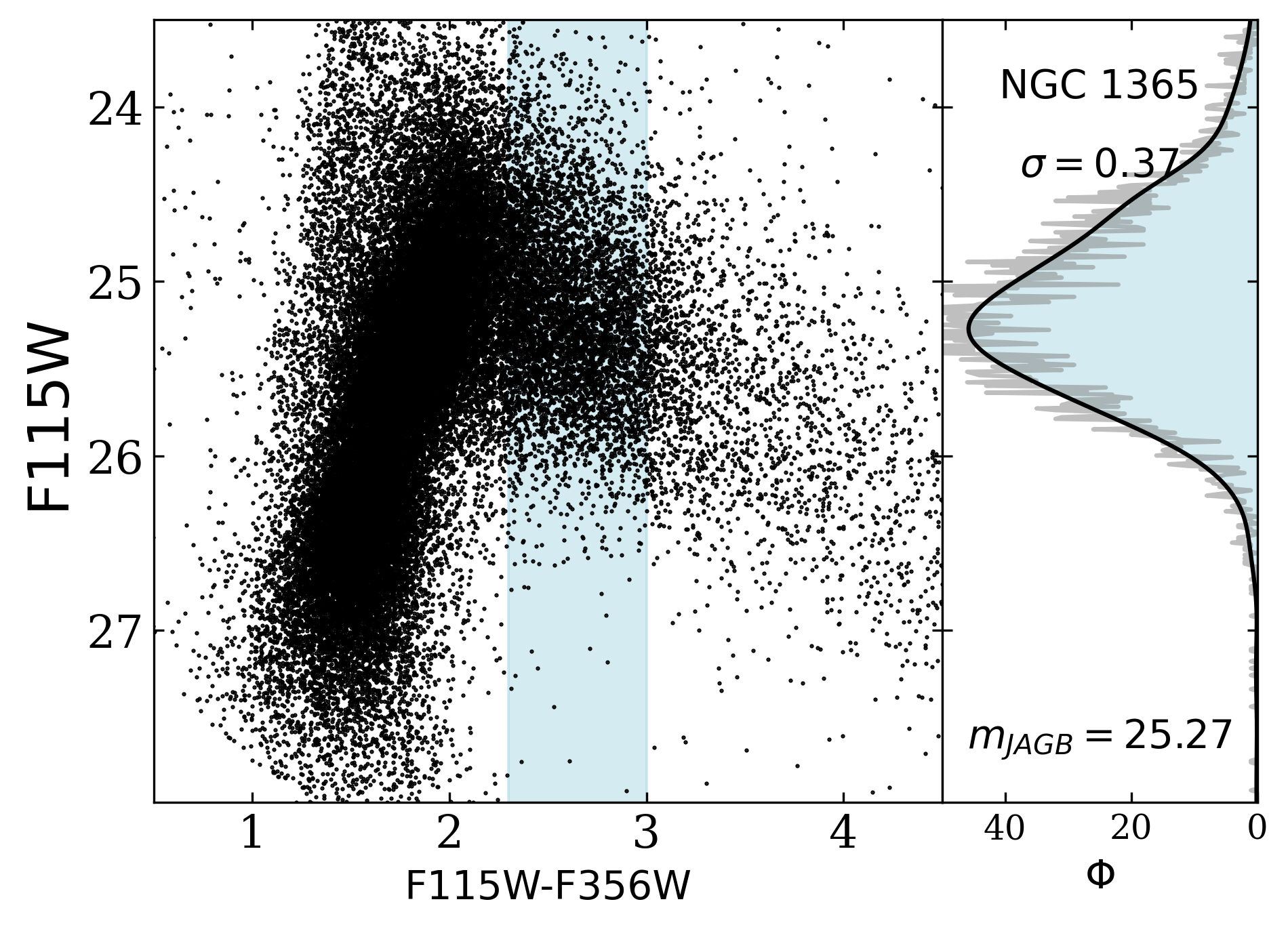}{0.333\textwidth}{}
                    \fig{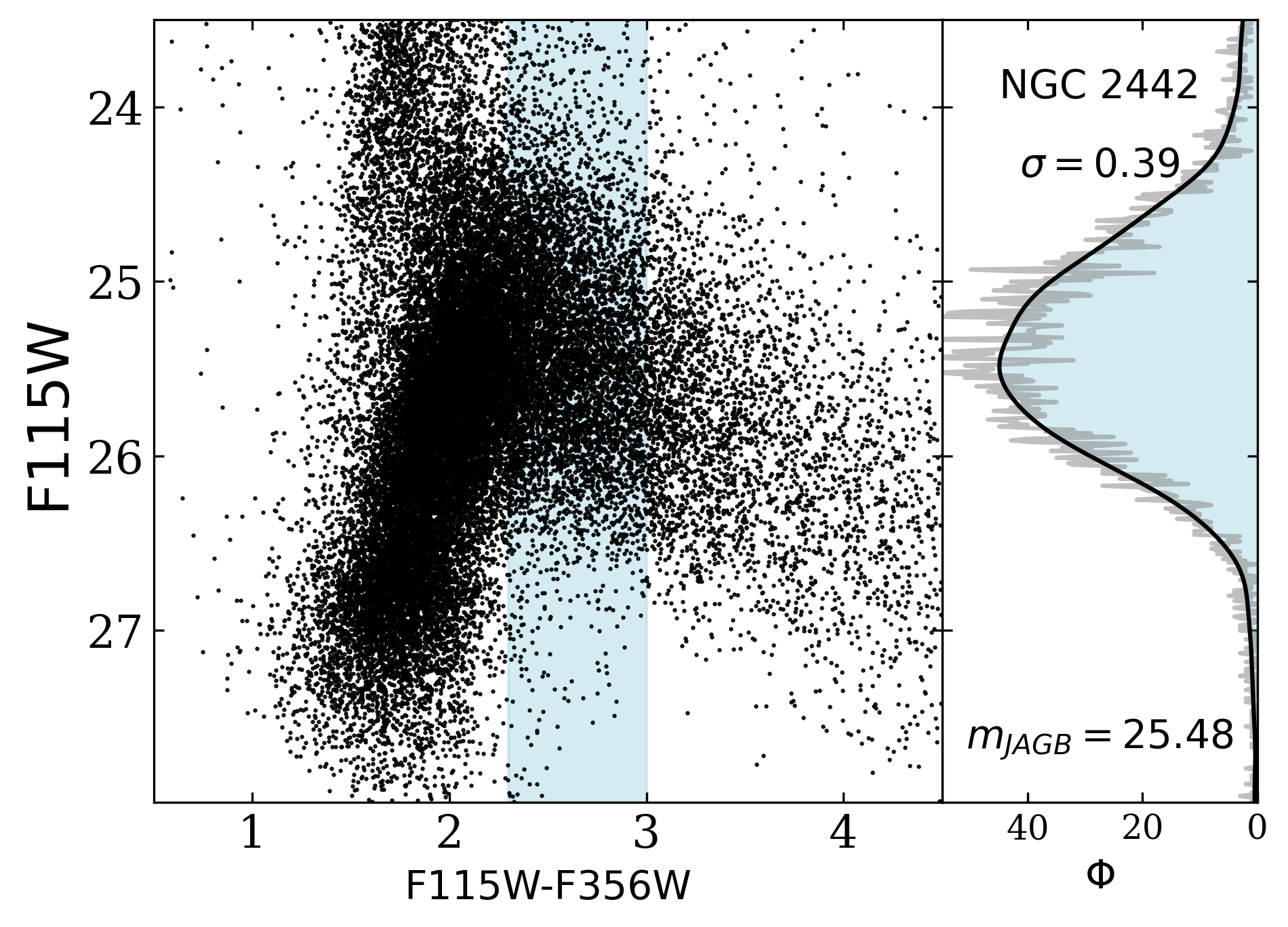}{0.333\textwidth}{}
          }
\gridline{\fig{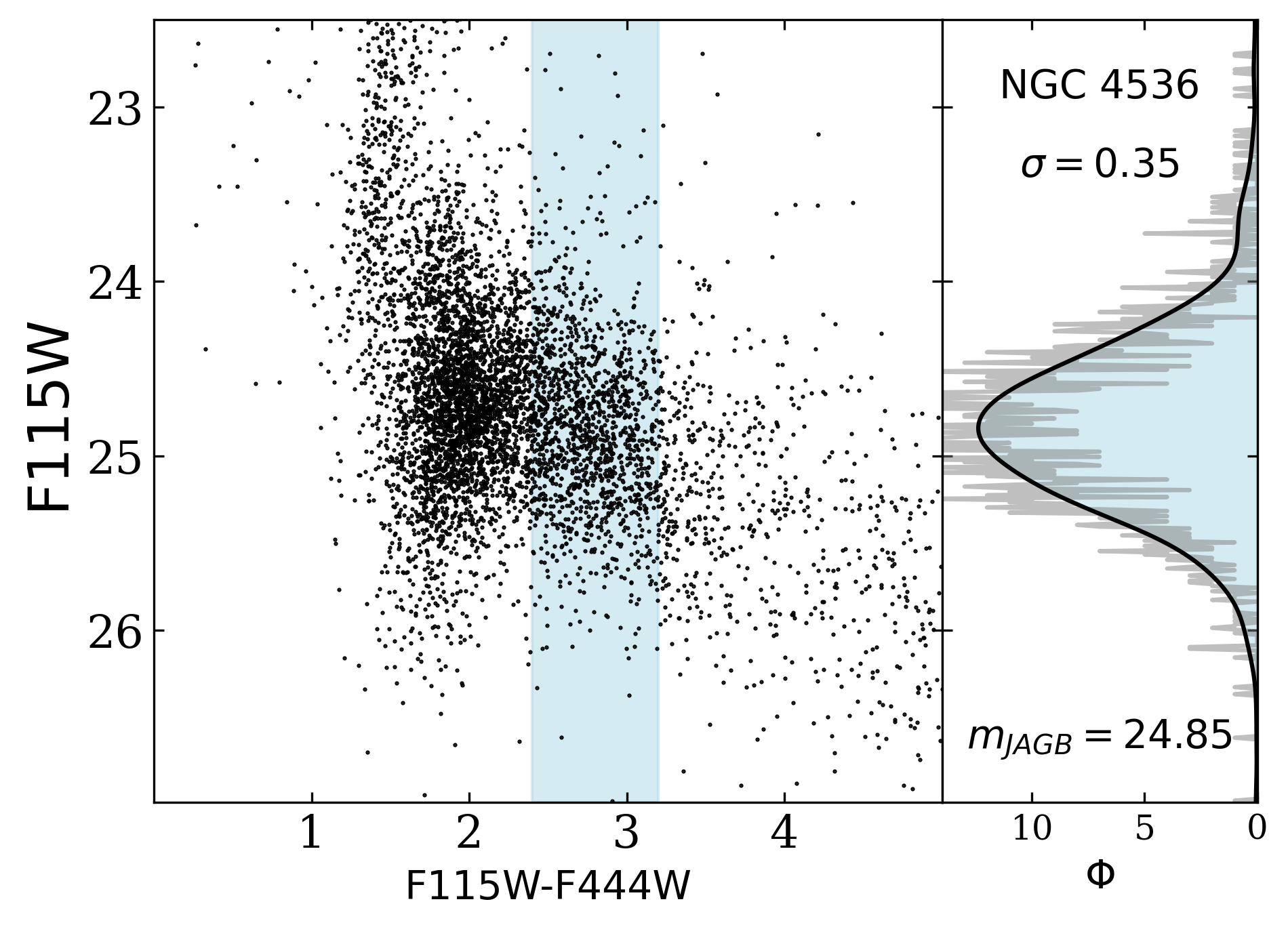}{0.333\textwidth}{}
          \fig{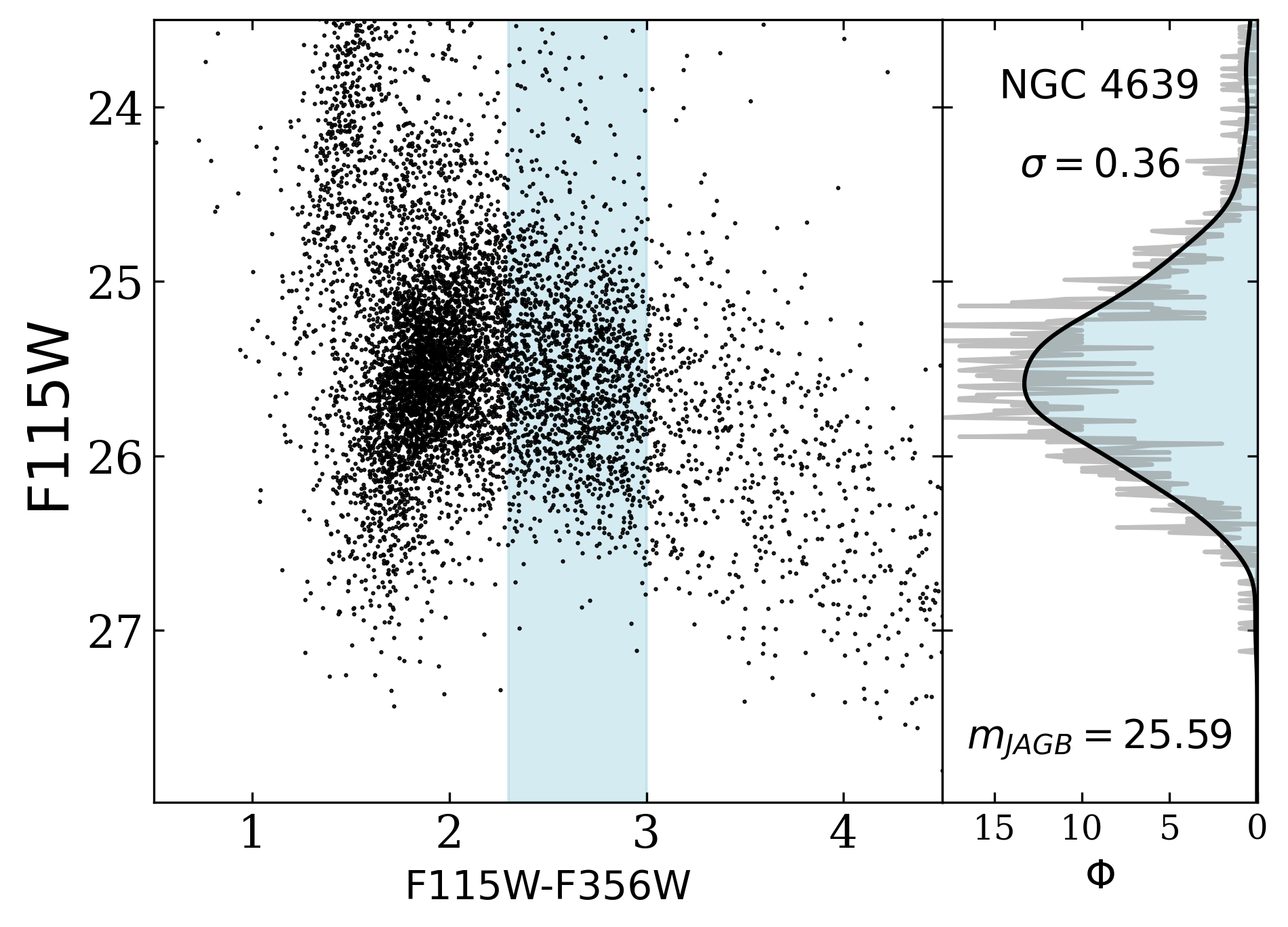}{0.333\textwidth}{}
                    \fig{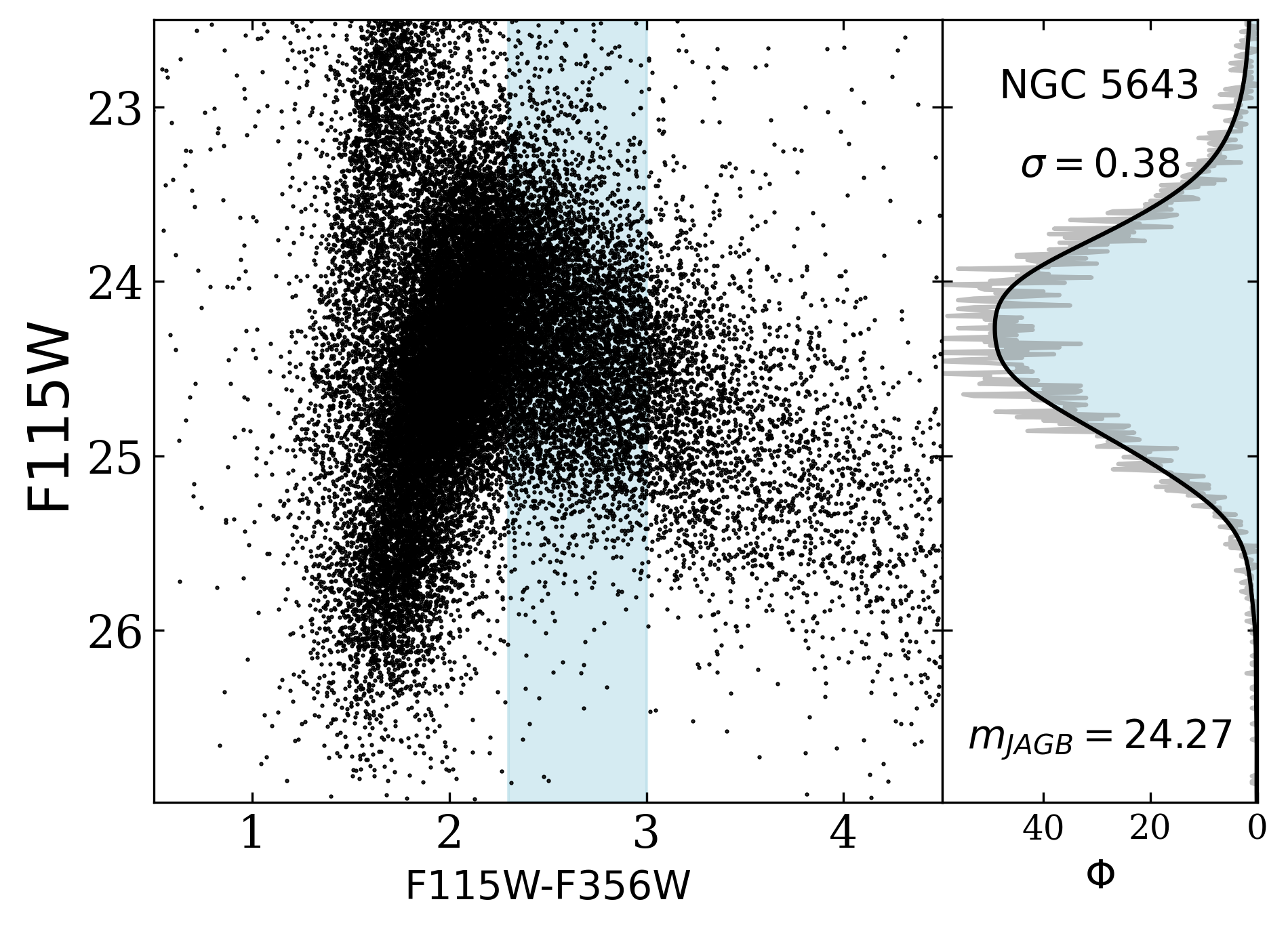}{0.333\textwidth}{}
          }
\gridline{\fig{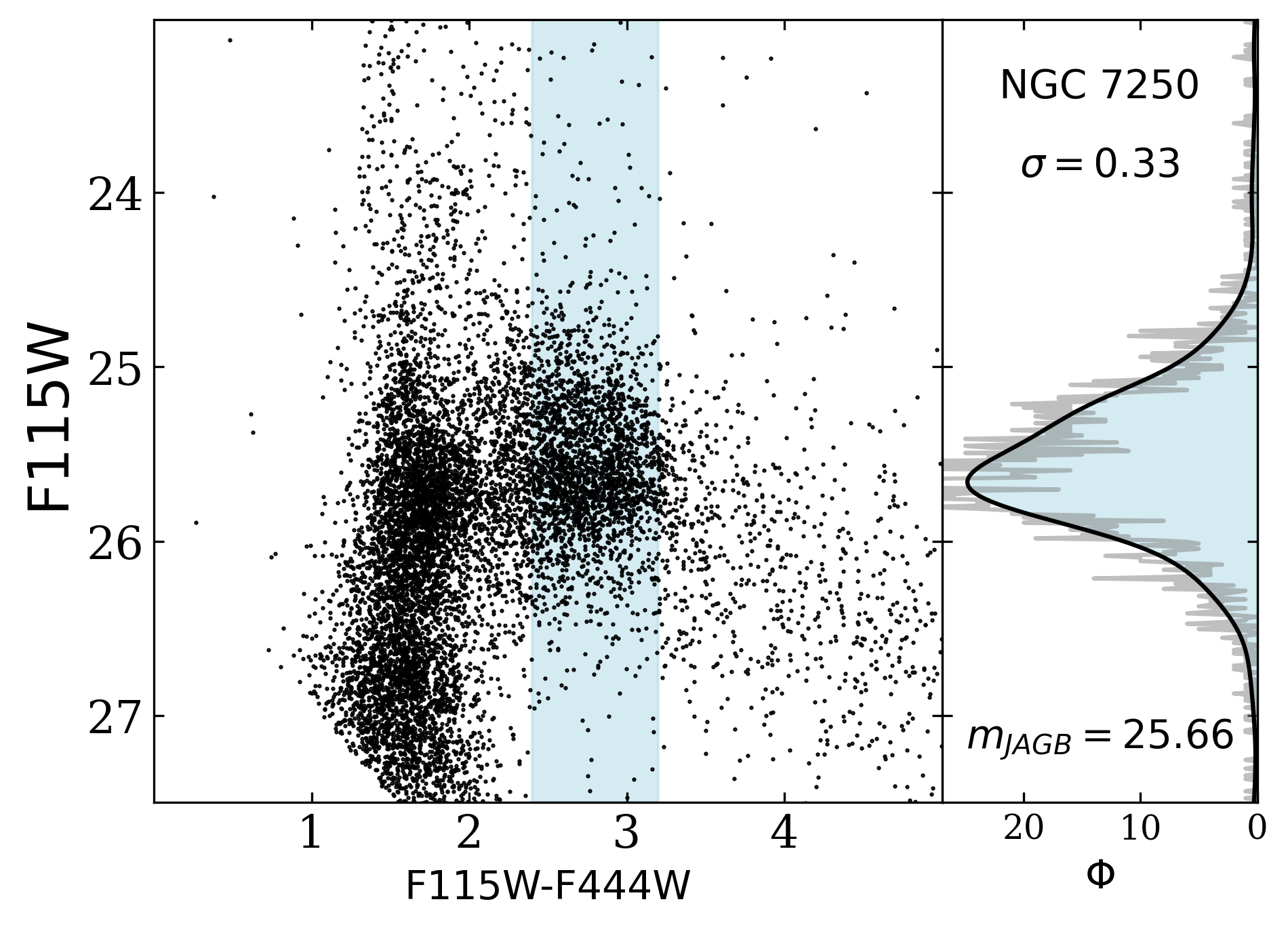}{0.333\textwidth}{}
          } \figurenum{A1}
\caption{CMDs (left panels) and JAGB LFs (right panels) for the inner regions of the SN Ia host galaxies, i.e. the data that were excluded from our analysis. 
 The JAGB stars were selected within the light blue shaded regions. The y-axis range is 4.5~mag for all galaxies.
The measured JAGB magnitude for each galaxy is also shown in the bottom righthand of each plot.
The JAGB luminosity functions in the inner regions have larger dispersions and are often skewed, likely due to a confluence of reddening, crowding, and blending effects. 
}\label{fig:CMD_inner}
\end{figure*}

In this section we show CMDs for the photometry in the `inner disk' region of the seven SN Ia host galaxies, i.e. the photometry \textit{inside} the dotted ellipses in Figure \ref{fig:images}.
These CMDs are shown in Figure \ref{fig:CMD_inner}.

The JAGB LFs exhibit more skew (relative to the JAGB LFs in the outer regions)  in the inner regions of M101, NGC 1365, NGC 2442, and NGC 7250.
This effect  was also observed by \cite{2023arXiv230502453L}, who hypothesized  skew in the JAGB LF may result from a confluence of reddening, crowding, and blending effects. Furthermore, in all the galaxies in our sample, the dispersion of the JAGB LF increases in the inner regions. For these reasons, we continue to emphasize the JAGB magnitude is most accurate and precise as a standard candle when measured in the outer disks of galaxies.

\section{Mean vs. Median vs. Mode}\label{sec:mean_median}

In this section, we test how choosing the mode, median, or mean as the JAGB magnitude affected the final measured distances and therefore our measurement of $H_0$. 
We adopted the mode as the JAGB magnitude because it is the most robust to outliers.
Other independent groups have chosen different statistics. \cite{2021arXiv210502120Z} use the mean value of a superimposed Gaussian function and quadratic function fit to the JAGB LF. \cite{2020MNRAS.495.2858R, 2021MNRAS.501..933P, 2023MNRAS.tmp..926P} employ the median value of a modified Lorentzian function fit to the JAGB LF. They then use either the LMC or SMC as a calibrator depending on the skew of the target galaxy. \cite{2024ApJ...966...20L} compared all three of these aforementioned statistics along with the straight mean and median, finding discrepancies of up to 0.2 mag between the various methods within the same host galaxy (which therefore yielded differences in $H_0$ of up to 9\%). However, if the JAGB LF is closely Gaussian and symmetric (which we found for all the galaxies in our sample), the mean, median, and mode of the LF should agree well. Therefore, the choice of JAGB statistic should not significantly change $H_0$. 
We now test how our choice of the mode over the mean and median affected our final measured $H_0$.

\begin{figure*}
\gridline{\fig{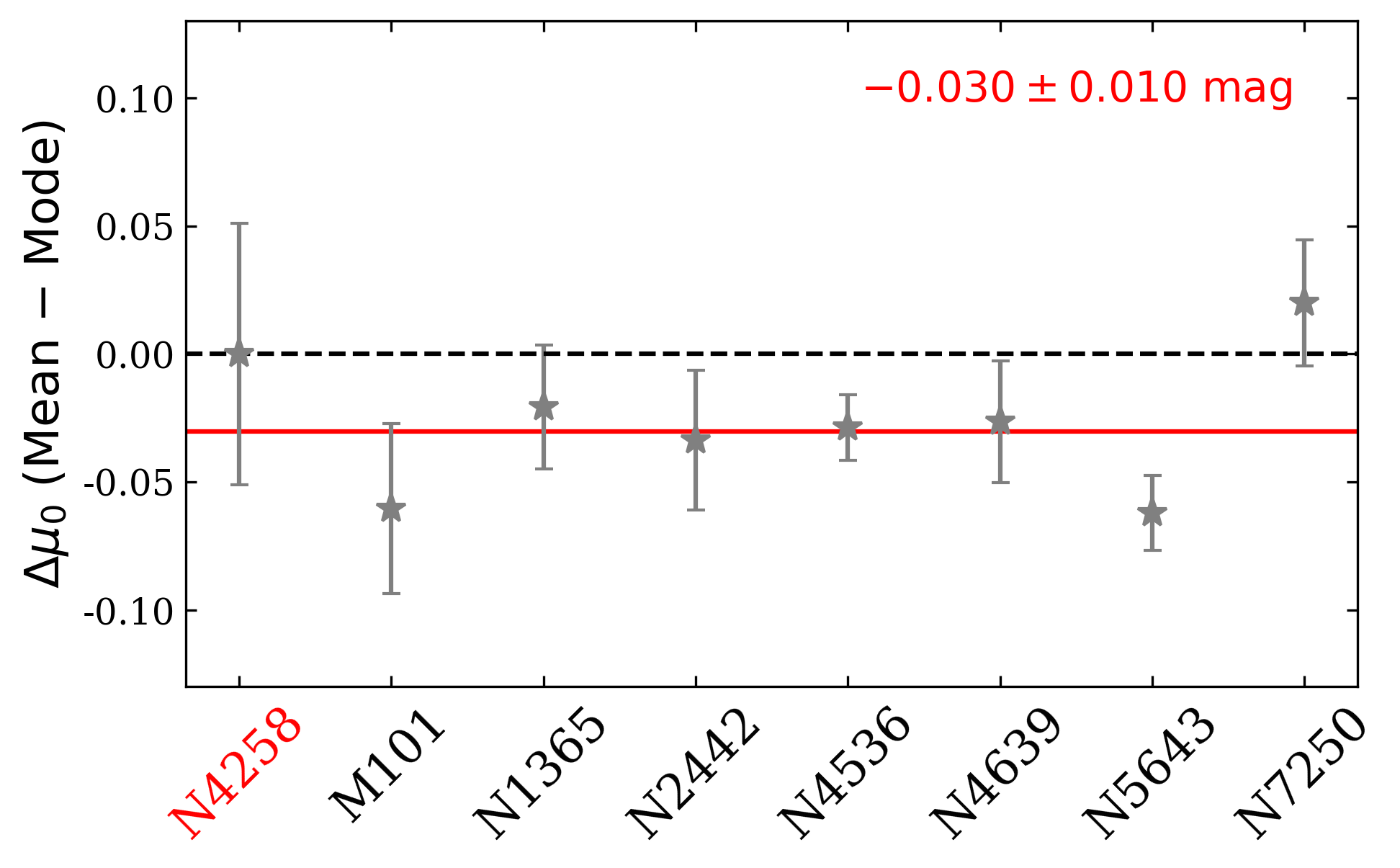}{0.5\textwidth}{}
          \fig{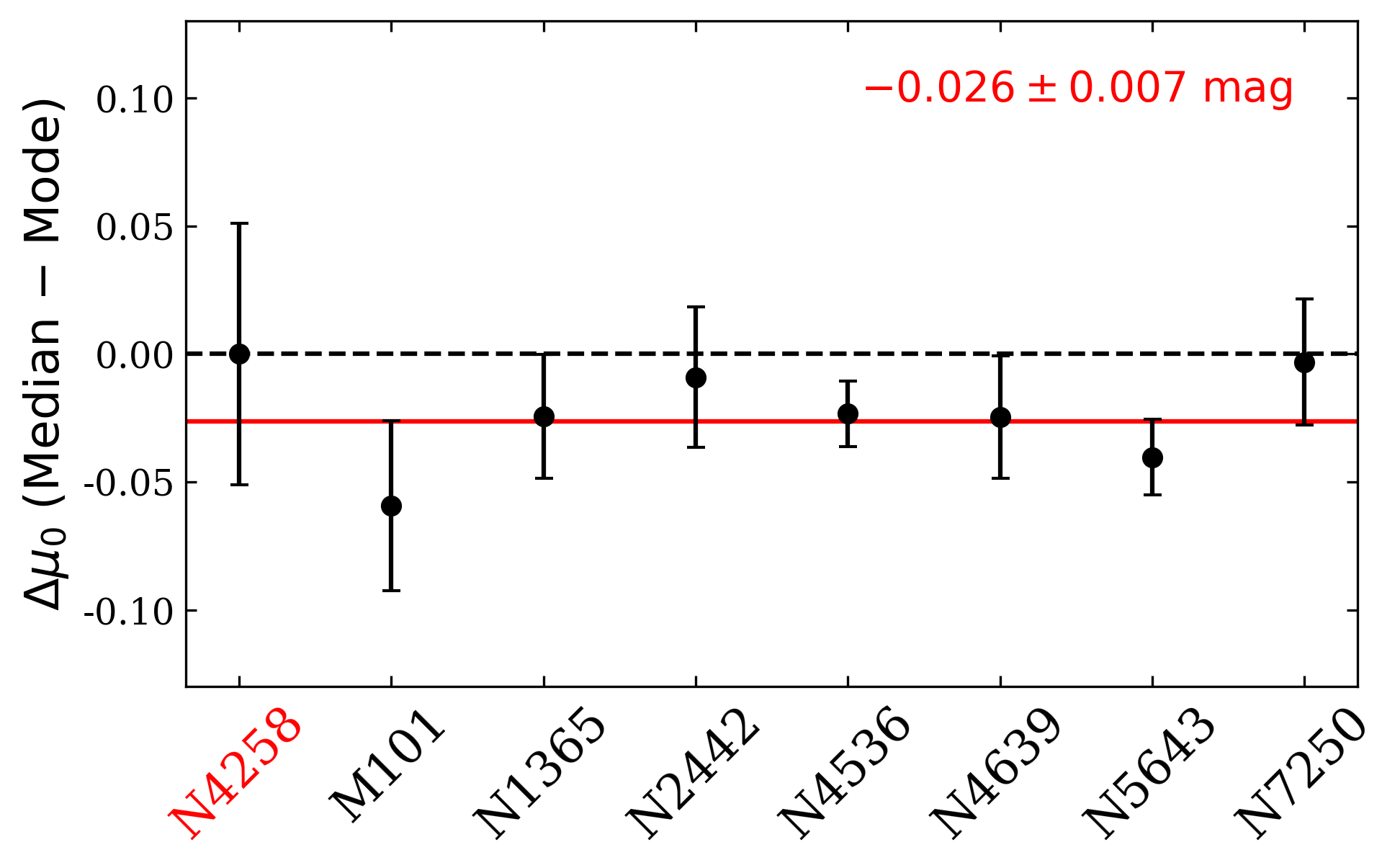}{0.5\textwidth}{}
          }
 \figurenum{B1}
\caption{Difference in the measured distance moduli between the mean vs. mode (left) and median vs. mode (right) for the seven SN Ia host galaxies. All three statistics were first calibrated in NGC 4258. We then measured the difference in distance moduli for the seven JAGB calibrator galaxies.  The average difference for all seven galaxies is shown in the upper right and is also marked by the red line. Both the mean and median delivered distances that were approximately 0.03~mag closer than distances measured using the mode. Error bars were calculated from adding the total statistical errors in quadrature. }\label{fig:mean_median}
\end{figure*}

To measure the mean and median JAGB magnitudes, we first selected JAGB stars within a magnitude range of $m_{JAGB}\pm0.75$~mag, where $m_{JAGB}$ was the measured mode. This corresponds to selecting stars approximately $\pm2\sigma$ from the mode. 
Then, we calibrated the JAGB method  using the  mean/median in NGC 4258, and measured distances to the seven SN Ia host galaxies. In Figure \ref{fig:mean_median}, we show the measured differences in distance moduli for if we used the mode versus the mean/median for our sample of galaxies.

Adopting the mean resulted in distance moduli that were $0.030\pm0.010$~mag closer than distance moduli measured using the mode. This corresponds to a 1.4\% larger $H_0$. 
Adopting the median resulted in distance moduli that were $0.026\pm0.007$~mag closer than distance moduli measured using the mode. This corresponds to a 1.2\% larger $H_0$.
 These systematic differences were fully encapsulated within
 our minimum adopted smoothing parameter error of $0.04$~mag because measuring the mode of an increasingly smoothed distribution will eventually converge to measuring the mean of that distribution, as discussed in Section \ref{subsubsec:statunc}.  Therefore, using the mode as the JAGB statistic is equivalent to using the mean/median as long as an appropriate smoothing parameter error is adopted. In conclusion, the differing shapes of the JAGB LFs in our sample propagate to a $\sim1.4$\% error to the final measured $H_0$, which we accounted for by adopting a minimum 2\% uncertainty in our error budget.

\end{document}